\documentclass[12pt]{article}

\usepackage[margin=2.5cm]{geometry}
\usepackage[T1]{fontenc}

\usepackage{amsmath}
\usepackage{
    accents, adjustbox, afterpage, amssymb, amsfonts, amsthm, array, booktabs, cancel, caption, color, comment, datetime, dcolumn, dsfont, eurosym, float, geometry, graphicx, lipsum, longtable, mathtools, mathbbol, multirow, pdflscape, pdfpages, ragged2e, rotating, setspace, subcaption, subfiles, textpos, threeparttable, soul, xcolor,
}

\usepackage[hang, flushmargin]{footmisc}

\usepackage{enumitem}
\usepackage{tikz}

\usepackage[authoryear]{natbib}
\setcitestyle{authoryear,aysep={},open={(},close={)}}
\bibpunct{(}{)}{;}{a}{,}{,}

\usepackage[hypertexnames=false,colorlinks]{hyperref}
\hypersetup{
    colorlinks,
    linkcolor={red!50!black},
    citecolor={blue!50!black},
    urlcolor={blue!50!black}
}

\usepackage{etoolbox}
\usepackage{xparse}
\NewDocumentCommand{\hyref}{m O{}O{}}{\hyperref[#1]{#2 \ref{#1}#3}}

\newdateformat{DDMonthYYYY}{%
  \THEDAY\, \monthname[\THEMONTH] \THEYEAR
  }

\DeclareMathOperator*{\argmax}{arg\,max}
\DeclareMathOperator*{\argmin}{arg\,min}
\DeclareMathOperator*{\supp}{supp}
\DeclareMathOperator*{\interior}{int}
\theoremstyle{plain}
\newtheorem{theorem}{Theorem}
\newtheorem{proposition}{Proposition}
\newtheorem{lemma}{Lemma}
\newtheorem{claim}{Claim}
\newtheorem{corollary}{Corollary}

\theoremstyle{definition}
\newtheorem{definition}{Definition}

\newtheorem{remark}{Remark}

\newcommand*\diff{\mathop{}\!\mathrm{d}}

\usepackage{dsfont}
\usepackage{fouriernc}
\usepackage{charter}
\usepackage{libertine}

\newcommand\blfootnote[1]{
  \begingroup
  \renewcommand\thefootnote{}\footnote{#1}
  \addtocounter{footnote}{-1}
  \endgroup
}

\usepackage[ruled,vlined]{algorithm2e}
\usepackage{bookmark}

\usepackage[sf,sl,outermarks]{titlesec}
\newcommand{\addperiod}[1]{#1.}

\titleformat{\section}[block]
{\normalfont\Large\bfseries}{\thesection.}{.5em}{\Large\bfseries}
\titlespacing*{\section}{0pt}{*1.3}{*0.2}

\titleformat{\subsection}[block]
{\normalfont\large\bfseries}{\thesubsection.}{.5em}{\large\bfseries}
\titlespacing*{\subsection}{0pt}{*1}{*0}

\titleformat{\subsubsection}[runin]
{\normalfont\bfseries}{}{0em}{\normalsize\bfseries\addperiod}
\titlespacing*{\subsubsection}{0pt}{*1}{*1}

\definecolor{mathematica1}{rgb}{0.368417, 0.506779, 0.709798}
\definecolor{mathematica2}{rgb}{0.880722, 0.611041, 0.142051}

\hypersetup{
    pdffitwindow=false,
    pdfstartview={XYZ null null 1.00},
    pdfnewwindow=true,
    colorlinks=true,
    linkcolor={red!50!black},
    citecolor={blue!50!black},
    urlcolor=black,
	linkbordercolor={white},
    pdfpagemode = UseNone,
}

\makeatletter
\DeclareRobustCommand\citepos													
  {\begingroup\def\NAT@nmfmt##1{{\NAT@up##1's}}%
   \NAT@swafalse\let\NAT@ctype\z@\NAT@partrue
   \@ifstar{\NAT@fulltrue\NAT@citetp}{\NAT@fullfalse\NAT@citetp}}

\pretocmd{\NAT@citex}{%
  \let\NAT@hyper@\NAT@hyper@citex
  \def\NAT@postnote{#2}%
  \setcounter{NAT@total@cites}{0}%
  \setcounter{NAT@count@cites}{0}%
  \forcsvlist{\stepcounter{NAT@total@cites}\@gobble}{#3}}{}{}
\newcounter{NAT@total@cites}
\newcounter{NAT@count@cites}
\def\NAT@postnote{}

\def\NAT@hyper@citex#1{
  \stepcounter{NAT@count@cites}%
  \hyper@natlinkstart{\@citeb\@extra@b@citeb}#1%
  \ifnumequal{\value{NAT@count@cites}}{\value{NAT@total@cites}}
    {\if*\NAT@postnote*\else\NAT@cmt\NAT@postnote\global\def\NAT@postnote{}\fi}{}%
  \ifNAT@swa\else\if\relax\NAT@date\relax
  \else\NAT@@close\global\let\NAT@nm\@empty\fi\fi								
  \hyper@natlinkend}
\renewcommand\hyper@natlinkbreak[2]{#1}

\patchcmd{\NAT@citex}															
  {\ifNAT@swa\else\if*#2*\else\NAT@cmt#2\fi
   \if\relax\NAT@date\relax\else\NAT@@close\fi\fi}{}{}{}
\patchcmd{\NAT@citex}
  {\if\relax\NAT@date\relax\NAT@def@citea\else\NAT@def@citea@close\fi}
  {\if\relax\NAT@date\relax\NAT@def@citea\else\NAT@def@citea@space\fi}{}{}
\patchcmd{\NAT@cite}{\if*#3*}{\if*\NAT@postnote*}{}{}
\makeatother

\setlist[enumerate]{leftmargin=*,wide=0pt,itemsep=0pt,topsep=2pt}
\AtBeginDocument{
  \setlength\abovedisplayskip{4pt}
  \setlength\belowdisplayskip{4pt}
}
\title{Sequential Sampling Equilibrium}
\author{Duarte Gonçalves}

\allowdisplaybreaks

\begin{document}
\thispagestyle{empty}
\setcounter{page}{0}

\setcounter{footnote}{0}
\renewcommand{\thefootnote}{\fnsymbol{footnote}}
~\vspace*{-2cm}\\
\begin{center}\Large
    {\noindent
    \bfseries  Sequential Sampling Equilibrium
    }
\end{center}
\vspace*{1em}

\makebox[\textwidth][c]{
    \begin{minipage}{1.2\linewidth}
        \Large\centering
        Duarte Gonçalves\footnotemark
    \end{minipage}
}
\setcounter{footnote}{1}\footnotetext{
    \setstretch{1} Department of Economics, University College London; \hyperlink{mailto:duarte.goncalves@ucl.ac.uk}{\color{black}duarte.goncalves@ucl.ac.uk}.
} 
\blfootnote{
    I am very grateful to Yeon-Koo Che, Mark Dean, and Navin Kartik for the continued encouragement and advice.
	I also thank 
	Larbi Alaoui, C\'{e}sar Barilla, Martin Cripps, Tommaso Denti, Teresa Esteban-Casanelles, Evan Friedman, Drew Fudenberg, Philippe Jehiel, Elliot Lipnowski, Qingmin Liu, Antonio Penta, Jacopo Perego, Philip Reny, Ariel Rubinstein, Evan Sadler, Ran Spiegler, Jakub Steiner, Yu Fu Wong,
	and the audiences at various seminars and conferences 
    for valuable feedback.\\
	\emph{First posted draft}: 25 November 2020. \emph{This draft}: \DDMonthYYYY\today.
}

\setcounter{footnote}{0} \renewcommand{\thefootnote}{\arabic{footnote}}

\begin{center} \textbf{\large Abstract} \end{center}
\noindent\makebox[\textwidth][c]{
    \begin{minipage}{.85\textwidth}
        \noindent
        This paper introduces an equilibrium framework based on sequential sampling in which players face strategic uncertainty over their opponents' behavior and acquire informative signals to resolve it.
        Sequential sampling equilibrium delivers a disciplined model featuring an endogenous distribution of choices, beliefs, and decision times, that not only rationalizes well-known deviations from Nash equilibrium, but also makes novel predictions supported by existing data.
        It grounds a relationship between empirical learning and strategic sophistication, and generates stochastic choice through randomness inherent to sampling, without relying on indifference or choice mistakes. 
        Further, it provides a rationale for Nash equilibrium when sampling costs vanish.
        ~
    \end{minipage}
}
\newpage

\section{Introduction}
\label{sec:introduction}

When faced with a choice, such as which TV show to watch, snack to buy, or savings account to open, decision-makers spend time and cognitive effort to resolve uncertainty about which alternative is best. 
This gives rise to a fundamental trade-off between \emph{speed}, how long it takes to choose, and \emph{accuracy}, the likelihood of choosing the best alternative. 

Rooted in \citeauthor{Wald1947Ecta}'s seminal work \citeyearpar{Wald1947Ecta}, sequential sampling emphasizes this trade-off between speed and accuracy.
It models the decision-maker's reasoning as sampling informative signals, reflecting the notion that individuals draw upon memory to guide their choices, even in novel settings---a premise supported by neurological evidence.\footnote{
    See e.g. \citet{ShadlenShohamy2016Neuron}, \citet{DuncanShohamy2020Ch}, and \citet{BidermanBakkourShohamy2020TrendsCognSci}.
} 
Initially adopted by cognitive sciences to explain reaction times in perception problems \citep{Ratcliff1978PsyRev,Luce1986Book}, sequential sampling stands as a cornerstone for understanding the relationship between choice and decision time in various domains, from consumer behavior to risky or intertemporal choice.\footnote{
    \citet{FehrRangel2011JEP}, \citet{KrajbichOudFehr2014AERPP}, \citet{Clithero2018JEPsy}, and \citet{SpiliopoulosOrtmann2018EE} provide reviews of the existing literature in economics. \citet{ForstmannRatcliffWagenmakers2016ARPsy} surveys the literature in cognitive sciences.
} 
Its success stems from its ability to rationalize established empirical regularities, particularly a \emph{time-revealed indifference}\footnote{
    Early evidence can be found in \citet{MostellerNogee1951JPE}; \citet{KrajbichArmelRangel2010NatNeuro}, \citet{Clithero2018JEBO}, and \citet{Alos-FerrerGaragnani2022JRU} provide more recent experimental evidence.
} whereby slower choices are associated with weaker preference intensity and greater choice randomness \citep{FudenbergStrackStrzalecki2018AER,Alos-FerrerFehrNetzer2022JPE}.

When making decisions in strategic settings like online auction bidding, protest participation decisions, or stock market transactions, individuals often grapple with uncertainty about others' behavior, and spend time and cognitive effort to address this strategic uncertainty.
And since the cost and benefit of reasoning may vary across situations, the time and effort committed to resolving this uncertainty is itself endogenous.
These observations resonate with experimental evidence from strategic environments: 
Existing work has shown that in dominance-solvable games, faster decisions are associated with less strategically sophisticated actions \citep{AgranovCaplinTergiman2015JESA,Rubinstein2016QJE,RecaldeRidelVesterlund2018JPubE}, and that stronger incentives entail longer decision times and greater strategic sophistication \citep{AlaouiPenta2016REStud,AlaouiJanezicPenta2020JET,Alos-FerrerBruckenmaier2021EE,Esteban-CasanellesGoncalves2020WP}. 
Moreover, in binary-action games longer decision times tend to be associated with indifference \citep{SchotterTrevino2021EE,FrydmanNunnari2023WP}.
While these findings agree with our understanding of individual decision-making, they are difficult to reconcile with existing equilibrium concepts.

This paper introduces an equilibrium framework based on sequential sampling in which players face strategic uncertainty and acquire informative signals to resolve it.
Players have a prior belief about others' distribution of actions and, before taking an action, they acquire costly signals about others' behavior.
These signals are assumed to be informative of others' behavior and therefore to have informational value.
Then, strategic uncertainty drives information acquisition and each player optimally trades-off the cost and benefit to sample.
As optimal sequential sampling renders players' action distributions dependent on their opponents', a \emph{sequential sampling equilibrium} corresponds to a fixed-point, a consistent distribution of actions of all players.
This delivers a disciplined model featuring an endogenous distribution of choices, beliefs, and decision times, 
that not only rationalizes empirical patterns relating choices and decision times and well-known deviations from Nash equilibrium, but also makes novel predictions supported by existing data. 
Moreover, it provides a rationale for Nash equilibrium as a limit case when sampling costs vanish.

The solution concept builds on an individual decision-making model of sequential sampling in a rich environment of choice under uncertainty.
Players effectively act as decision-makers: each takes as given others' uncertain behavior, characterized by an unknown distribution.
Sequential sampling serves as a stylized model of stepwise reasoning about others' behavior, occurring prior to making a decision.\footnote{
    See \citet{AlaouiPenta2022JPE} for an axiomatization of the value of stepwise reasoning as the value of sampling information, analogous to our model.
}
Each player faces an optimal stopping problem, trading-off informational gains and costs.
Upon stopping, players choose an action to maximize their expected payoff, given their posterior beliefs.
Optimal sequential sampling yields stochastic choice through the randomness inherent to sampling, without relying on indifference or choice mistakes.
The actions chosen upon stopping depend on posterior beliefs, informed by the sampled observations, whose distribution depends on others' behavior.
For simplicity, we focus on the case in which players can sample at a cost from their opponents' choice distribution and defer the discussion of more general information structures. 

A sequential sampling equilibrium determines an endogenous joint distribution of actions, beliefs, and stopping time.
Equilibrium emerges as a consistency condition on the distribution over players' actions arising from the fact that signals sampled are informative. 
We show a sequential sampling equilibrium always exists.
The proof follows from the novel observation that each player's optimal stopping time is uniformly bounded with respect to opponents' distribution of actions---which renders this into a computationally tractable finite-horizon problem. 
While our equilibrium definition is in terms of action distributions, a player's optimal sequential sampling together with the \emph{equilibrium} distribution of opponents' actions pins down an endogenous distribution of actions, beliefs, and stopping time.
Furthermore, as a player is uncertain about others' (equilibrium) distribution of actions, and the latter corresponds to but one of the possible action distributions the player entertains, equilibrium stopping time does not correspond to the players' expected costs of information.

Sequential sampling equilibrium provides a rationale for the relationship between higher incentives, longer decision times, and more sophisticated play.
While sequential sampling equilibrium does allow for players to choose non-rationalizable with positive probability, 
we prove that when scaling up players' payoffs, only $k$-rationalizable actions are chosen with positive probability at any equilibrium, where the order of rationalizability depends monotonically on the scaling factor.
This connection between empirical learning and strategic sophistication arises directly from the fact that higher payoffs induce longer decision times, leading players to sample sufficiently to learn and choose only k-rationalizable actions.

We then turn to binary action games and examine how relative payoffs influence the joint distribution of choices and decision time.
We establish comparative statics results for sequential sampling.
First, increasing the payoffs to a given action leads to that action being chosen more often and faster and the other less often and slower, a finding that generalizes beyond two-action settings.
Second, an increase in the underlying probability that an action is optimal leads to an analogous result.
These results allow us to prove two behavioral implications of sequential sampling equilibrium: that the frequency with which an action is chosen increases in its payoffs, and that the opponent chooses the best response to that action more often and \emph{faster}.
If the former corresponds to a well-documented deviation from Nash equilibrium in experimental literature \citep[e.g.][]{GoereeHolt2001AER}, the latter provides a novel prediction on how time relates to choice, which we find to be borne out by existing experimental evidence.

Sequential sampling equilibrium also has implications for players' equilibrium beliefs.
Experimental evidence has suggested that beliefs about others' behavior are often biased \citep{Costa-GomesWeizsacker2008REStud}, appear stochastic \citep{FriedmanWard2022WP}, and depend on own incentives even when others' behavior is held fixed \citep{Esteban-CasanellesGoncalves2020WP}.
All these patterns are implied by sequential sampling, where beliefs upon stopping will typically be \emph{biased} toward the prior, \emph{stochastic}, as they depend on the realized observations players sample, and \emph{payoff-dependent}, given these affect when players stop sampling.
To go beyond these general properties, we consider the case of Beta-distributed priors and prove that sequential sampling equilibrium provides a foundation for time-revealed indifference observed in binary action games.
Specifically, we show that the longer the decision time, the closer is the player to being indifferent between taking either action---a game-theoretic counterpart to \citet{FudenbergStrackStrzalecki2018AER}.
Furthermore, we uncover monotone comparative statics on how beliefs respond to payoff changes.
In a nutshell---and recalling that stopping beliefs are stochastic---when payoffs to a player's action increase, the opponent's equilibrium beliefs shift (in a first-order stochastic dominance) toward assigning a higher probability to that action being chosen.

Sequential sampling also provides a new rationale for Nash equilibrium, based on costly information acquisition.
While optimal stopping implies that conditional on stopping observations are neither independent nor identically distributed, it is possible to use martingale theory to show that sequential sampling equilibria nevertheless converge to Nash equilibria.
However, not all Nash equilibria can be reached through this approach: those involving weakly dominated actions cannot, while pure strategy Nash equilibria that don't involve such actions can.

We also explain how this solution concept can be seen as a steady state of a dynamic process. 
Specifically, we show how sequential sampling equilibria coincide with the steady states of the distribution of play of short-lived players who sequentially sample from data on past play, and obtain global asymptotic convergence results for a generic class of $2\times2$ games.
This parallels the role of Nash equilibria in scenarios where these short-lived players have frictionless access to the entirety of past data \citep{FudenbergKreps1993GEB,FudenbergLevine1998Book}.

Finally, we conclude with a discussion of variations to the model, including extensions to incomplete information games and more general information structures.
It is straightforward to adjust the solution concept to Bayesian games by having samples include information on the realized actions as well as the state.
This can be interpreted as making inferences about a specific context by also relying on information about behavior in similar settings.
An analogous result to that of convergence to Nash equilibrium ensues: limit points of Bayesian sequential sampling equilibria as sampling costs vanish are Bayesian Nash equilibria.
Furthermore, we provide an extension to more general information structures, accommodating situations in which recollections are noisy, some players' signals are silent about a subset of their opponents, or where in general players' ability to distinguish between opponents' actions or types is limited.
In the latter case, indistinguishable action profiles constitute an analogy class and it is shown that limit points of a sequence of equilibria with vanishing costs are analogy-based expectations equilibria \citep{Jehiel2005JET,JehielKoessler2008GEB}.

To summarize, sequential sampling equilibrium constitutes a flexible equilibrium framework for analyzing strategic interaction.
It provides a rationale for standard solution concepts, accounts for several behavioral patterns that have been documented in experiments, and makes novel predictions not just regarding choices that individuals make in strategic settings, but for timed-stochastic choice data, the joint distribution of choices, beliefs, and decision times.

\subsection{Related Literature}
\label{sec:introduction:literature}
This paper is related to two broad literatures: sequential sampling and information acquisition in games.

\subsubsection*{Sequential Sampling}
The study of optimal sequential sampling can be traced back to the seminal works of \citet{Wald1947Ecta} and \citet{ArrowBlackwellGirshick1949Ecta}.
Sequential sampling has since been used as a modeling device in cognitive psychology and neuroscience to ground a relation between choice and decision time,\footnote{
    The classic reference is \citet{Ratcliff1978PsyRev}. 
    See \citet{RatcliffSmithBrownMcKoon2016TrendsCognSci} and \citet{ForstmannRatcliffWagenmakers2016ARPsy} for a review of the literature and \citet{KrajbichLuCamererRangel2012FrontiersPsy,SpiliopoulosOrtmann2018EE,Clithero2018JEPsy,ChiongShumWebbChen2020WP} for economic applications.
}
and, in particular, to model choice based on memory retrieval \citep{GoldShadlen2007ARN,ShadlenShohamy2016Neuron,DuncanShohamy2020Ch,BidermanBakkourShohamy2020TrendsCognSci}.
\citet{AlaouiPenta2022JPE} provide an axiomatic foundation of sequential sampling as a representation of iterative of reasoning.
\citet{FudenbergStrackStrzalecki2018AER} consider a binary-action problem in which a decision-maker sequentially acquires information the payoff difference.
They show that at longer stopping times, the agent is closer to being indifferent between the two actions.\footnote{
    A related literature on optimal sequential information acquisition studies the dynamic \emph{choice} of information, be it deciding about its intensity \citep{MoscariniSmith2001Ecta}, selecting across sources \citep{CheMierendorff2019AER,LiangMuSyrgkanis2022Ecta}, or choosing it in a fully flexible manner \citep{SteinerStewartMatejka2017Ecta,Zhong2022Ecta}.
}
\citet{Alos-FerrerFehrNetzer2022JPE} examine the general relation between time-revealed indifference and stochastic choice primitives.

The individual decision-making framework of sequential sampling motivated the experimental study of decision times in games.
This led to establishing a number of regularities, such as a positive association between decision times and the strategic sophistication of actions chosen---as given by level-$k$ model \citep{Nagel1995AER,StahlWilson1995GEB} or the highest level of $k$-rationalizability---in dominance-solvable games \citep[see e.g.][]{AgranovCaplinTergiman2015JESA,Rubinstein2016QJE,AlaouiJanezicPenta2020JET,Alos-FerrerBruckenmaier2021EE,GillProwse2023EJ}, and that scaling up incentive levels causally increases decision times and leads to more sophisticated play \citep{Esteban-CasanellesGoncalves2020WP}.
Additionally, as in individual decision-making, response times also reveal indifference in global games \citep{SchotterTrevino2021EE,FrydmanNunnari2023WP}.

Sequential sampling equilibrium adopts sequential sampling to model belief formation in strategic settings, providing a relation between stopping time, beliefs, and choices.
It contributes to this theoretical literature with novel results in problems with multiple available actions: comparative statics results on how choices and stopping time relate to payoffs in general decision-problems with arbitrary payoff correlation across actions.
It rationalizes the relation between incentives, decision times, and strategic sophistication of choices that has been documented in the experimental literature,
Further, in the binary-action case that has been the focus of much of the literature, our model relates the true data generating process to the distribution of choices, stopping times, and posterior beliefs, and obtain a time-revealed indifference prediction in a tractable discrete-time environment, which we show bears out in \citepos{FriedmanWard2022WP} data.

\subsubsection*{Information Acquisition in Games}
There is a growing literature on equilibrium solution concepts featuring information acquisition.
\citet{OsborneRubinstein2003GEB} suppose each player observes a fixed number of samples from their opponents' equilibrium distribution of actions, and the mapping from samples to actions is exogenously specified.
\citet{SalantCherry2020Ecta} study a special case of this solution concept in mean-field games with binary actions, while keeping the sampling procedure exogeneous: players employ unbiased estimators and best-respond to the obtained estimate.\footnote{
    Related are solution concepts with noisy but unbiased beliefs, e.g. \citet{FriedmanMezzetti2005GEB,Friedman2022AEJMicro}.
}
\citet{OsborneRubinstein1998AER} examine a similar notion of equilibrium, where players receive a fixed number of samples from the payoffs of each of their actions and choose the action with the highest average payoff in the sample.
More broadly, these correspond to a form of self-confirming equilibrium \citep{FudenbergLevine1993Ecta,BattigalliGilliMolinari1992RE} in which the feedback function is fixed.

In contrast to these, sequential sampling equilibrium (1) endogenizes the sampling and (2) adds a time dimension via its sequential nature, enabling results regarding the joint distribution of stopping time and choices that cannot be captured with exogenous sampling.
If the latter provides the basis for the time-revealed indifference, the former grounds the relationship between payoffs, decision times, and the level of strategic sophistication (in the sense of $k$-rationalizability) of the actions chosen in equilibrium.

The literature also studied solution concepts with \emph{costly} information acquisition.
\citet{Yang2015JET} examines a coordination game in which players acquire flexible but costly information about an exogenous payoff-relevant parameter.
As in much of the rational inattention literature \citep{Sims2003JME,MatejkaMcKay2015AER}, the cost of information is given by prior entropy reduction.
\citet{MatejkaMcKay2012AERPP} and \citet{Martin2017GEB} study pricing games with a similar approach. 
\citet{Denti2022TE} allows for players to obtain correlated information under general information cost \citep[as in][]{CaplinDean2015AER}.
\citet{HebertLaO2020NBERWP} study this solution concept in mean-field games.

Our paper provides the first solution concept in which the cost of information acquisition is experimental \citep{DentiMarinacciRustichini2022AER}, with information acquisition corresponding to costly sequentially sampling from an information structure.
While the sequential information acquisition can be studied from a static, ex-ante perspective \citep{MorrisStrack2019WP,BloedelZhong2021WP,HebertWoodford2022WP}, there are two conceptual features distinguishing sequential sampling equilibrium---beyond, of course, results specific to stopping time.
First, in these papers players hold beliefs and can learn about their opponents' \emph{action realizations}. 
Second, players' equilibrium beliefs are correct, and so, absent uncertainty about exogenous parameters, equilibria correspond to Nash equilibria of the underlying normal-form game.

In our framework, players are uncertain---neither correct or incorrect---about the prevailing \emph{distribution} of actions of their opponents.
While actions yet to be taken are not learnable, a prevailing stable distribution of opponents' actions is.
Further, if the choice of an information structure by our players bears a cost proportional to the expected stopping time, our analysis speaks to the joint distribution of stopping times and choices as determined in equilibrium---noting that the equilibrium stopping time is distributed not by the measure induced by players' prior beliefs, but by that arising from the equilibrium distribution of their opponents' actions (which has measure zero according to their prior).

Finally, we comment on the relation to the work on learning with misspecification, chiefly \citepos{EspondaPouzo2016Ecta} Berk-Nash equilibrium.
This solution concept allows for general forms of misspecification of the players' prior beliefs and is not restricted to either normal-form or complete information games.
There, players best-respond to their equilibrium beliefs, those in the support of players' priors that minimize the Kullback--Leibler divergence to equilibrium gameplay, which can be taken as arising as the limit case of Bayesian learning with potentially misspecified priors \citep[see][]{FudenbergLanzaniStrack2021Ecta}.

\section{Sequential Sampling Equilibrium}
\label{sec:sse}

\subsection{Setup}
\label{sec:sse:setup}
\subsubsection*{Preliminaries}
Let $\Gamma=\langle I,A,u\rangle$ denote a normal-form game, where
$I$ denotes a finite set of players or roles, with generic elements $i, j$ and where $-i$ denotes $I\setminus i$;
$A:=\times_{i \in I} A_i$, where $A_i$ is $i$'s finite set of feasible actions;
and $u:={(u_i)}_{i\in I}$, with $u_i:A_i \times \Delta(A_{-i}) \to \mathbb R$ denoting player $i$'s payoff function, where $u_i$ is continuous, $\Delta(A_{-i})$ being endowed with the Euclidean norm.\footnote{
    We will use $\|\cdot\|_p$ to denote the $p$-norm and $\|\cdot\|_\infty$ for the sup-norm.
}
We extend $u_i$ to the space of probability distributions over actions with $u_i(\sigma_i,\sigma_{-i})= \mathbb E_{\sigma_i} [u_i(a_i,\sigma_{-i})]$, where $\mathbb E_{\sigma_i}[\cdot]$ corresponds to the expectation taken with respect to $\sigma_i$.
While we focus throughout on normal-form games, having payoffs directly depend on opponents' distribution of actions will render proofs readily adaptable to Bayesian games, a setting to which our framework extends naturally as discussed in \hyref{sec:extensions}[Section].

\subsubsection*{Beliefs}
In contrast to other solution concepts, each player $i$ is uncertain about others' true distribution of actions, $\sigma_{-i}$. 
While this implies uncertainty about the actions that others ultimately take, $a_{-i}$, the main conceptual difference is that, if others' actions are only observable after these are taken, players may still reason and learn about the prevailing stable action distribution prior to others choosing an action---e.g. the likelihood a restaurant is too crowded, the probability that others abstain in an election, or the distribution of prices of a product across different platforms.
Such uncertainty captures the players' limited experience and imperfect memory. 
Expressing this uncertainty, each player $i$ holds beliefs about $\sigma_{-i}$, given by a Borel probability measure $\mu_i \in \Delta(\Delta(A_{-i}))$, where $\Delta(\Delta(A_{-i}))$ is endowed with the topology of weak$^*$ convergence, metricized by L\'{e}vy-Prokhorov metric $\|\cdot\|_{LP}$.
We require player $i$'s beliefs to have as support, $\supp(\mu_i)$, the set of all distributions, \emph{allowing for correlation}---$\supp(\mu_i)=\Delta(A_{-i})$---or the set of all distributions \emph{assuming independence} across opponents, in which case beliefs are given by a product measure $\mu_i=\times_{j \in -i}\mu_{ij}$, where each $\mu_{ij}$ is a probability measure on $\Delta(A_j)$ with full support.
Results will hold in either unless explicitly mentioned.\footnote{
    It is also possible to extend this framework to accommodate other cases potentially of interest, e.g. ruling out opponents play strictly dominated actions; we omit these cases to simplify the presentation.
}

\subsubsection*{Information and Sequential Sampling}
We model a player's costly reasoning about others' behavior via a sequential sampling stage that takes place prior to choice, as in sequential sampling models used to describe reasoning in individual decision-making settings \citep{FudenbergStrackStrzalecki2018AER,ForstmannRatcliffWagenmakers2016ARPsy}.
That is, prior to making a choice, player $i$ can acquire signals about the unknown distribution $\sigma_{-i}$ in a sequential and costly manner.
Sequential sampling then captures the recollection of past experiences in similar situations (others' behavior, payoffs to tried out actions), asking friends, or more general stepwise reasoning processes---see \citet{AlaouiPenta2022JPE} for an axiomatic foundation of stepwise reasoning in this guise.
Alternatively, it can be taken in more literal fashion, with players acquiring or parsing through existing data.

A key assumption is that this reasoning is \emph{informative} about others' behavior, that is, that each player $i$ has access to an information structure $\pi_i:\Delta(A_{-i}) \to \Delta(Y_i)$, where $Y_i$ is a finite signal space.
Throughout the main text, we restrict attention to the case in which these signals are observations drawn from $\sigma_{-i}$, i.e. $\pi_i$ corresponds to the identity.
More general signal structures are considered in \hyref{appendix:information-structures}.

As mentioned, information acquisition is sequential.
In other words, prior to taking an action, player $i$ can sequentially observe signals $y_{i,t}$ and decide when to stop sampling.
Players' \emph{stopping time} will be interpreted as their decision time, and thus take a prominent role in our endeavor to ground the relation between incentives, choices, and decision time in strategic settings.
The sequentiality in sampling marks an important distinction relative to other models of sampling in games \citep[e.g.][]{OsborneRubinstein1998AER,OsborneRubinstein2003GEB}, allowing rich joint distributions of actions and decision time to emerge, in which particular actions can be associated with lengthier decision times, and others with shorter.

We will write $y_i^t={(y_{i,\ell})}_{\ell\in [1\,..\,t]}$ to stand for the sample path up to time $t$, where $[n\,..\,n+k]=\{n,n+1,...,n+k\}$ and each realization $y_{i,\ell}$ is distributed according to $\sigma_{-i}$, with the understanding that $y_i^0 = \emptyset$.
Formally, we denote $\mathbf y_i =\left\{y_{i,t}\right\}_{t \in \mathbb N}$ as a stochastic process defined on the probability space $(\Omega, \mathcal F, \mathbb P)$ with $\mathbb F_i$ denoting the natural filtration of $\mathbf y_i$.
The set of sample paths of length $t$ is denoted by $Y_i^t$ and the set of all finite sample path realizations is denoted by $\mathcal Y_i:=\bigcup_{t \in \mathbb N} Y_{i}^t$.
Upon observing a given sample path up to time $t$, $y_i^t$, player $i$ updates beliefs about $\sigma_{-i}$ according to Bayes' rule, denoted by $\mu_i |y_i^t$.\footnote{
    Note that $\mu_i$ induces a measure on $\Delta(A_{-i}) \times \mathcal Y_i$.
}

\subsubsection*{Sampling Costs}
Naturally, sampling is costly, capturing the effort involved in reasoning.
For convenience, we will throughout assume that player $i$'s cost of each observation is given by $c_i>0$.
It is straightforward to adjust the model in order to accommodate costs that depend on the number of observations, insofar as they are eventually bounded away from zero from below,\footnote{
    \label{footnote:cost-restrictions}
    Formally: there is some $N$ and $\underline c_i>0$ such that player $i$'s cost for any observation following the $N$-th is greater than $\underline c_i$.
}
which, for all purposes, subsumes cases in which there is an upper bound on the number of observations.

\subsubsection*{Extended Games}
An \emph{extended game} $G$ is then a tuple comprising an underlying normal-form game $\Gamma$, each players' prior beliefs $\mu={(\mu_i)}_{i\in I}$, and sampling costs $c={(c_i)}_{i\in I}$.

\subsection{Equilibrium}
\label{sec:sse:equilibrium}
Having introduced all the primitives of the model, we now turn to defining equilibrium.

\subsubsection*{Choice}
Given a belief $\mu_i'\in \Delta(\Delta(A_{-i}))$, player $i$ upon stopping sampling chooses an action in order to maximize their expected utility.
We denote the player's maximized utility by $v_{i}:\Delta(\Delta(A_{-i}))\to \mathbb R$
\[v_{i}(\mu_i'):=\max_{\sigma_i \in \Delta(A_i)}\mathbb E_{\sigma_i}[\mathbb E_{\mu_i'}[u_i(a_i,\sigma_{-i})]],\]
where $\mathbb E_{\sigma_i}[\mathbb E_{\mu_i'}[u_i(a_i,\sigma_{-i})]]=\int_{A_i}\int_{\Delta(A_{-i})}u_i(a_i,a_{-i})\diff\mu_i'(\sigma_{-i})\diff\sigma_i(a_i)$. 
We write $\sigma_i^*:\Delta(A_{-i})\to \Delta(A_i)$ to denote a selection of optimal choices given beliefs, $\sigma_i^*(\mu_i')\in \argmax_{\sigma_i \in \Delta(A_i)}\mathbb E_{\mu_i}[u_i(\sigma_i,\sigma_{-i}')]$.

\subsubsection*{Optimal Stopping}
Player $i$ samples optimally in order to maximize expected payoffs.
That is, each player $i$ faces an optimal stopping problem:
based on the expected value of future reasoning, decide whether to stop and make a choice or obtain another signal.
Formally, player $i$ chooses a \emph{stopping time} $t_i$ in the set $\mathbb T_i$ of all stopping times taking values in $\mathbb N_0\cup \{\infty\}$ and adapted with respect to natural filtration associated to $\mathbf y_i$.

Given a prior $\mu_i \in \Delta(\Delta(A_{-i}))$, player $i$'s value function $V_i:\Delta(\Delta(A_{-i}))\to \overline{\mathbb R}$ can be written as
\begin{align*}
    V_{i}(\mu_i):=\sup_{t_i \in \mathbb T_i}\mathbb E_{\mu_i}[v_{i}(\mu_i\mid y_i^{t_i})- c_i \cdot t_i],
\end{align*}
where $\mu_i\mid y_i^{t_i}$ denotes the player's posterior belief when, upon stopping according to stopping time $t_i$, the sample $y_i^{t_i}$ was observed.

It will be useful to consider the dynamic programming formulation of the optimal stopping problem, with $V_i$ corresponding to a fixed point of an operator $B_i:C_b(\Delta(\Delta(A_{-i})))\to C_b(\Delta(\Delta(A_{-i})))$, 
\[
    B_i(\tilde V_i)(\mu_i')=\max\{v_i(\mu_i'),\mathbb E_{\mu_i'}[\tilde V_i(\mu_i'\mid y)]-c_i\},
\]
which will be equivalent for our purposes. 
This lends our value function a clear interpretation:
\begin{align*}
    \underbrace{V_{i}(\mu_i')}_{\text{value at belief }\mu_i'}=\max\{\underbrace{v_{i}(\mu_i')}_{\text{value of stopping}}\,,\,\underbrace{\mathbb E_{\mu_i}[V_{i}(\mu_i'\mid y_i)]-c_i}_{\text{expected value of continuing sampling}}\}.
\end{align*}

We focus on the earliest optimal stopping time
\[
    \tau_i(\omega):=\min\{t \in \mathbb N_0 \mid V_i(\mu_i'\mid y_i^t(\omega))=v_i(\mu_i'\mid y_i^t(\omega))\},
\]
where its optimality follows by standard arguments \citep[Ch. 3, Theorem 3]{Ferguson2008}; while omitted, we note the dependence of $\tau_i$ on the prior $\mu_i'$.

For ease of reference, we summarize properties of optimal sequential sampling in this proposition: 
\begin{proposition}
    \label{proposition:properties-value-stopping}
    The following properties hold:
    (1) $v_i$ and $V_i$ are bounded, convex, and uniformly continuous.
    (2) For any prior $\mu_i$, player $i$'s optimal stopping time is finite $\mu_i$-a.s. and satisfies
    $\mathbb P_{\mu_i}(\tau>T)\leq 2 \|u_i\|_\infty/c_i T$.
\end{proposition}
This and the remaining omitted proofs are in \hyperref[appendix:proofs:proposition:properties-value-stopping]{Appendix A}.

\subsubsection*{Equilibrium Definition}
We now close the model by considering equilibrium behavior.
Players' strategic uncertainty about others' action distribution motivates their sequential sampling behavior.
The randomness inherent to sampling entails randomness in players' optimal actions given their posterior beliefs.
Equilibrium is then a fixed point, a consistency condition based on the premise that each player's signals are \emph{informative} about others' distribution of actions.

Formally, each player acquires information on their opponents' action distribution and their optimal stopping policy, $\tau_i$, determines the sequences of signals following which they optimally stop at take an action:
\[
        \mathcal Y_{i}^{\tau_i}:=\left\{
            y_i^t \in \mathcal Y_i:
            V_i(\mu_i\mid y_i^t)=v_i(\mu_i\mid y_i^t) 
            \text{ and } 
            \forall 0\leq \ell<t, V_i(\mu_i\mid y_i^\ell)>v_i(\mu_i\mid y_i^\ell)
        \right\}.
    \]
Different opponent action distributions $\sigma_{-i}$ induce different distributions over the signals acquired $y_i^{\tau_i}$.
Since different sequences of signals $y_i^{\tau_i}$ induce different posteriors $\mu_i\mid y_i^{\tau_i}$ at which different actions $a_i$ may be optimal, sequential sampling implies a mapping from opponents' action distributions to the player's distribution of actions, 
\[\mathbb E_{\sigma_{-i}}[\sigma_i^*(\mu_i\mid y_i^{\tau_i})]
=
\underbrace{\sum_{y_i^t \in \mathcal Y_{i}^{\tau_i}}}_{
    \substack{
        \text{set of stopping }\\
        \text{sequences }\vphantom{y_i^t}
    }
}
\underbrace{\prod_{\ell \in[1\,..\,t]}\sigma_{-i}(y_{i,\ell})\vphantom{\sum_{y_i^t \in Y_{i,\tau_i}}}}_{
    \substack{
        \text{probability }\\
        \text{observing }y_i^t
    }
}
\underbrace{\sigma_i^*(\mu_i\mid y_i^t)\vphantom{\sum_{y_i^t \in Y_{i,\tau_i}}}}_{
    \substack{
        \text{best response }\\
        \text{at posterior }\mu_i\mid y_i^t
    }
}
\]
That is, the probability of player $i$ taking action $a_i$ is given by 
the probability of taking such an action once player $i$ after observing $y_i^t$, $\sigma_i^*(\mu_i\mid y_i^t)$,
considering every sequence of signals $y_i^t$ following which player $i$ optimally stops, $y_i^t \in \mathcal Y_i^{\tau_i}$, 
and weighting it by the probability of its occurrence.
The probability that a sequence of signals $y_i^t$ is observed---
--- that is, the true probability of optimally stopping following $y_i^t$, $\mathbb P_{\sigma_{-i}}(y_i^{\tau_i}=y_i^t)$---is then given by $1_{y_i^t \in \mathcal Y_i^{\tau_i}}\prod_{\ell \in [1\,..\,t]}\sigma_{-i}(y_{i,\ell})$, as each observation corresponds to an action profile $y_{i,\ell} \in A_{-i}$, sampled independently from $i$'s opponents' action distribution, $\sigma_{-i}$.

Equilibrium then follows as a consistency condition between players' action distributions:
\begin{definition}
    \label{definition:equilibrium}
    A sequential sampling equilibrium of an extended game $G=\langle \Gamma, \mu,c\rangle$ is a profile of action distributions $\sigma$ such that, for every $i$, 
    $\sigma_i=\mathbb E_{\sigma_{-i}}[\sigma_i^*(\mu_i\mid y_i^{\tau_i})],$ 
    where $\tau_i$ is player $i$'s earliest optimal stopping time
    and $\sigma_i^*(\mu_i')$ is optimal given belief $\mu_i'$.
\end{definition}

\subsubsection*{Equilibrium Stopping Time}
It is important to emphasize that a sequential sampling equilibrium implies a novel relationship between actions and time.
While a sequential sampling equilibrium is defined in the space of action distributions, note that, in a given extended game, the equilibrium distribution of opponents' actions $\sigma_{-i}$ completely pins down the joint distribution of choices and (optimal) stopping time for player $i$ ($a_i, \tau_i$).
Such joint distribution is determined \emph{in equilibrium}, as it crucially depends on the true (equilibrium) action distribution of player $i$'s opponents about which player $i$ is reasoning.
Although one could think about $c\mathbb E_{\mu_i}[\tau_i]$ as a static cost of information and rephrase our equilibrium notion from a static viewpoint \citep[e.g.][]{Yang2015JET,Denti2022TE,HebertLaO2020NBERWP}, a major distinctive feature of this model relative to equilibrium models of costly information acquisition in games is that it enables one to speak not of the subjectively expected cost of information acquisition, but of the joint distribution of realized actions and time.
As also mentioned earlier, a second difference is that uncertainty here refers to distribution of actions, and therefore it does not require---but can accommodate---exogenous sources of uncertainty.

\subsubsection*{Interpretation}
Sequential sampling equilibrium can be interpreted as positing that, prior to taking an action, players reason through others' behavior to better ground their choices.
In this sense, players' sequential sampling reflects an underlying introspective process whereby players reason about how others may act by reaching back in their memory and past experiences.
This interpretation is motivated by literature in cognitive science \citep{ShadlenShohamy2016Neuron}, which has made use of sequential sampling models to ground the relation between time and choices not only in association problems and perceptual tasks,\footnote{
    See \citet{Ratcliff1978PsyRev} for a pioneering study of the use of sequential sampling models in cognitive sciences and \citet{ForstmannRatcliffWagenmakers2016ARPsy,RatcliffSmithBrownMcKoon2016TrendsCognSci} for recent review articles.
} but also in domains where choices are guided by individual preferences \citep[e.g.][]{Clithero2018JEBO}, with patterns being consistent across these different domains \citep{SmithKrajbich2021PsyBullRev}.
Recent literature provided neurological evidence of memory guiding preference-based choice, conforming with sampling from memory \citep[see][]{BakkourZylberbergShadlenShohamy2018WP,DuncanShohamy2020Ch,BidermanBakkourShohamy2020TrendsCognSci,BidermanShohamy2021NatComm}.

A second, complementary and more literal, interpretation of sequential sampling is to see it as acquiring hard information---such as data, experts' opinions, or reviews.
For instance, a seller doing market research to better price its product, consumers parsing reviews on a product's quality, voters learning about candidates' platforms through their statements about different issues, or infrequent bidders in online auctions looking at data from other past auctions to reason how to bid.

At the heart of our model is the assumption that sampling is informative about others' behavior, so as to render it valuable.
Equilibrium delivers this consistency by positing stationarity of the environment---\hyref{sec:sse:learning-foundation}[Section] illustrates how it coincides with steady states of a particular dynamic process.
Further, it also allows us to obtain comparative statics predictions on how payoffs affect players' choices both by changing the trade-offs in players' information acquisition decisions and the signal realizations.
Naturally, stationarity of the environment is a strong assumption that may be unwarranted in situations where non-equilibrium is more apt to describe behavior---see \citet{AlaouiPenta2016REStud} for one such model.

In order to obtain sharper predictions, we will focus on a simple form of sequential sampling.
Players' sampling is then represented in a stylized manner, with signals directly sampled from the prevailing distribution of actions of opponents.
In \hyref{sec:extensions}[Section] we discuss how sequential sampling equilibrium can be easily extended to richer settings and information structures---e.g. noisy recollections, inability to learn about some players',  or allowing signals about others' behavior (or types) in one setting to inform players in the situation at hand.

Sequential sampling equilibrium also provides a particular way to relax the implicit epistemic assumption in Nash equilibrium that, in equilibrium, players come to know their opponents' distribution of actions.
If players' priors did assign probability one to the same Nash equilibrium of the underlying game, that Nash equilibrium will coincide with a sequential sampling equilibrium of the game.\footnote{
    As it is implicit in this statement, even though beliefs are degenerate and coincide on the same Nash equilibrium, not all best responses need to coincide with that same Nash equilibrium, which explains why there may be multiple sequential sampling equilibria instead of there being a unique equilibrium coinciding with the Nash equilibrium players believe to occur.
    Such non-uniqueness can occur even when the game has a unique Nash equilibrium,
    echoing \citepos{AumannBrandenburger1995ECTA} results on the epistemic characterization of Nash equilibrium, whereby conjectures---and not choices---are found to coincide with Nash equilibrium.
}
In our model, however, players are uncertain about the prevailing distribution of actions, and it is this uncertainty that drives their sequential sampling behavior.
Further, it dispenses with the assumption of mutual knowledge of the game and of others' rationality, since all learning is driven by the procured information and players need not know others' payoff functions.

\subsubsection*{Existence}
We briefly note that a sequential sampling equilibrium exists in all extended games.
\begin{theorem}
    \label{theorem:existence}
    Every extended game has a sequential sampling equilibrium.
\end{theorem}
The proof proceeds by verifying that, for every player $i$, $\sigma_{-i}\mapsto b_i(\sigma_{-i}):=\mathbb E_{\sigma_{-i}}[\sigma_i^*(\mu_i\mid y_i^{\tau_i})]$ (1) maps to a well-defined probability distribution of player $i$'s actions, and (2) such mapping is continuous.
The main difficulty is that, while we know that, by \hyref{proposition:properties-value-stopping}[Proposition], $\tau_i$ is finite with probability 1 \emph{with respect to the player's prior} ($\mathbb P_{\mu_i}(\tau_i<\infty)=1$), 
we need player $i$'s optimal stopping time to be finite with probability 1 \emph{with respect to the actual distribution of opponents' actions}, ($\mathbb P_{\sigma_{-i}}(\tau_i<\infty)=1$), as otherwise $b_i(\sigma_{-i})\notin \Delta(A_i)$.
If player $i$ never stops sampling with positive probability (with respect to the true distribution of opponents' actions), then $b_i$ does not define a probability distribution over player $i$'s actions and no equilibrium exists.\footnote{
    We provide an example in \hyref{appendix:misspecification}[Online] to illustrate the potential non-existence of equilibria when priors do not have full support.
}
The following lemma demonstrates that this condition on stopping time is also sufficient to guarantee the desired properties on $b_i$:
\begin{lemma}
    \label{lemma:stopping-implies-continuity}
    The following two statements are equivalent: 
    (1) player $i$'s optimal stopping time is finite with probability 1 with respect to $\sigma_{-i}$, $\mathbb P_{\sigma_{-i}}(\tau_i<\infty)=1$ for any $\sigma_{-i}\in \Delta(A_{-i})$; (2) $b_i(\sigma_{-i})\in \Delta(A_i)$ $\forall \sigma_{-i}\in \Delta(A_{-i})$.
    Moreover, if (1) holds, then $b_i$ is also continuous.
\end{lemma}
The proof can be found in \hyperref[proposition:convergence-dynamic-ss]{Appendix A}.
With \hyref{lemma:stopping-implies-continuity}[Lemma] in hand, it is then straightforward to show existence of a sequential sampling equilibrium.
\begin{proof}
    Given \hyref{lemma:stopping-implies-continuity}[Lemma], if, for any $\sigma_{-i}$, $\tau_i$ is finite with probability one with respect to $\sigma_{-i}$, then $b_i$ is a continuous mapping from $\Delta(A_{-i})$ to $\Delta(A_i)$, and existence follows from Brouwer's fixed point theorem.
    By assumption, $\text{supp}(\mu_i)=\Delta(A_{-i})$ and, for any $\sigma_{-i}$,
    \[
        \mathbb P_{\sigma_{-i}}(\tau_i(\omega)\leq T)=\mathbb P_{\sigma_{-i}}(\left\{\omega:\,\inf\{t \mid \mathbb E_{\mu_{i,t}(\omega)}[V_i(\mu_{i,t}(\omega)|y_{i,t+1}]-V_i(\mu_{i,t}(\omega))\leq c_i\}\leq T\right\} ).
    \]
    As $V_i$ is uniformly continuous, there is $\delta>0$ such that, $\forall \mu_i,\mu_i' \in \Delta(P_i)$ satisfying $\|\mu_i-\mu_i'\|_{LP}<\delta$, $|V_i(\mu_i)-V_i(\mu_i')|<c$.
    Since each observation is subjectively iid, by \citet{Berk1966AnnMathStat}, $\mu_{i,t}$ weak$^*$ converges to a Dirac on $\sigma_{-i}$, $\sigma_{-i}$-a.s.,
    $\mathbb P_{\sigma_{-i}}(\lim_{t \to \infty}\mathbb E_{\mu_{i,t}}[V_i(\mu_{i,t}|y_{i,t+1}]-V_i(\mu_{i,t})> c_i )=0$.
\end{proof}

In fact, we can obtain an upper bound on the stopping time by combining uniform continuity of $V_i$ and the fact that $\mu_i$ uniformly accumulates around the empirical frequency:
\begin{remark} 
    \label{remark:bounded-stopping}
    For every player $i$, $\exists \overline T_i<\infty$ such that $\tau_i\leq \overline T_i$, where $\overline T_i$ depends on $u_i$, $\mu_i$, and $c_i$.
\end{remark}
This transforms optimal stopping into a finite horizon problem, a useful result that not only simplifies the analysis, but also makes our solution concept amenable to computational applications.

\section{Behavioral Implications}
\label{sec:implications}
In this section, we characterize different behavioral implications of sequential sampling equilibrium.
First, we explore the relation between stopping time and action sophistication.
Then, we will relate incentives to the joint distribution of choices and stopping time.
Finally, we focus on players' beliefs and their relation with stopping time.

\subsection{Rationality and Sequential Sampling}
\label{sec:implications:rationalizability}

Existing evidence points toward an association between longer decision times and choices reflecting greater sophistication \citep[e.g.][]{AgranovCaplinTergiman2015JESA,Rubinstein2016QJE,Alos-FerrerBruckenmaier2021EE}, which may express heterogeneity in individual costs of reasoning.
However, when facing higher stakes, decision times increase first-order stochastic dominance sense, and choices do reflect higher sophistication, as given by their level of rationalizability \citep{Esteban-CasanellesGoncalves2020WP}.
Existing benchmark models like level-$k$ and cognitive hierarchy \citep{Nagel1995AER,CamererHoChong2004QJE} are unable to deliver such comparative statics, since choices are invariant with respect to incentive levels.\footnote{
    A notable exception is the non-equilibrium model by \citet{AlaouiPenta2016REStud}, which endogenizes level-$k$ via a cost-benefit analysis.
    Differently from our model, though, it predicts degenerate stopping times.
}

This section shows how sequential sampling equilibrium can provide a rationale for such an association by relating higher incentives, longer decision times, and a lower bound on the level of rationalizability of action chosen in equilibrium.
Further, this establishes a relation between empirical learning (as given by sequential sampling) and introspective learning (as given by rationalizability).

We first observe that, in our context, higher incentives as given by scaling up a player's payoffs, is equivalent to scaling down the cost to sampling.
Optimal sequential sampling naturally predicts an inverse relation between sampling cost and stopping time:
\begin{remark}
    \label{remark:stopping-costs}
    For lower sampling costs $c_i$, player $i$'s optimal stopping time increases in first-order stochastic dominance with respect to the prior $\mu_i$ and to any true distribution of opponents' actions $\sigma_{-i}\in \Delta(A_{-i})$; that is, both $\mathbb P_{\mu_{i}}(\tau_i\leq t)$ and $\mathbb P_{\sigma_{-i}}(\tau_i\leq t)$ increase for any $t$.
\end{remark}

Our main result of this section goes further in determining a relation between cost and the level of sophistication of actions chosen in equilibrium.
Let us recall the definition of rationalizable actions.
\begin{definition}
    \label{definition:rationalizability}
    An action $a_i\in A_i$ is \emph{1-rationalizable} if there is some $\sigma_{-i}\in \Delta(A_{-i})$ such that, $\forall \sigma_i \in \Delta(A_i)$, $u_i(a_i,\sigma_{-i})\geq u_i(\sigma_i,\sigma_{-i})$.
    An action $a_i \in A_i$ is \emph{$k$-rationalizable}, for $k\geq 2$, if there is some $\sigma_{-i}\in \Delta(A_{-i}^{k-1})$ such that, $\forall \sigma_i \in \Delta(A_i)$, $u_i(a_i,\sigma_{-i})\geq u_i(\sigma_i,\sigma_{-i})$, where $A_{-i}^{k-1}:=\times_{j \ne i}A_j^{k-1}$ denotes the set of $(k-1)$-rationalizable action profiles of player $i$'s opponents.
    An action $a_i$ is \emph{rationalizable} if $a_i \in \cap_{k \in \mathbb N}A_i^k$.
\end{definition}
For presentation purposes---as implied by the above definition---we focus on a definition of rationalizability allowing for correlation among opponents' actions, and require priors to have full support on $\Delta(A_{-i})$.
The below result holds as well when considering a definition of rationalizable actions that requires independence across opponents' action distributions, provided beliefs also do not allow for correlation.

We now show that scaling up incentives enough---or, equivalently, for low enough sampling costs---only $k$-rationalizable actions are chosen at sequential sampling equilibria:
\begin{theorem}
    \label{theorem:rationalizability}
    For any normal-form game $\Gamma$, priors $\mu$, and $k \in \mathbb N$, there are cost thresholds $\overline c^k_i>0$ such that, for any extended game $G=\langle \Gamma, \mu,c\rangle$ in which $c_i< \overline c^k_i$ for all $i$, in any sequential sampling equilibrium $\sigma$ of $G$ only $k$-rationalizable actions are chosen with positive probability.
\end{theorem}

The result is obtained by combining three observations---the proofs for which can be found in \hyperref[appendix:proofs:lemma:rationalizable-best-response]{Appendix A}.
First, if player $i$ believes that, with high enough probability, their opponents only choose $(k-1)$-rationalizable actions, then player $i$ will choose a $k$-rationalizable action:
\begin{lemma}
    \label{lemma:rationalizable-best-response}
    For any $k\geq 2$, there are $\epsilon,\delta>0$, such that, if $\mu_i(B_{\delta}(\Delta(A_{-i}^{k-1}))>1-\epsilon$, then \\
    $\argmax_{a_i \in A_i}\mathbb E_{\mu_i}[u_i(a_i,\sigma_{-i})]\subseteq A_i^k$.
\end{lemma}

Second, that if player $i$'s opponents do indeed only choose $(k-1)$-rationalizable actions, then player $i$'s beliefs uniformly accumulate on the event that opponents only choose $(k-1)$-rationalizable actions:
\begin{lemma}
    \label{lemma:uniform-concentration-k-rationalizable}
    For any $\mu_i \in \Delta(\Delta(A_{-i}))$ with full support, and all $\epsilon,\delta>0$, there is $t$ such that, for any sequence of observations $y_i^t$ for which $y_{i,\ell}\in A_{-i}^{k-1}$ for $\ell \in [1\,..\,t]$, $\mu_i\mid y_i^t(B_\delta(\Delta(A_{-i}^{k-1})))>1-\epsilon$.
\end{lemma}

And third, that, when not all of player $i$'s actions are rationalizable, it suffices that sampling costs are low enough to ensure that the player acquires a minimum number of signals:
\begin{lemma}
    \label{lemma:lower-bound-stopping}
    Suppose that there is no action $a_i$ that is a best response to all distribution of opponents' actions $\sigma_{-i} \in \Delta(A_{-i})$. 
    Then, for any $T\in \mathbb N_0$ and any full support prior $\mu_i \in \Delta(A_{-i})$, 
    there is $\overline c_i>0$ such that for any sampling cost $c_i< \overline c_i$, the associated earliest optimal stopping time $\tau_i\geq T+1$.
\end{lemma}

The proof of \hyref{theorem:rationalizability}[Theorem] then proceeds easily:
\begin{proof}
    The proof follows an induction argument.
    First, observe that no player will choose actions that are not 1-rationalizable.
    Now, for $k\geq 1$, assume that players choose only $(k-1)$-rationalizable actions with positive probability.
    From \hyref{lemma:uniform-concentration-k-rationalizable}[Lemma], for any $\delta,\epsilon>0$ there is a $T$ such that, for all $t\geq T$, all $i\in I$, and any $y_i^t \in A_{-i}^{k-1}$, $\mu_{i,t}(B_\delta(\Delta(A_{-i}^{k-1})))\geq \mu_{i,t}(B_\delta(\delta_{\overline y_i^t}))>1-\epsilon$.
    By \hyref{lemma:rationalizable-best-response}[Lemma], this implies that if all players sample for at least $T$ periods, they will only choose $k$-rationalizable actions with positive probability.
    \hyref{lemma:lower-bound-stopping}[Lemma] ensures that we can find $c^k>0$ such that, if $c_i\leq c^k$ $\forall i$, all players sample at least $T$ periods, i.e. that each player's earliest optimal stopping time is bounded below by $T$, $\tau_i\geq T$.
    This concludes the proof.
\end{proof}

\begin{remark}
    It is possible to generalize the result to classes of priors that satisfy a condition akin to a lower bound on density:
    \begin{definition}[\citealt{DiaconisFreedman1990AnnStat}]
        Let $\phi:\mathbb R_{++}\to \mathbb R_{++}$.
        The set of $\phi$-positive distributions on $\Delta(A_{-i})$ is given by $\mathcal M_i(\phi):=\{\mu_i \in \Delta(\Delta(A_{-i}))\mid \inf_{\sigma_{-i} \in \Delta(A_{-i})}\mu_i(B_\epsilon(\sigma_{-i}))\geq \phi(\epsilon),\, \forall \epsilon>0\}$.
    \end{definition}
    Since it is possible to obtain a uniform rate of accumulation around the empirical mean for any prior $\mu_i \in \mathcal M_i(\phi)$ that depends only on $\phi$, we can then extend \hyref{theorem:rationalizability}[Theorem] so that the same cost thresholds holds for all $\phi$-positive priors $\mu_i$.
    Further, these cost thresholds $c_i^k$ can be made tight and non-increasing in $k$ by considering all priors within such class.
\end{remark}

In short, \hyref{theorem:rationalizability}[Theorem] uncovers a relationship between sampling costs and degree of sophistication of equilibrium choices.
Since scaling up payoffs is tantamount to scaling down costs, providing---via sequential sampling---a rationale for the link from incentive levels to action sophistication, beliefs about opponents, and decision times documenetd in dominance-solvable games e.g. in \citet{Esteban-CasanellesGoncalves2020WP}.

\subsection{Comparative Statics in Binary Action Games}
\label{sec:implications:comparative-statics-actions}

\subsubsection*{Binary Action Games} 
In this section we provide comparative statics results for binary actions games: normal-form games $\Gamma=\langle I,A,u\rangle$ with two players, $|I|=2$, and such that each player $i \in I$ has two actions $A_i=\{0,1\}$ and $u_i(1,\sigma_{-i})-u_i(0,\sigma_{-i})$ is strictly monotone and continuous in $\sigma_{-i}$, the probability that $i$'s opponent chooses action 1.\footnote{
    Monotonicity in $\sigma_{-i}$ is automatically satisfied when $u_i(a_i,\sigma_{-i})=\mathbb E_{\sigma_{-i}}[u_i(a_i,a_{-i})]$.
    We require strict monotonicity to prevent the case in which players are always indifferent between both actions ($u_i(1,\sigma_{-i})=u_i(0,\sigma_{-i})$, $\forall \sigma_{-i}$), which is a trivial case.
    Since we will allude to extensions that may require non-linearity in $\sigma_{-i}$, we impose only minimal conditions on payoffs.
}
Consequently, we identify $\mu_i \in \Delta(\Delta(A_{-i}))$ with a distribution on the unit interval.
An extended binary action game $G$ is an extended game for which $\Gamma$ is a binary action game.

\subsubsection*{Actions and Stopping Time}
Our object of interest will be the probability (according to $\sigma_{-i}$) that player $i$ stops before time $t$ and, upon stopping, action $a_i$ is optimal, that is,
\[\mathbb P_{\sigma_{j}}(a_i \in A_i^*(\mu_i\mid y_i^{\tau_i}) \text{ and }\tau_i\leq t),\]
where $A_i^*(\mu_i'):=\argmax_{a_i \in A_i}\mathbb E_{\mu_i'}[u_i(a_i,\sigma_{-i}')]$ denotes the set of optimal choices at a given belief $\mu_i'$.
Our main result characterizes how the joint distribution of player $i$'s choices and their stopping times changes along to three dimensions: (1) the player's payoffs, $u_i$, (2) their beliefs, $\mu_i$, and (3) the true (unknown) distribution of their opponent's actions, $\sigma_{-i}$, taken as exogenous.

\subsubsection*{Ordering Payoffs and Beliefs}
Let us introduce a partial order on player $i$'s utility functions:
\begin{definition}
    Let $u_i,u_i' : A_i\times \Delta(A_{-i})\to \mathbb R$.
    $u_i'$ is said to has higher incentives to action $a_i$ than $u_i$, $u_i'\geq_{a_i} u_i$, if and only if there is $g:\Delta(A_{-i})\to \mathbb R_{+}$ such that $u_i'(a_i',\sigma_{-i}') = u_i(a_i',\sigma_{-i}') + 1_{a_i'=a_i}g(\sigma_{-i})$.
\end{definition}

Beliefs are ordered according to a generalized version of the monotone likelihood ratio property \citep[cf.][]{LehrerWang2020WP}:
\begin{definition}
    Let $\mu_i,\mu_i' \in \Delta([0,1])$.
    $\mu_i'$ is said to strongly stochastic dominate $\mu_i$, $\mu_i'\geq_{SSD} \mu_i$,
    if 
    $\mu_i'\mid y_i^t$ first-order stochastically dominates $\mu_i\mid y_i^t$ for any $y_i^t \in \mathcal Y_i$.
\end{definition}
Note that, when $\mu_i$ and $\mu_i'$ are mutually absolutely continuous, $\geq_{SSD}$ corresponds to the monotone likelihood ratio property, i.e. $d\mu_i'/d\mu_i$ is increasing $\mu_i'$-a.e.

\subsubsection*{Monotone Comparative Statics}
The next result characterizes the behavior induced by optimal sequential information acquisition taking $\sigma_{-i}$ are exogenous:
\begin{theorem}
    \label{theorem:monotonicity-actions}
    Let $G$ be an extended binary action game and let $a_i\in \argmax_{a_i' \in A_i}u_i(a_i',1)$.
    Then,
    $\mathbb P_{\sigma_{-i}}(a_i \in A_i^*(\mu_i\mid y_i^{\tau_i}) \text{ and }\tau_i\leq t)$
    increases 
    (i) in $u_i$ with respect to $\geq_{a_i}$,
    (ii) in $\mu_i$ with respect $\geq_{SSD}$, and
    (iii) in $\sigma_{-i}$.
    Moreover, it is $\mathcal C^\infty$ in $\sigma_{-i}$.
\end{theorem}
Let us discuss the intuition behind the theorem (the proof is deferred to the \hyperref[appendix:proofs:theorem:monotonicity-actions]{Appendix A}).

Claim (i) shows that increasing the payoff associated to action $a_i$, $u_i'\geq_{a_i}u_i$, makes the player not only more likely to take that action under the true distribution of actions of the opponent, but to take it \emph{faster} and to choose the other action less often and \emph{slower}.
While an increase in payoffs does increase the value of sampling at some posterior beliefs---which could lead the player to learn more about the true $\sigma_{-i}$ and find out that perhaps action $a_i$ is not optimal after all---this additional information acquisition occurs only when before the player was stopping and taking an action other than $a_i$.
In other words, player $i$ requires now less information to be convinced to stop and take action $a_i$ and more information to stop and choose another action.
This result is not particular to binary action games: claim (i) is shown for general settings with arbitrary finitely many actions and general payoff functions.\footnote{
    See \hyref{proposition:monotonicity-actions-payoffs}[Proposition] in the \hyref{appendix:proofs}.
}

Claim (ii) can be interpreted as stating that player $i$ is more likely to stop earlier and take action $a_i$ the greater the probability their prior assigns to action $a_i$ being optimal.
The main difficulty is again to show that this seemingly tautological statement holds with respect to the actual, unknown, distribution of the opponent's actions; importantly, note the claim does not depend on whether or how correct player $i$'s beliefs are.
Such monotonicity in beliefs allows one to make predictions on how behavior changes with, for instance, the provision of information that shifts beliefs in the stochastic dominance order (e.g. $\mu_i|1\geq_{SSD}\mu_i|0$).

Finally, the argument for why claim (iii) should hold is straightforward: higher $\sigma_{-i}$ means that player $i$ is more likely to observe higher signals and therefore becoming convinced that action $a_i$ is the better alternative.
The proof follows from claim (ii) and an induction argument.
The fact that the probability of action $a_i$ being optimal when stopping before time $t$ is a polynomial with respect to $\sigma_{-i}$ implies the claim on differentiability.

\hyref{theorem:monotonicity-actions}[Theorem] provides comparative statics on the optimality of a given action, but leaves open the possibility that more than one action is optimal.
The next lemma closes this gap by showing that, in binary action games, a player is never indifferent between the two actions at any belief held upon stopping, provided the player samples at least once or is not indifferent under the prior $\mu_i$.
\begin{lemma}
    \label{lemma:never-stop-indifferent}
    Let $G$ be an extended binary action game.
    Then, for any player $i$,
    $A_i^*(\mu_i\mid y_i^{\tau_i})$ is a singleton or $\tau_i=0$.
    Moreover, if $\tau_i>0$,
    $A_i^*(\mu_i\mid y_i^{\tau_i})= \argmax_{a_i \in A_i}u_i(a_i,y_{i,\tau_i})$.
\end{lemma}

The reasoning underlying the proof is simple.
Without loss of generality, assume that player $i$'s best-response to $a_{-i}$ is to choose $a_i=a_{-i}$.
Suppose that player $i$ stops sampling after observing a 0-valued signal leaving player $i$ indifferent between the two actions (the argument is symmetric if the last signal is 1-valued).
Then, before sampling the last observation, action 1 was already optimal under player $i$'s prior, as observing a 0-valued observation induces a lower belief mean.
Moreover, if the last observation had instead realized to be 1-valued, player $i$ would still want to choose action 1.
This implies that if player $i$ stops sampling when indifferent between the two actions, whichever action was optimal before taking the last signal is still optimal regardless of the realization of the signal.
Therefore, given that the player will not sample any further, the last signal has no informational value.
As the signal is costly, it is suboptimal to take it.

\subsubsection*{Applications}
One immediate implication of \hyref{theorem:monotonicity-actions}[Theorem] and \hyref{lemma:never-stop-indifferent}[Lemma] is in establishing a strong connection between uniqueness of sequential sampling equilibrium in an extended game and uniqueness of a Nash equilibrium of the underlying binary action game:
\begin{proposition}
    \label{proposition:unique-sse-ne}
    A binary action game $\Gamma$ has a unique Nash equilibrium if and only if 
    any extended game $G=\langle \Gamma, \mu,c\rangle$ in which players with no weakly dominant actions sample at least once there is a unique sequential sampling equilibrium.
\end{proposition}
An analogous result holds when, in symmetric extended binary action games (same payoff functions, same prior, same sampling cost), one restricts to symmetric Nash equilibria and symmetric sequential sampling equilibria.
While uniqueness of a Nash equilibrium implies uniqueness of a sequential sampling equilibrium, it is not the case that the two coincide.

\begin{figure}[tp]
    \setstretch{1.25}
    \centering
    \begin{tabular}{lc|cc}
        \multicolumn{2}{c}{~} &\multicolumn{2}{c}{{\color{mathematica2}$C$}lasher}\\
        & & \color{mathematica2}$a$ & \color{mathematica2}$b$ \\
        \cline{2-4}
        \multirow{2}{*}{{\color{mathematica1}$M$}atcher}
        & \color{mathematica1}$a$ & {\color{mathematica1}$\delta_M$} , {\color{mathematica2}0} & {\color{mathematica1}0} , {\color{mathematica2}1} \\
        & \color{mathematica1}$b$ & {\color{mathematica1}0} , {\color{mathematica2}$\gamma_C$} & {\color{mathematica1}1} , {\color{mathematica2}0} 
    \end{tabular}
    \caption{Generalized Matching Pennies}
    \label{figure:matching-pennies-matrix}
    \begin{minipage}{.5\textwidth}
    \footnotesize{
        Note: $\delta_M,\gamma_C>0$.
    }
    \end{minipage}
\end{figure}

A well-known and counter-intuitive prediction of Nash equilibrium pertains to generalized matching pennies, that is, $2 \times 2$ games with a unique Nash equilibrium in fully mixed strategies, whose structure is illustrated in \hyref{figure:matching-pennies-matrix}[Figure].
When the payoffs to action $a_i$ of player $i$ increase, Nash equilibrium predicts that the probability with which action $a_i$ is chosen remains the same and it is, instead, the opponent's mixed strategy that adjusts to make player $i$ indifferent between choosing any of the two actions---what one could call the \emph{opponent-payoff choice effect}.
However, experimental evidence shows that increasing player $i$'s payoffs to an action leads that player to choose that action more often, an \emph{own-payoff choice effect}.\footnote{
    This finding has been replicated several times, namely by \citet{Ochs1995GEB}, \citet{McKelveyPalfreyWeber2000JEBO} and \citet{GoereeHolt2001AER}.\label{footnote:opce}
}
This motivated the emergence of different models, one of the most successful of which quantal response equilibrium \citep{McKelveyPalfrey1995GEB}, which directly embeds monotonicity of choices with respect to payoffs in the assumptions for players' behavior \citep{GoereeHoltPalfrey2005EE}.

Sequential sampling equilibrium not only rationalizes this empirical regularity via comparative statics pertaining to behavior induced by optimal information acquisition, it delivers novel behavior implications regarding stopping times.
Increasing player $i$'s payoffs to action $a_i$, (1) increases the \emph{equilibrium} probability that player $i$ chooses action $a_i$, and (2) leads their opponent, player $j$, in equilibrium, choosing the best response to action $a_i$ more often and faster, and their other action less often and slower, in the sense of \hyref{theorem:monotonicity-actions}[Theorem].
If the first observation states sequential sampling equilibrium predicts the own-payoff choice effect,\footnote{
    A similar result holds in my model with respect to symmetric anti-coordination (extended) games.
    In such case, the unique symmetric sequential sampling equilibrium exhibits the own-payoff effect under the same conditions as in generalized matching pennies.
    This matches gameplay patterns documented in experimental settings by \citet{ChierchiaNagelCoricelli2018SciRep} in the context of symmetric two-player anti-coordination games.
} the second uncovers an entirely novel prediction relating equilibrium choices and stopping time.
Both follow directly from combining \hyref{proposition:unique-sse-ne}[Proposition], \hyref{theorem:monotonicity-actions}[Theorem], and \hyref{lemma:never-stop-indifferent}[Lemma].

\subsubsection*{Supporting Evidence}
To investigate whether these predictions find support in existing data, we rely on experimental data generously made available by \citet{FriedmanWard2022WP} who collected data on choices and decision times for six different generalized matching pennies games.
The goal of this exercise is not to fit data or claim that sequential sampling equilibrium perfectly describes subjects' behavior or that it does so better than other existing models, but rather to present suggestive evidence supporting its novel behavioral implications.
No feedback or information was provided throughout the experiment; details on the experiment, the data, and further analysis can be found in \hyref{appendix:experimental-analysis}.

As shown in \hyref{figure:figure-clasher-time-elicitbeliefs0}[Figure], if one is to interpret stopping time as a proxy for decision time, the data supports our predictions: when increasing $\delta_M$ subjects in the Clasher's role do tend to choose action $b$ not only more often but also faster.
Moreover, they choose action $a$ less often and slower.

\begin{figure}[t]
    \centering
    \includegraphics[width=.8\linewidth]{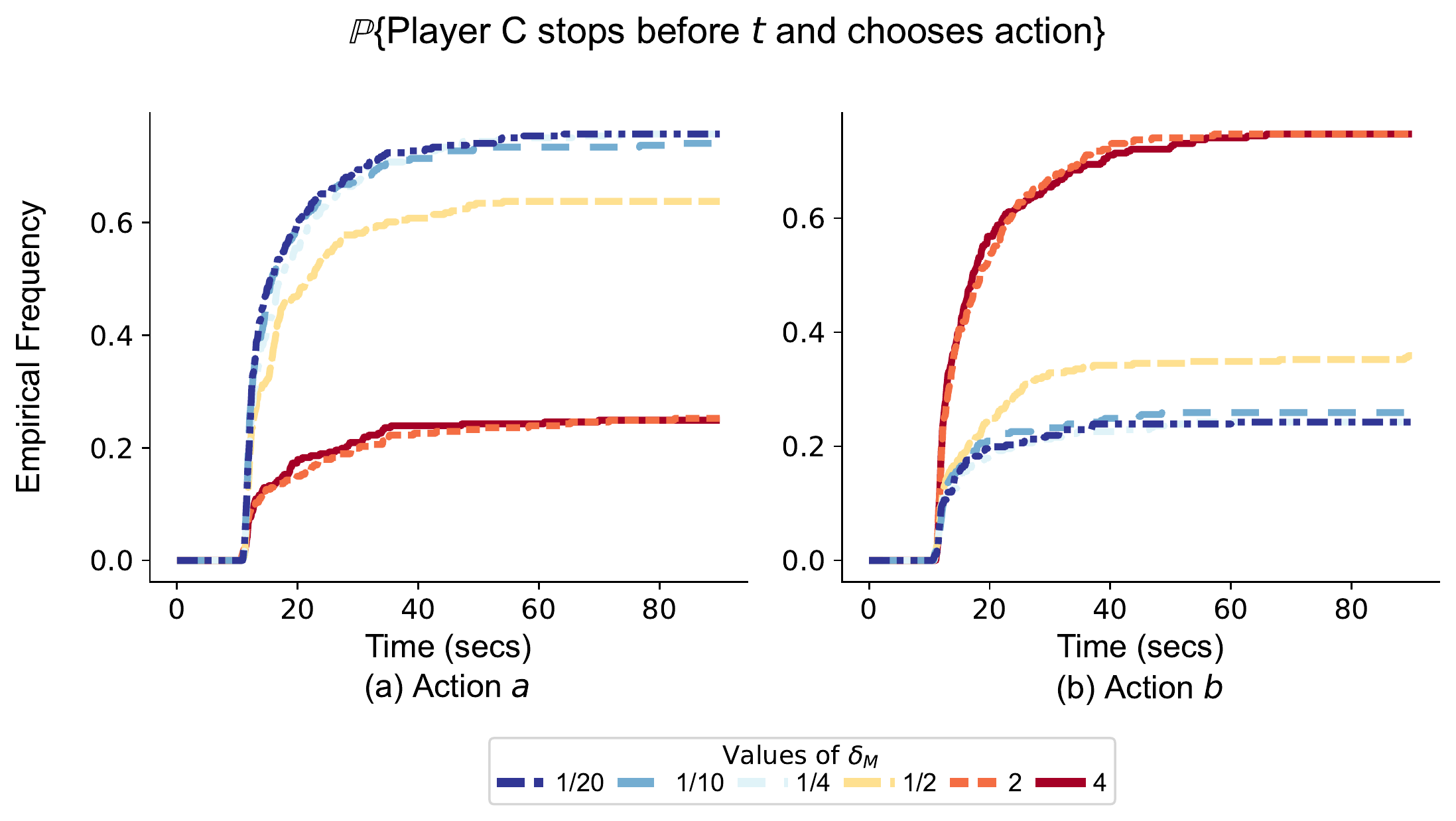}
    \begin{minipage}{1\linewidth}
        \small
        \vspace*{.5em}
        \caption{Opponent-Payoff Time Effect}
        \label{figure:figure-clasher-time-elicitbeliefs0}
        \vspace*{-1em}
        \singlespacing \emph{Notes}: 
        The figure compares choices and decision times in generalized matching pennies games as given in \hyref{figure:matching-pennies-matrix}[Figure], for $\gamma_C=1$ (and scaled by 20).
        The data is from \citet{FriedmanWard2022WP}.
        The panels exhibit the frequency with which subjects in the player $C$'s role take 
        a given action ($a$ in panel (a); $b$ in panel (b)) before time $t$ (in seconds).
        Different lines correspond to games in which the player $M$ has different payoffs to action $a$.
        This figure uses only choice data for instances where beliefs were not elicited.
        The same patterns are present when beliefs are elicited.
        See \hyref{appendix:experimental-analysis} for further details on the data.
    \end{minipage}
\end{figure}

\subsection{Time-Revealed Preference Intensity}
\label{sec:implications:time-revealed-intensity}
In this section we characterize how stopping time relates to players' posterior beliefs by considering a general family of priors in binary action games.
For this section, we restrict attention to games in which payoffs are linear in the opponent's distribution of actions, i.e. $u_i(a_i,\sigma_{-i})=\mathbb E_{\sigma_{-i}}[u_i(a_i,a_{-i})]$.

\subsubsection*{Beta Beliefs}
For tractability, we consider priors that are linear in new information in a manner that mimics Bayesian updating for Gaussian priors:
\begin{definition}
    \label{definition:linear-info}
    A prior $\mu_i$ is said to be {linear in the accumulated information} if it is non-degenerate and there are constants $a_t,b_t \in \mathbb R$ such that for any $y_i^t\in \mathcal Y_i$ the posterior mean satisfies $\mathbb E_{\mu_i}[\sigma_{-i} \mid y_i^t]=a_t \sum_{\ell=1}^t y_{i,\ell}^t + b_t$.
\end{definition}

This property, together with the fact that beliefs are a martingale and some algebraic manipulation, allows us to write the posterior mean as a convex combination of the prior mean and the empirical mean of the accumulated information, $\mathbb E_{\mu_i}[\sigma_{-i} \mid y_i^t]= \alpha_t/t\cdot \sum_{\ell=1}^t y_{i,\ell}^t + (1-\alpha_t)\cdot \mathbb E_{\mu_i}[\sigma_{-i}]$, where $\alpha_t/t=1/({(1-\alpha_1)/\alpha_1+t}) \in (0,1)$.
This is extremely convenient as, by linearity of expected utility, one can then analyze optimal stopping just relying on the belief mean and the number of samples.
In fact, as shown by \citet[Theorem 5]{DiaconisYlvisaker1979AnnStats}, identifies a specific parametric class of priors: a prior $\mu_i$ is linear in the accumulated information if and only if it is a Beta distribution.

\subsubsection*{Collapsing Boundaries}
When beliefs are linear in the accumulated information, we have the following characterization of the set of beliefs at which player $i$ optimally stops:
\begin{proposition}
    \label{proposition:collapsing-bounds}
    Let $\Gamma$ be a binary action game.
    Suppose that there is $\tilde \sigma_{-i}\in \Delta(A_{-i})$ such that $u_i(1,\tilde \sigma_{-i})=u_i(0,\tilde \sigma_{-i})$.
    For any $c_i>0$, there are continuous functions $\overline \sigma_{-i},\underline \sigma_{-i}:\mathbb R_{++}\to [0,1]$ such 
    for any Beta distributed prior $\mu_i$ with parameters $(\alpha,\beta)$
    player $i$ does not optimally stop at $\mu_i$ if and only if
    $\mathbb E_{\mu_i}[\sigma_{-i}]\in (\underline \sigma_{-i}(\alpha+\beta),\overline \sigma_{-i}(\alpha+\beta))$.
    Furthermore,
    $\overline \sigma_{-i}$ is decreasing and $\underline \sigma_{-i}$ is increasing, and $\exists T_i$ such that $\forall t\geq T_i$ $\overline \sigma_{-i}(t)=\underline \sigma_{-i}(t)=\tilde\sigma_{-i}$.
\end{proposition}
The proof of the result is in \hyperref[appendix:proofs:proposition:collapsing-bounds]{Appendix A}.

\hyref{proposition:collapsing-bounds}[Proposition] shows that when beliefs are linear in accumulated information, it is sufficient to consider the posterior mean to characterize the beliefs at which player $i$ continues sampling at any given moment as is illustrated in \hyref{figure:figure-beta-stopping-paper}[Figure].
Note that if $\mu_i$ is a Beta distribution with parameters summing to $t$, then $\mu_i\mid y_i$ has parameters summing to $t+1$.
The continuation region is then characterized by an upper and lower threshold that delimit a decreasing interval that ``collapses'' to a single point: the distribution at which player $i$ is indifferent between either action.
This translates to our setting what is commonly known in the neuroscience literature as ``collapsing boundaries''.\footnote{
    See \citet{HawkinsForstmannWagenmakersRatcliffBrown2015JNeuro} for a discussion on the evidence of collapsing boundaries and \citet{Bhui2019CBB} for supporting experimental evidence in an environment in which, as in our model, there is uncertainty about the difference in the binary actions' expected payoffs.
}

One can then interpret the stopping time as an indicator of the intensity of player $i$'s preference for one action over another:
player $i$ samples for longer if and only if the player is sufficiently close to being indifferent between the two alternatives, a phenomenon that resembles existing experimental evidence in individual decision-making \citep[e.g.][]{KonovalovKrajbich2019JDM}.
In other words, \hyref{proposition:collapsing-bounds}[Proposition] entails a behavior marker in the form of time-revealed preference intensity, akin to results in \citet{Alos-FerrerFehrNetzer2022JPE}.

\begin{figure}[t]
    \centering
    \includegraphics[width=.65\linewidth]{Figures/figure-beta-stopping-paper.pdf}
    \begin{minipage}{1\linewidth}
        \small
        \vspace*{.5em}
        \caption{Stopping Regions for Beta Priors}
        \label{figure:figure-beta-stopping-paper}
        \vspace*{-1em}
        \singlespacing \emph{Notes}: 
        The figure exhibits the continuation region (shaded area) and the stopping thresholds (darker blue lines) for posterior means at which player $i$ with a Beta stops.
        The figure also illustrates the possible realizations of the sampling process for a player with a uniform prior (Beta(1,1)), with the posterior means indicated by circles.
    \end{minipage}
\end{figure}

\begin{table}[!thp]\setstretch{1.1}
	\vspace*{-2em}
    \centering
	\small
    {\setstretch{1.0}\hspace*{-1em}\begin{tabular}{l@{\extracolsep{4pt}}ccc@{}}
\hline\hline
&\multicolumn{3}{c}{Distance to Indifference} \\
                  &       Player M &       Player C &               Both \\
                  \cline{2-2} \cline{3-3} \cline{4-4}
& (1) & (2) & (3) \\
\hline
Log Decision Time \hspace*{.25em} & -3.682$^{***}$ &  -2.021$^{**}$ &     -2.961$^{***}$ \\
                  &        (1.225) &        (0.881) &            (0.790) \\ [.25em]
        Intercept & 42.314$^{***}$ & 45.365$^{***}$ &     48.185$^{***}$ \\
                  &        (4.181) &        (2.670) &            (2.435) \\[3pt]
Fixed Effects     &           Game &           Game & Role $\times$ Game \\

\hline
        R-Squared &           0.08 &           0.27 &               0.18 \\
     Observations &         1620 &         1680 &             3300 \\
\hline\hline
\multicolumn{4}{l}{\footnotesize Heteroskedasticity-robust standard errors in parentheses. }\\[0pt]
\multicolumn{4}{l}{\footnotesize $^{*}$ \(p<0.1\), $^{**}$ \(p<0.05\), $^{***}$ \(p<0.01\).}\\
\end{tabular}}
    \begin{minipage}{1\textwidth}
        \small
        \vspace*{.5em}
        \captionof{table}{Decision Time and Reported Beliefs: Time-Revealed Preference Intensity} 
        \label{table:table-time-distance-indifference}
    \end{minipage}
    \begin{minipage}{1\textwidth}
        \small
        \vspace*{-1.5em}
        \singlespacing \emph{Notes}: 
        This table presents regression results on the relation between log decision times (in seconds) and the distance between reported beliefs to indifference points.
        Reported beliefs refer to elicited beliefs about the probability the opponents plays action $a$; the indifference point refers to the probability that makes the player indifferent between taking either action. 
        Columns (1) and (2) only use data for subjects in the roles of player $M$ and $C$, respectively; column (3) uses both. 
        The games are generalized matching pennies games as given in \hyref{figure:matching-pennies-matrix}[Figure], for $\gamma_C=1$ (scaled by 20); the data is from \citet{FriedmanWard2022WP}.
        \vspace*{.5em}
    \end{minipage}
    \includegraphics[width=.55\linewidth]{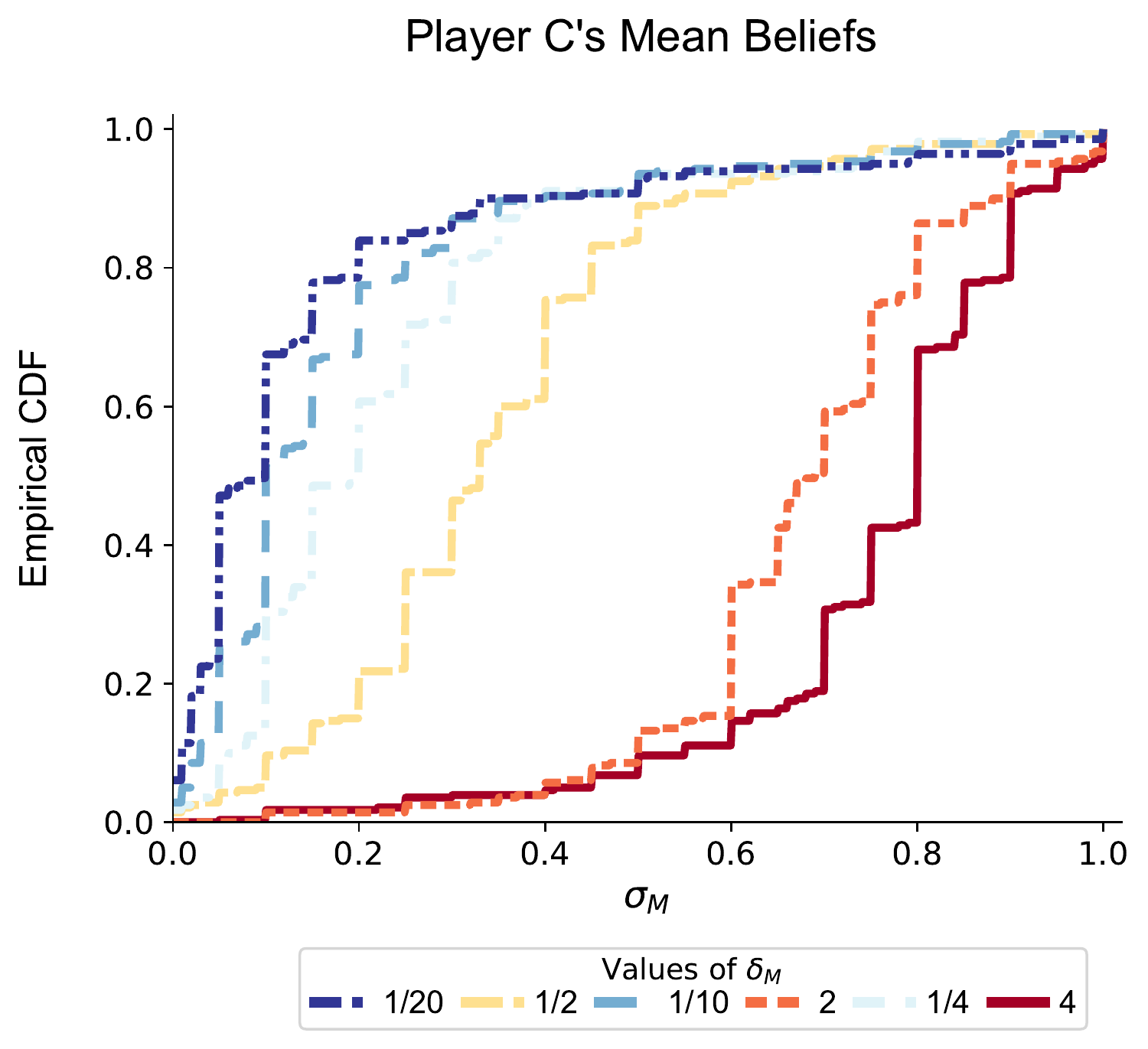}    
    \begin{minipage}{1\textwidth}
        \small
        \vspace*{.5em}
        \captionof{figure}{Opponent Payoff and Beliefs}
        \label{figure:figure-fosd-beliefs-clasher}
    \end{minipage}
    \begin{minipage}{1\textwidth}
        \small
        \vspace*{-1.5em}
        \singlespacing \emph{Notes}: 
        This figure exhibits empirical CDF of reported (mean) beliefs about the probability with which subjects in the role of player $C$ believe their opponent (in the role of player $M$) will take action $a$; different lines correspond to games with different payoffs to action $a$ for player $M$ as parametrized by $\delta_M$. 
        The games are generalized matching pennies games as given in \hyref{figure:matching-pennies-matrix}[Figure], for $\gamma_C=1$ (scaled by 20); the data is from \citet{FriedmanWard2022WP}.
    \end{minipage}
\end{table}

When the absolute difference in the expected payoffs is known---the case where the prior's support is a doubleton---the stopping region is characterized by fixed bounds in terms of the posterior means as shown by \citet{ArrowBlackwellGirshick1949Ecta}.
In contrast, when there is richer uncertainty about the difference in expected payoffs, as when the prior is given by a Beta distribution, the stopping region is characterized by bounds that collapse to the posterior mean that makes the individual indifferent between the two alternatives.
A clear parallel emerges between our setup and that in \citet{FudenbergStrackStrzalecki2018AER}, where the individual infers the difference in payoffs of two alternatives from the drift of a Brownian motion and a similar contrast between known and unknown payoff differences gives rise to, respectively, fixed and collapsing stopping bounds.

\subsubsection*{Comparative Statics in Stopping Beliefs}
From \hyref{proposition:collapsing-bounds}[Proposition] and \hyref{theorem:monotonicity-actions}[Theorem], we obtain that the distribution of beliefs shifts monotonically with respect to the true distribution.
Specifically, approximating the stopping posterior mean by the threshold, $\mathbb E_{\mu_i}[\sigma_{-i}\mid y_i^{\tau_i}]$,\footnote{
    This is so as to avoid discreteness issues inherent to the sampling procedure.
}
and labeling actions so that $u_i(1,\sigma_{-i})-u_i(0,\sigma_{-i})$ is increasing in $\sigma_{-i}$, then player $i$'s stopping (threshold) beliefs increase in a first-order stochastic dominance sense as $\sigma_{-i}$ increases.
This is because a higher $\sigma_{-i}$ leads to a higher probability that player $i$ chooses action 1 more often and faster (resp. action 0 less often and slower), implying that the posterior mean has to exceed a higher threshold when the player stops earlier (resp. later), as the upper (resp. lower) bound characterizing the continuation region is decreasing (resp. increasing) in the stopping time.

\subsubsection*{Supporting Evidence}
Relying again on \citepos{FriedmanWard2022WP} data, we find support for both these predictions:
(1) decision time is significantly negatively related to the distance between the reported mean belief and the indifference point (\hyref{table:table-time-distance-indifference}[Table]), and (2) increasing a player's payoff to an action significantly shifts the opponent's beliefs in the predicted first-order stochastic dominance sense (\hyref{figure:figure-fosd-beliefs-clasher}[Figure])---see \hyref{appendix:experimental-analysis}[Online] for additional statistical tests.

\section{Relation to Nash Equilibrium}
\label{sec:relation-ne}
One initial interpretation of Nash equilibrium posits that equilibrium beliefs are reached as players ``accumulate empirical information'' \citep[p. 21]{Nash1950Thesis}.
In a sequential sampling equilibrium, players accumulate empirical information but at a cost.
A natural question is whether, as these costs vanish, sequential sampling equilibria converge to a Nash equilibrium.
In this section we show this is the case.
Formally, 
\begin{theorem}
    \label{theorem:convergence-ne}
    Let $\Gamma$ be a normal-form game, $\mu$ a collection of priors, and $\{c^n\}_n$ be a sequence of sampling costs such that $c^n \to 0$.
    For any sequence $\{\sigma^n\}_{n}$ such that each $\sigma_n$ is a sequential sampling equilibrium of extended game $G^n=\langle \Gamma, \mu, c^n\rangle$, the limit points of $\{\sigma^n\}_n$ are Nash equilibria of $\Gamma$.
\end{theorem}
The claim is conventional in form: players best-respond to their beliefs and their beliefs converge to the true distribution of actions of their opponents.\footnote{
    This will hold regardless of whether players' prior beliefs allow or not for opponent action correlation.
}
The main complication comes from the fact that, conditional on stopping, the observations $y_i^{\tau_i}$ are not independent nor independently distributed according to player $i$'s opponents' action distribution.
To overcome this issue, the proof (see \hyperref[appendix:proofs:theorem:convergence-ne]{Appendix A}) relies on three arguments.
First, from \hyref{lemma:lower-bound-stopping}[Lemma] one has that as sampling costs vanish, players acquire a minimum number of observations $T$, and, for that minimum number, each observation $y_i^{T}$ is iid according to the opponents' action distribution.
Second, we note beliefs accumulate at a uniform rate around the empirical mean of the observed signals.
Finally, we use the optional stopping theorem to show that beliefs upon stopping converge to the true underlying distribution in an appropriate manner.

Some comments on which Nash equilibria can be selected in this manner are in order.
First, let us define the concept of reachability of a Nash equilibrium:
\begin{definition}
    \label{definition:selection-NE}
    A Nash equilibrium $\sigma$ of a normal-form game $\Gamma$ is \emph{reachable} if there is a collection of priors $\mu$, a sequence of costs $\{c^n\}_n\subset\mathbb R_{++}$ such that $c^n \to 0$, and a sequence $\{\sigma^n\}^n$, where for each $n$, $\sigma^n$ is a sequential sampling equilibrium of the extended game $G^n=\langle \Gamma, \mu, c^n\rangle$, 
    such that $\sigma^n\to \sigma$.
    A Nash equilibrium if \emph{robustly reachable} if it is reachable for any collection of priors $\mu$.
\end{definition}

In the remainder of the section, we will restrict player's payoffs to be linear in distributions as usual.
In other words, we require that, for every player $i$, 
$u_i(a_i,\sigma_{-i})=\mathbb E_{\sigma_{-i}}[u_i(a_i,a_{-i})]$, as conventional.
This will be a maintained assumption throughout the rest of this section.

Our first result provides, separately, necessary and sufficient conditions for reachability of a Nash equilibrium.
\begin{proposition}
    \label{proposition:reachability}
    Let $\Gamma$ be a normal-form game.
    (1) If $\sigma$ is a Nash equilibrium of $\Gamma$ involving weakly dominated actions, then $\sigma$ is not reachable.
    (2) If $a$ is a pure-strategy Nash equilibrium of $\Gamma$ not involving weakly dominated actions, then $a$ is reachable.
\end{proposition}
Part (1) holds since for any prior, no player will ever choose weakly dominated actions---recall that priors have full support.
For (2), note that if $a$ does not involve weakly dominated strategies, then, by \citepos{Pearce1984Ecta} Lemma 4, for each player $i$ there is $\sigma_{-i}^0\in \interior \Delta(A_{-i})$ such that $a_i$ is a best response to $\sigma_{-i}^0$.
If we endow each player $i$ with prior $\mu_i\in \Delta(\Delta(A_{-i}))$ corresponding to a Dirichlet distribution with mean $\sigma_{-i}^0$, then $a_i$ is a best response to any posterior belief $\mu_i\mid y_i^t$ when $y_{i,t} = a_{-i}$.
Hence, for any costs $c^n$, $\sigma$ is sequential sampling equilibrium of $\langle \Gamma,\mu, c^n\rangle$.
Note that we require the Nash equilibrium to be in pure-strategies in order to control posterior beliefs exactly, as otherwise, with some probability, $\sigma_i$ may not be a best response to the posterior belief held upon stopping.

For a Nash equilibrium to be reachable with any priors, we obtain a sufficient condition:
\begin{proposition}
    \label{proposition:robust-reachability}
    If $a$ is a pure-strategy Nash equilibrium in undominated strategies of the normal-form game $\Gamma$ such that, for any player $i$, 
    $a_i$ is a best response to any $\sigma_{-i}'\in B_{\epsilon_i}(\delta_{a_{-i}})$ for some $\epsilon_i>0$, then $a$ is robustly reachable.
\end{proposition}
The intuition for the proof (in \hyperref[appendix:proofs:proposition:robust-reachability]{Appendix A}) is as follows:
for any prior, if player $i$ samples enough $a_{-i}$ observations, their posterior mean will lie within $\epsilon_i$ of $a_{-i}$ and choosing $a_i$ is optimal.
\hyref{lemma:lower-bound-stopping}[Lemma] guarantees that players do sample enough.
Our requirement that $a_i$ is a best response to any distribution of opponents' actions assigning high enough probability to $a_{-i}$ is at the same time more relaxed than strict Nash equilibrium, and more restrictive than trembling hand perfection.

\section{A Dynamic Formulation}
\label{sec:sse:learning-foundation}
One can view sequential sampling equilibrium as a steady state of a dynamic process in whereby agents sequentially sample from past realizations.
This section formalizes that argument.

\subsubsection*{Dynamic Sequential Sampling}
To fix ideas, consider a simple dynamic process, similar to fictitious play.
Fix an extended game $G$.
Every period, $n=1,2,...$, a unit measure of agents plays the extended game $G$, evenly divided across the different roles $I$.
Each agent believes they face a stationary distribution of opponents' actions, matching the empirical frequency of past actions, $\sigma^{n-1}\in \Delta(A)$, not knowing calendar time.

Within period $n$, each agent with role $i$ leans about $\sigma_{-i}^{n-1}$ by optimally sequentially samples according to $\tau_{i}$.
Upon stopping, the agent best responds to their posterior beliefs.\footnote{
    We keep fixed a selection of best responses $\sigma_{i}^*$ used to break-ties.
}
This induces a distribution of actions and types in period $n$ given by $b(\sigma^{n-1})$, where $b: \Delta(A) \to \Delta(A)$ is such that $b(\sigma)(a):=\times_{i\in I}b_i(\sigma_{-i})(a_i)$, with $b_i(\sigma_{-i}):=\mathbb E_{\sigma_{-i}}[\sigma_i^*(\mu_i\mid y_i^{\tau_i})]$, as before.
After taking an action, agents then exit and are replaced by a new population as is standard in evolutionary models of learning in strategic settings.
At the start of the following period, the empirical frequency is then $\sigma^{n}=\frac{1}{n+1}b(\sigma^{n-1})+\frac{n}{n+1}\sigma^{n-1}$, with $\sigma_0$ given.
Call any such $\{\sigma^n\}_n$ a \textbf{dynamic sequential sampling process} of $G$.

While akin to fictitious play \citep{Brown1951}, under dynamic sequential sampling, each agent observes but a sample of past play realizations and the sample itself is an endogenous object.

\subsubsection*{Equilibria and Steady States}
We now show an equivalence between sequential sampling equilibria and steady states of dynamic sequential sampling processes.

\begin{theorem}
    \label{theorem:sse-is-steady-state}
    Let $G$ be an extended game. 
    $\sigma$ is a sequential sampling equilibrium of $G$ if and only if there is some dynamic sequential process $\{\sigma^n\}_n$ of $G$ such that $\sigma^n \to \sigma$.
\end{theorem}
\begin{proof}
    We restrict attention to the if part, since the converse is immediate.
    Let $\bar \sigma$ denote the limit of $\sigma^n$.
    Then,
    \begin{align*}
        0&=\lim_{n\to \infty}\left\lVert \sigma^n-\bar \sigma \right\rVert_{\infty}
        =\left\lVert\lim_{n\to \infty} \sigma^n-\bar \sigma \right\rVert_{\infty}
        =\left\lVert\lim_{n\to \infty} \frac{1}{n+1}\sigma^0 + \frac{n}{n+1}\left(\frac{1}{n}\sum_{\ell\in[0\,..\,n-1]}b(\sigma^{\ell})\right)-\bar \sigma \right\rVert_{\infty}.
    \end{align*}
    As $\sigma^n \to \bar \sigma$ and $b$ is continuous, then $b(\sigma^n)\to b(\bar \sigma)$.
    Consequently, the Ces\`{a}ro mean $\frac{1}{n}\sum_{\ell\in[0\,..\,n-1]}b(\sigma^{\ell})$ also converges to $b(\bar \sigma)$ and therefore
    $0=\lVert b(\bar \sigma)-\bar \sigma\rVert_\infty \Longrightarrow b(\bar \sigma)=\bar \sigma$.
\end{proof}

The steady-state characterization of sequential sampling equilibria in \hyref{theorem:sse-is-steady-state}[Theorem] provides a clear analogue to the characterization of Nash equilibria as steady-states of fictitious play in \citet{FudenbergKreps1993GEB}.
The main difference between fictitious play and the dynamic process analyzed is that, whereas data is freely observable in fictitious play, sequential sampling players face information acquisition costs.
Moreover, as we have seen (\hyref{theorem:convergence-ne}[Theorem]), as these costs vanish, limiting sequential sampling equilibria correspond to Nash equilibria.
Below we discuss two ways in which the dynamic process can be generalized.

\begin{remark}
    Often it may be the case that information about more recent events is more easily accessible.
    This can be modeled as a giving a different weight to each period, for instance, exponential discounting past data: $\sigma^n = \beta\, \sigma^{n-1}+(1-\beta)\,b(\sigma^{n-1})$, $\beta \in (0,1)$.
    \hyref{theorem:sse-is-steady-state}[Theorem] also holds under this alternative definition:
    as $\sigma^n \to \bar \sigma \implies b(\sigma^n)\to b(\bar \sigma)$ and, for any fixed $\ell$, $\beta^{n-1-\ell} b(\sigma^{\ell})\to 0$, we have $\sigma^n=\beta^n \sigma^0+(1-\beta)\cdot \sum_{\ell\in [0\,..\,n-1]}\beta^{n-1-\ell}\cdot b(\sigma^\ell) \to \bar \sigma = b(\bar \sigma)$.
\end{remark}
\begin{remark}
    The assumption that there is a continuum of agents for each role is also not essential: a similar result holds when the populations are finite.
    Write $a^n$ for the realized actions in period $n$ and $\sigma^n$ for their empirical frequency (given $a^0$), with $a^n \sim b(\sigma^{n-1})$.\footnote{
        If agents directly sample data with past actions, $\{a^\ell\}_{\ell <n}$, one may worry that about whether sampling without replacement affects the result; this is not the case---provided, of course, the starting dataset large enough (but still finite; cf. \hyref{remark:bounded-stopping}[Remark]) so that sequential sampling without replacement is well defined.
    }
    Note that $\sigma^n \to \bar \sigma$ still implies that $b(\sigma^n)\to b(\bar \sigma)$, and the arguments above remain the same, with $a^n$ converging in distribution to a sequential sampling equilibrium.
\end{remark}

\subsubsection*{Convergence}
While in general we cannot exclude dynamic sequential sampling from cycling and failing to converge---similarly to what occurs with fictitious play\footnote{
    Classical references are \citet{Shapley1964} and \citet{Jordan1993GEB}.
    Cycling can occur even with stochastic fictitious play: see \citet{HommesOchea2012GEB}.
} ---
in specific classes of games, convergence and asymptotic stability are guaranteed.\footnote{
    An equilibrium $\sigma$ is asymptotically stable if for all $\epsilon>0$, there is a $\delta>0$ such that for any $\sigma^0: \| \sigma^0- \sigma\|_{\infty}<\delta$, $\| \sigma^n- \sigma\|_\infty<\epsilon$ for all $n$.
    That is, if the dynamic sequential sampling process starting close enough to the equilibrium remains closeby thereafter.
}
This next proposition shows this is the case for binary action games, which we will discuss in more depth in the next section.
\begin{proposition}
    \label{proposition:convergence-dynamic-ss}
    Let $G=\langle \Gamma, \mu, c\rangle$ be a two-player extended game.
    If $\Gamma$ has a unique Nash equilibrium, the limit of dynamic sequential sampling is a globally asymptotically stable sequential sampling equilibrium.
\end{proposition}
The proof is deferred to the appendix.

\section{Extensions and Discussion}
\label{sec:extensions}
We conclude with a discussion of possible extensions of sequential sampling equilibrium.

\subsubsection*{Types and Bayesian Games}
Sequential sampling equilibrium can be easily extended to accommodate Bayesian games.
This aligns with the idea that sequential sampling equilibria corresponds to the case in which players don't know their opponents' payoffs, as given by their type.
Alternatively, one can consider that different types characterize different settings, and players are using information about behavior in similar settings to make inferences about behavior in the particular game they are facing.

In particular, consider games described by $\Gamma:=\langle I, A, \Theta, \rho, u\rangle$, such that $I$ denote the finite set of players, $A:=\times_{i \in I}A_i$ the (finite) set of action profiles, $\Theta=\times_{i\in I}\Theta_i$ the (finite) set of type profiles, where $\Theta_i$ are player $i$'s possible types, $u_i: A\times \Theta\to \mathbb R$ payoff functions, and $\rho\in \Delta(\Theta)$ a distribution over types.
Endowing each player $i$ with a prior $\mu_i\in \Delta(\Delta(A_{-i}\times \Theta))$ and a sampling cost $c_i$ as before we have extended games $G=\langle \Gamma, \mu,c\rangle$.
Each player $i$ with type $\theta_i$ now learns about the joint distribution of opponents' action profiles and type profiles, $q_i\in \Delta(A_{-i}\times \Theta)$, sequentially sampling from $q_i$ at cost $c_i$ and stopping according to the earliest optimal stopping time $\tau_{i,\theta_i}$.

A sequential sampling equilibrium $\sigma$ would then correspond to a fixed point such that
$\sigma_{i,\theta_i}=\mathbb E_{q_{i}}[\sigma_{i,\theta_i}^*(\mu_i\mid y_i^{\tau_{i,\theta_i}})]$,
where $\sigma_{i,\theta_i}^*(\mu_i)$ is a selection of best responses given belief $\mu_i$, and
$q_{i,\theta_i}(a_{-i},\theta)=\rho(\theta)\times_{j \ne i}\sigma_{j,\theta_j}(a_j)$ for every $a_{-i}={(a_j)}_{j\ne i}\in A_{-i}$ and $\theta={(\theta_j)}_{j \in I}\in \Theta$.

Different assumptions on players beliefs will give rise to different equilibria.
To apply similar arguments to obtain existence of an equilibrium, we need but to require that players know the distribution of their own types and that $\mu_i$ has full support on the set of distributions $q_i \in \Delta(A_{-i}\times \Theta)$ satisfying $q_i(\theta_i)=\rho(\theta_i)$ for any $\theta_i \in \Theta_i$.\footnote{
    This renders their expected payoff given their type, $\mathbb E_{q_i}[u_i(a_i,a_{-i},\theta)\mid \theta_i]$,  to be continuous in $q_i\in \supp \mu_i$.
}
Differently, one could assume players know the true distribution of types, or, when types are independent, that they know so.

With the proper adjustments, behavioral implications can also be obtained, now comparing across types.
For instance, if the payoff to action $a_i$ is higher for type $\theta_i$ than for $\theta_i'$, everything else equal, in every sequential sampling equilibrium type $\theta_i$ chooses action $a_i$ more often and faster (in the sense of \hyref{theorem:monotonicity-actions}[Theorem]) than type $\theta_i'$, a result that can be exploited and tested in a number of traditional settings, from global games to voting.
Finally, convergence to Bayesian Nash equilibria when sampling costs vanish can be similarly obtained.

\subsubsection*{General Information Structures and Analogy Partitions}
Throughout, it was assumed that players observe action profiles drawn from a steady state distribution.
Often, of course, information---and even memory---is fuzzier, and it is not possible to perfectly distinguish between certain actions taken by others, or even to observe what some other players do at all.
In \hyref{appendix:information-structures}[Online], we provide sufficient conditions under which is it possible to generalize sequential sampling equilibrium to cases under which players observe not action profiles of their opponents, but a garbling, thereby accommodating situations such as noisy recollections, or missing or misrecorded data.
This formalism can also be used to formalize the idea that players' payoffs may depend on the behavior of others about which they are unable to obtain information.
For instance, it may be possible to obtain information about behavior within the same firm, but impossible to learn what people do in other firms.
For the special case in which players are unable to distinguish between specific action profiles (or types) of their opponents, as sampling costs vanish, sequential sampling equilibria reach not Nash equilibria but analogy-based expectation equilibria \citep{Jehiel2005JET,JehielKoessler2008GEB}.

\subsubsection*{Sampling Costs and Discounting}
In this paper, we considered a constant additive cost per observation.
One could have defined this cost of information in a more general manner, allowing it to depend on the number of observations already acquired, or allowing a finite number of observations at no cost.
Alternatively, one could rely on discounting payoffs instead.
It is indeed possible to extend the setup to accommodate either, posing no problem for existence of an equilibrium.

\subsubsection*{Misspecified Priors}
Another maintained assumption was that priors are not misspecified, i.e. players are able to learn the true data generating process.
This assumption was crucial to obtain existence of an equilibrium: it is the full support of players' prior that guarantees that, as they acquire more and more observations, their beliefs accumulate around a degenerate distribution.
When, instead, priors are misspecified, it is possible that players never stop sampling (according to the true data generating process), even though they believe they will (according to their posterior beliefs).
We provide one such example of nonexistence in \hyref{appendix:misspecification}[Online].

\subsubsection*{Empirical Estimation}
Empirical estimation of the baseline model is facilitated by utilizing Dirichlet beliefs, a flexible class of priors with parameters $\alpha^i =(\alpha_1^i,...,\alpha_n^i)$, where $n=|A_{-i}|$.
When $n=2$, it corresponds to the Beta distribution.
It can be shown that, for this rich class of conjugate priors, given player $i$'s payoffs and sampling costs, there is a compact set of parameters $\alpha^i$ for which player $i$ would sample.
This also implies the existence of a tight upper bound for the stopping time $\overline T$ \emph{for any Dirichlet prior}, converting the infinite horizon problem in a finite horizon one and rendering it a computationally tractable problem.
Assuming payoffs are known, a particular parameter and sampling cost would map to a given joint distribution of actions and stopping time.
Maximum likelihood and other estimation procedures analogous to those relied upon for sequential sampling models in individual decision-making tasks can then be used \citep[see e.g.][]{MyersInterianMoustafa2022FrontiersPsy}.\footnote{
    In order to map theoretical stopping times and empirical decision times, one can use quantiles instead of the values.
}

\subsubsection*{Myopic Sequential Sampling}
Finally, a comment on a simpler, alternative, version of sequential sampling equilibrium, in which sequential sampling behavior is myopic.
This is closely related to a recent paper by \citet{AlaouiPenta2022JPE}, which provides an axiomatization of the deliberation process as one of myopic sequential information acquisition (cf. their Theorem 4).
In other words, the myopic stopping time would be given by $\tau_i^M(\omega):=\min\{t\mid \mathbb E_{\mu_i}[v_i(\mu_i\mid y_i^{t+1})\mid y_i^t(\omega)]-v_i(\mu_i\mid y_i^t(\omega))\leq c_i\}$, with players stopping whenever the expected value of sampling information is smaller than the cost of one more observation.

One could then define myopic sequential sampling equilibria simply by replacing the optimal stopping time $\tau_i$ with the myopic one $\tau_i^M$.
While appealing for its simplicity, one immediate implication is that myopic sequential sampling equilibria do not generically reach Nash equilibria as the sampling costs vanish.
This is because the expected value of sampling information is zero whenever the optimal choices under belief $\mu_i$ are still optimal under the posterior $\mu_i\mid y_i$, no matter the realization of the observation.
It is therefore immediate that any prior that is sufficiently concentrated around a distribution of opponents' actions for which $a_i$ is a strict best response, \emph{for any} $c_i>0$, player $i$ will not see it worthwhile to acquire information.
Thus, if in no Nash equilibrium it is optimal to play such an action with probability 1, myopic sequential sampling will fail to converge to a Nash equilibrium.

\section*{References}
{
    \setstretch{1}
    \setlength\bibhang{0pt}
    \setlength{\bibsep}{0em plus 0ex}
    \bibliographystyle{econ-aea}
    \bibliography{sse.bib}
}
\setcounter{section}{0}
\renewcommand{\thesection}{Appendix \Alph{section}}
\renewcommand{\thesubsection}{\Alph{section}.\arabic{subsection}}

\section{Omitted Proofs}
\label{appendix:proofs}
\subsubsection{Proof of \hyref{proposition:properties-value-stopping}[Proposition]}
\label{appendix:proofs:proposition:properties-value-stopping}
For ease of notation, I'll write $Y_i:=A_{-i}$, $P_i:=\Delta(Y_i)$, and $\mu_i \in \Delta(P_i)$.
We consider general continuous utility functions $u_i:A_i\times P_i\to \mathbb R$.
Boundedness of $V_i$ and $v_i$ follows immediately from boundedness of $u_i$.
Below we prove the remaining properties.

\begin{claim} For any $\mu_i\in \Delta(P_i)$, and any optimal stopping time $\tau_i$, $\mathbb P_{\mu_i}(\tau_i>T)\leq 2\|u_i\|_\infty /c_i T$. \end{claim}
\vspace*{-1em}
\begin{proof}
    Let $\|u_i\|_\infty=\max_{(a_i,p_i)\in A_i\times P_i}|u(a_i,p_i)|<\infty$.
    We then have that, for any $\mu_i \in \Delta(P_i)$ and $T \in \mathbb N$,
    \begin{align*}
        &-\|u_i\|_\infty \leq V_i(\mu_i) \leq \mathbb P_{\mu_i}(\tau_i\leq T)\|u_i\|_\infty +\mathbb P_{\mu_i}(\tau_i>T)(\|u_i\|_\infty-c_iT)\Longrightarrow 
        \mathbb P_{\mu_i}(\tau_i>T)\leq 2\|u_i\|_\infty /c_i T.
    \end{align*}
\end{proof}
\vspace*{-1.5em}
\begin{claim} $v_i$ is uniformly continuous. \end{claim}
\vspace*{-1em}
\begin{proof}
    $\mathbb E_{\mu_i}[u_i(a_i,p_i)]$ is jointly continuous in $(a_i,\mu_i)$ with respect to product topology.
    Let $\mu_{i,n} \to \mu_i$, $a_{i,n} \to a_i$.
    Note that $\forall \epsilon>0$, $\exists N$ such that $\forall n\geq N$, $\forall p_i$, $|u_i(a_{i,n},p_i)-u_i(a_i,p_i)|<\epsilon/2$
    and $|\mathbb E_{\mu_{i,n}}[u_i(a_i,p_i)]-\mathbb E_{\mu_i}[u_i(a_i,p_i)]|<\epsilon/2$.
    Hence, 
    \[
    |\mathbb E_{\mu_{i,n}}[u_i(a_{i,n},p_i)]-\mathbb E_{\mu_i}[u_i(a_i,p_i)]|
    \leq |\mathbb E_{\mu_{i,n}}[u_i(a_{i,n},p)-u_i(a_i,p_i)]|+|\mathbb E_{\mu_{i,n}}[u_i(a_i,p_i)]-\mathbb E_{\mu_i}[u_i(a_i,p_i)]|
    < \epsilon.
    \]
    Continuity of $v_i$ follows from Berge's maximum theorem and uniform continuity from Heine--Cantor theorem.
\end{proof}
\vspace*{-.5em}
\begin{claim} $V_i$ is uniformly continuous. \end{claim}
\vspace*{-1em}
\begin{proof}
    Let $\mathbb T_{i,T}$ denote the set of stopping times $\tau' \in \mathbb T_i$ that are bounded above by $T$ and, for every $T \in \mathbb N$, $V_{i,T}:\Delta(P_i)\to \mathbb R$ be given by
    \[V_{i,T}(\mu):=\sup_{\tau' \in \mathbb T_{i,T}} \mathbb E_{\mu_i}[v_i(\mu_i\mid y_i^{\tau'})-c_i \cdot \tau'].\]
    Note that, as $\mathbb T_{i,T}$ is finite, it is compact with respect to the discrete topology, and an application of Berge's maximum theorem implies $V_{i,T}$ is continuous.
    
    Note that, for any $\mu_i \in \Delta(P_i)$, $T \in \mathbb N$,
    $0\leq V_i(\mu_i)-V_{i,T}(\mu_i)\leq \mathbb P_{\mu_i}(\tau_i>T)\|u_i\|_\infty\leq 2\|u_i\|_\infty^2/c_iT.$
    Hence, 
    $\|V_i-V_{i,T}\|\leq 2\|u_i\|_\infty^2/c_iT,$ and $V_{i,T}$ converges uniformly to $V_i$.
    Since for any $T$, $V_{i,T}$ is in the space of bounded continuous functions $C_b^0(\Delta(P_i))$, which, endowed with the sup-norm is a Banach space, $V_i$ is continuous; by the Heine--Cantor theorem, it is uniformly continuous.
\end{proof}
\vspace*{-.5em}
\begin{claim} $v_i$, $V_i$, and $V_{i,T}$ are convex, for any $T\in \mathbb N$. \end{claim}
\vspace*{-1em}
\begin{proof}
    This follows since each of these can be seen as the pointwise supremum over a family of convex functions over $\Delta(P_i)$, which is compact with respect to $\|\cdot\|_{LP}$.
\end{proof}

\subsubsection{Proof of \hyref{lemma:stopping-implies-continuity}[Lemma]}
\label{appendix:proofs:lemma:stopping-implies-continuity}
By contrapositive, that (2) implies (1) is straightforward.
We prove a more general claim that implies the converse.
Let $\Sigma_{-i}^{\tau_i}:=\{\sigma_{-i}\in \Delta(A_{-i})\mid \mathbb P_{\sigma_{-i}}(\tau_i<\infty)=1\}$ denote the opponents' distribution actions with respect to which player $i$'s optimal stopping time is finite with probability 1.
\begin{lemma}
    On $\Sigma_{-i}^{\tau_i}$, $b_i$ is a continuous mapping to $\Delta(A_i)$.
\end{lemma}
\vspace*{-1em}
\begin{proof}
    Fix a selection of optimal choices $\sigma_i^*(\mu_i')\in \argmax_{\sigma_i \in \Delta(A_i)}\mathbb E_{\mu_i'}[u_i(\sigma_i,\sigma_{-i}')]$.
    For $t \in \mathbb N$, let $b_{i,t}: \Sigma_{-i}^{\tau_i}\to [0,1]^{|A_i|}$ be given by $b_{i,t}(\sigma_{-i}):=\mathbb E_{\sigma_{-i}}[1_{\tau_i<t}\sigma_i^*(\mu_i\mid y_i^{\tau_i})]$, and $p_{i,t}: \Sigma_{-i}^{\tau_i}\to [0,1]$ denote $p_{i,t}(\sigma_{-i}):=\mathbb P_{\sigma_{-i}}(\tau_i<t)$.
    Note that, for every $t'\geq t$, $b_{i,t}\leq b_{i,t'}$ and $p_{i,t}\leq p_{i,t'}$, ensuring pointwise convergence to $b_i$ and $1$, respectively.
    Further, both $b_{i,t}$ and $p_{i,t}$ are continuous.
    
    Since $p_{i,t}\to 1$, by Dino's theorem $p_{i,t}$ converges uniformly on $\Sigma_{-i}^{\tau_i}$.
    We then have $\|b_i(\sigma_{-i})-b_{i,t}(\sigma_{-i})\|_1\leq 1-\mathbb p_{i,t}(\sigma_{-i})$, and $b_{i,t}$ also converges uniformly to $b_i\mid_{\Sigma_{-i}^{\tau_i}}:\Sigma_{-i}^{\tau_i}\to [0,1]^{|A_i|}$, ensuring that, on $\Sigma_{-i}^{\tau_i}$, $b_i$ is continuous.
    To see that $b_i(\sigma_{-i})\in \Delta(A_i)$ for any $\sigma_{-i} \in \Sigma_{-i}^{\tau_i}$, note that $1\geq \|b_i(\sigma_{-i})\|_1\geq \|b_{i,t}(\sigma_{-i})\|_1 \geq p_{i,t}(\sigma_{-i}) \to 1$.
\end{proof}

\subsubsection{Proof of \hyref{remark:bounded-stopping}[Remark]}
\label{appendix:proofs:remark:bounded-stopping}
Let $\sum_{\ell \in [1\,..\,t]}\delta_{y_{i,\ell}^t}/t=:\overline {y_i}^t \in \Delta(A_{-i})$ denote the empirical frequency.
When $\mu_i$ has full support on $\Delta(A_{-i})$, then there is $\phi:\mathbb R_{++}\to \mathbb R_{++}$ such that $\inf_{\sigma_{-i} \in \Delta(A_{-i})}\mu(B_\epsilon(\sigma_{-i}))\geq \phi(\epsilon)$.
Then, by \citet{DiaconisFreedman1990AnnStat}, 
for every $\epsilon>0$,
there is $T$ such that for all $t\geq T$, $\mu_{i,t}(B_{\epsilon/2}(\overline{y_i}^t))/(1-\mu_{i,t}(B_{\epsilon/2}(\overline{y_i}^t))\geq \frac{2-\epsilon}{\epsilon}$, which implies $\|\mu_{i,t}-\delta_{\overline{y_i}^t}\|_{LP}\leq \epsilon$.
Immediately, for every $\epsilon>0$,
there is $T$ such that for all $t\geq T$,
\begin{align*}
    \|\mu_{i,t}-\mu_{i,t+1}\|_{LP}\leq 
    \|\mu_{i,t}-\delta_{\overline{y_i}^t}\|_{LP}
    +
    \|\mu_{i,t+1}-\delta_{\overline{y_i}^{t+1}}\|_{LP}
    +
    \|\delta_{\overline{y_i}^t}-\delta_{\overline{y_i}^{t+1}}\|_{LP} 
    \leq 
    2/3\epsilon + 2/t
    \leq 
    \epsilon.
\end{align*}
As $V_i$ is uniformly continuous (cf. \hyref{proposition:properties-value-stopping}[Proposition]), this implies $\exists \overline T_i$ such that $\forall t\geq T$, 
$\forall y_i^{t+1}$,
$V_i(\mu_i\mid y_i^{t+1})-V_i(\mu_i\mid y_i^t)\leq c_i \Longrightarrow \mathbb E_{\mu_i}[V_i(\mu_i\mid y_i^{t+1})-V_i(\mu_i\mid y_i^t)\mid y_i^t]\leq c_i\Longrightarrow \tau_i\leq \overline T_i$.

When the prior does not allow for correlation, $\mu_i=\times_{j\ne i}\mu_{ij}$ and each marginal $\mu_{ij}$ uniformly accumulates around the empirical frequency projected on $\Delta(A_j)$. As $I$ is finite, one can similarly obtain a uniform rate of convergence that depends only on $t$.
It is then straightforward to adjust the proof to obtain the result.

\subsubsection{Proof of \hyref{proposition:convergence-dynamic-ss}[Proposition]}
\label{appendix:proofs:proposition:convergence-dynamic-ss}
Let $A_i=\{0,1\}$, $i\in I$; denote the probability that player $i$ chooses action $1$ by $\sigma_i$.
By manner of a continuous-time approximation as in \citet[Ch. 2]{FudenbergLevine1998Book}, the dynamic system can be written as $\dot \sigma_i=b_i(\sigma_j)-\sigma_i$, $i,j \in I$, $i\ne j$.
The Jacobian of the dynamic system is given by 
\begin{align*}
    \begin{pmatrix}
        -1&b_i'(\sigma_j)\\
        b_j'(\sigma_i)&-1
    \end{pmatrix}
\end{align*}
and its eigenvalues are given by $\lambda=-1\pm \sqrt{b_i'(\sigma_j)b_j'(\sigma_i)}$, where differentiability of $b_i,b_j$ is ensured by \hyref{theorem:monotonicity-actions}[Theorem].
If $b_i'(\sigma_j)b_j'(\sigma_i)\leq 0$, there is a unique $\sigma$ such that $b_i(\sigma_j)=\sigma_i$ 
and, since the real parts of the eigenvalues of the Jacobian matrix are strictly negative, by the Jacobian conjecture on global asymptotic stability---proved to hold on the plane \citep{ChenHeQin2001AMSinica}---$\sigma$ is globally asymptotically stable.

In particular, if there is a unique Nash equilibrium, either one player has a dominant strategy (and then $b_i'=0$ for some player $i$) or neither does.
If some player does not sample, i.e. $\tau_i=0$ for some player $i$, then again $b_i'=0$.
If both players sample, we must then have $b_i'(\sigma_j)b_j'(\sigma_i)<0$.
In any of these cases, $b_i'(\sigma_j)b_j'(\sigma_i)\leq 0$.

\subsubsection{Proof of \hyref{remark:stopping-costs}[Remark]}
\label{appendix:proofs:remark:stopping-costs}
Let $c_i'\geq c_i$ and denote $V_i'$ and $V_i$ the value functions associated with $c_i'$ and $c_i$, respectively.
Since $v_i(\mu_i)\leq V_i'(\mu_i)\leq V_i(\mu_i)$, it is immediate that $V_i(\mu_i)=v_i(\mu_i)\Longrightarrow V_i'(\mu_i)=v_i(\mu_i)$.
Let $\tau_i'$ and $\tau_i$ the earliest optimal stopping times associated with $c_i'$ and $c_i$, respectively.
Then
\begin{align*}
    \left\{\omega \in \Omega \;\bigg\vert\; \tau_i(\omega)\leq t\right\}
    &=
    \left\{\omega \in \Omega \;\Bigg\vert\; 
    \exists t'\leq t:
    \begin{array}{l}
    V_i(\mu_i \mid y_i^{t'}(\omega)) = v_i(\mu_i \mid y_i^{t'}(\omega))\\
    V_i(\mu_i \mid y_i^{\ell}(\omega)) > v_i(\mu_i \mid y_i^{\ell}(\omega)),\, \forall \ell<t'
    \end{array}
    \right\}\\
    &\subseteq
    \left\{\omega \in \Omega \;\Bigg\vert\; 
    \exists t'\leq t:
    \begin{array}{l}
    V_i'(\mu_i \mid y_i^{t'}(\omega)) = v_i(\mu_i \mid y_i^{t'}(\omega))\\
    V_i'(\mu_i \mid y_i^{\ell}(\omega)) > v_i(\mu_i \mid y_i^{\ell}(\omega)),\, \forall\ell<t'
    \end{array}
    \right\}
    =\left\{\omega \in \Omega \;:\; \tau_i'(\omega)\leq t\right\}.
\end{align*}

\subsubsection{Proof of \hyref{lemma:rationalizable-best-response}[Lemma]}
\label{appendix:proofs:lemma:rationalizable-best-response}
Take any $a_i \in A_i \setminus A_i^k$.
For $\delta\geq 0$ define 
\[\overline{B}_\delta(\Delta(A_{-i}^{k-1})):=\{\sigma_{-i} \in \Delta(A_{-i})\mid \exists \sigma_{-i}' \in \Delta(A_{-i}^{k-1}): \|\sigma_{-i}-\sigma_{-i}'\|\leq \delta\}\]
\[h_i^k(\delta):=\max_{\sigma_{-i} \in \overline{B}_\delta(\Delta(A_{-i}^{k-1}))}
\left(u_i(a_i,\sigma_{-i})-\max_{\sigma_i' \in \Delta(A_i)}u_i(\sigma_i',\sigma_{-i})\right).
\]
Since $u_i$ is continuous and $\delta \mapsto \overline{B}_\delta(\Delta(A_{-i}^{k-1}))$ is a continuous, convex-, compact-, and nonempty-valued correspondence, $h_i^k$ is continuous by Berge's maximum theorem.
As $\overline{B}_\delta(\Delta(A_{-i}^{k-1}))$ increases in subset order with $\delta$, $h_i^k$ is nondecreasing.

By definition of $k$-rationalizability, $h_i^k(0)<0$.
This implies that there is $\delta'>0$ such that, $\forall \delta\leq \delta'$, $h_i^k(\delta)\leq h_i^k(0)/2<0$.
Then, for any $\delta<\delta'$, and any $\sigma_{-i} \in {B}_\delta(\Delta(A_{-i}^{k-1}))$,
$u_i(a_i,\sigma_{-i})-\max_{\sigma_i' \in \Delta(A_i)}u_i(\sigma_i',\sigma_{-i})\leq h_i^k(0)/2<0$.
Finally, observe that
$\max_{\sigma_{-i}\in \Delta(A_{-i})\setminus\Delta(A_{-i}^{k-1})} \max_{\sigma_i \in \Delta(A_i)}u_i(a_i,\sigma_{-i})-u_i(\sigma_i,\sigma_{-i})<2\|u_i\|_\infty$.
Let 
$\epsilon<-h_i^k(0)/(4\|u_i\|_\infty-h_i^k(0))$ and $\mu_i({B}_\delta(\Delta(A_{-i}^{k-1})))>1-\epsilon$.
It then follows that 
\begin{align*}
    \mathbb E_{\mu_i}[u_i(a_i,\sigma_{-i})]-\max_{\sigma_i \in \Delta(A_i)}\mathbb E_{\mu_i}[u_i(\sigma_i,\sigma_{-i})]
    &\leq \mu_i(B_\delta(\Delta(A_{-i}^{k-1})))h_i^k(0)/2+(1-\mu_i(B_\delta(\Delta(A_{-i}^{k-1}))))2\|u_i\|_\infty\\
    &<(1-\epsilon)h_i^k(0)/2+\epsilon 2\|u_i\|_\infty<0.
\end{align*}

\subsubsection{Proof of \hyref{lemma:uniform-concentration-k-rationalizable}[Lemma]}
\label{appendix:proofs:lemma:uniform-concentration-k-rationalizable}
By assumption, $\overline y_i^t \in \Delta(A_{-i}^{k-1})$.
Since every player's prior has full support, by \citet{DiaconisFreedman1990AnnStat}, each player's prior concentrates on an $\delta$-ball around the empirical frequency of at a uniform rate.
This implies that, for any $\delta,\epsilon>0$ there is a $T$ such that, for all $t\geq T$ and any $y_i^t \in A_{-i}^{k-1}$, $\mu_{i,t}(B_\delta(\Delta(A_{-i}^{k-1})))\geq \mu_{i,t}(B_\delta(\delta_{\overline y_i^t}))>1-\epsilon$.

\subsubsection*{Auxiliary Results}
\begin{lemma}
    \label{lemma:bound-regret}
    For any $\epsilon>0$, $\exists \delta>0$ such that for any $T\in \mathbb N$, $c_i>0$,
    $\mathbb E_{\mu_i}[v_i(\delta_{\sigma_{-i}})]-V_i(\mu_i)\leq (1-2 \exp(-2T\delta^2))\epsilon/4 + 8 \exp(-2T\delta^2) \|u_i\|_\infty +c_i T$.
\end{lemma}
\vspace*{-1em}
\begin{proof}
    Let $\hat T_i$ be the stopping time such that player $i$ stops after $T$ periods and let $\overline y_i^T$ denote the empirical frequency of $y_i^T$, i.e. $\overline y_i^T:=\sum_{t \in [1 .. T]}\delta_{y_{i,t}}/T$.
    Let $\hat a_i:\Delta(A_{-i})\to A_i$ be such that $\hat a_i(\mu_i')\in \arg\max \mathbb E_{\mu_i'}[u_i(a_i,\sigma_{-i})]$.
    Since $u_i(a_i(\sigma_{-i}),\sigma_{-i})=\max_{a_i \in A_i}u_i(a_i,\sigma_{-i})$ is continuous by Berge's maximum theorem and uniformly so by Heine--Cantor theorem, let $\delta>0$ be such that for any $\|\sigma_{-i}-\sigma_{-i}'\|<\delta \Longrightarrow |u_i(a_i(\sigma_{-i}),\sigma_{-i})-u_i(a_i(\sigma_{-i}'),\sigma_{-i}')|+
    |u_i(a_i(\sigma_{-i}'),\sigma_{-i})-u_i(a_i(\sigma_{-i}'),\sigma_{-i}')|<\epsilon/4
    $.
    
    Since $\hat T_i$ and $\hat a_i(\overline y_i^{\hat T_i})$ are potentially suboptimal stopping time and choices for player $i$,
    $V_i(\mu_i)\geq \mathbb E_{\mu_i}[u_i(\hat a_i(\overline y_i^{\hat T_i}),\sigma_{-i})-c_i \hat T_i]$.
    For $c_i=k/(4 T)>0$, 
    \begin{align*}
        \mathbb E_{\mu_i}[v_i(\delta_{\sigma_{-i}})]-V_i(\mu_i)
        &\leq 
        \mathbb E_{\mu_i}[u_i(\hat a_i(\sigma_{-i}),\sigma_{-i})-u_i(\hat a_i(\overline y^{\hat T_i}),\sigma_{-i})+c_i \hat T_i]
        \\
        &=
        \mathbb E_{\mu_i}[u_i(\hat a_i(\sigma_{-i}),\sigma_{-i})-u_i(\hat a_i(\overline y^{\hat T_i}),\overline y_i^{\hat T_i})
        +u_i(\hat a_i(\overline y_i^{\hat T_i}),\overline y_i^{\hat T_i})-u_i(\hat a_i(\overline y_i^{\hat T_i}),\sigma_{-i})]+c_i T
        \\
        &\leq 
        (1-2 \exp(-2T\delta^2))\epsilon/4 + 8 \exp(-2T\delta^2) \|u_i\|_\infty 
        +c_i T,
    \end{align*}
    where the last inequality follows by the Dvoretzky--Kiefer--Wolfowitz--Massart inequality \citep{Massart1990AnnProb}, which delivers that, for any $\delta>0$ and $\sigma_{-i}$, $\mathbb P_{\sigma_{-i}}(\|\sigma_{-i}-\overline y_i^T\|>\delta)<2 \exp(-2T\delta^2)$.
\end{proof}

\subsubsection{Proof of \hyref{lemma:lower-bound-stopping}[Lemma]}
\label{appendix:proofs:lemma:lower-bound-stopping}
We first prove the weaker statement:
\begin{claim}
    Suppose that there is no action $a_i$ that is a best response to all distribution of opponents' actions $\sigma_{-i} \in \Delta(A_{-i})$. 
    Then, for any full support prior $\mu_i \in \Delta(A_{-i})$, there is a sampling cost $c_i>0$ such that the associated earliest optimal stopping time $\tau_i\geq 1$.
\end{claim}
\vspace*{-1em}
\begin{proof}
    Let us start by showing that $\mathbb E_{\mu_i}[v_i(\delta_{\sigma_{-i}})]-v_i(\mu_i)>0$.
    Since there is no action $a_i$ that is a best response to all distribution of opponents' actions $\sigma_{-i} \in \Delta(A_{-i})$, for any $\mu_i$ with full support and any $a_i(\mu_i)\in \argmax_{a_i \in A_i}\mathbb E_{\mu_i}[u_i(a_i,\sigma_{-i})]$, there is $\sigma_{-i}' \in \Delta(A_{-i})$ such that $v_i(\delta_{\sigma_{-i}'})=\max_{a_i \in A_i}u_i(a_i,\sigma_{-i}')>u_i(a_i(\mu_i),\sigma_{-i}')$.
    By continuity, there is an $\epsilon>0$ such that for all $\sigma_{-i}'' \in B_\epsilon(\sigma_{-i}')$, $v_i(\delta_{\sigma_{-i}''})-u_i(a_i(\mu_i),\sigma_{-i}'')\geq (v_i(\delta_{\sigma_{-i}}')-u_i(a_i(\mu_i),\sigma_{-i}'))/2>0$.
    Hence, $\mathbb E_{\mu_i}[v_i(\delta_{\sigma_{-i}})]-v_i(\mu_i)\geq\mu_i(B_\epsilon(\delta_{\sigma_{-i}'}))(v_i(\delta_{\sigma_{-i}}')-u_i(a_i(\mu_i),\sigma_{-i}'))/2>0$.
    
    Next, we show that, if $c_i$ is low enough, $V_i(\mu_i)>v_i(\mu_i)$ (implying $\tau_i\geq 1$) by proving that 
    $\mathbb E_{\mu_i}[v_i(\delta_{\sigma_{-i}})]-V_i(\mu_i)<\mathbb E_{\mu_i}[v_i(\delta_{\sigma_{-i}})]-v_i(\mu_i)=:k$.
    By \hyref{lemma:bound-regret}[Lemma], 
    for any $\epsilon>0$, $\exists \delta>0$ such that for any $T\in \mathbb N$, $c_i>0$,
    $\mathbb E_{\mu_i}[v_i(\delta_{\sigma_{-i}})]-V_i(\mu_i)\leq (1-2 \exp(-2T\delta^2))\epsilon/4 + 8 \exp(-2T\delta^2) \|u_i\|_\infty +c_i T$.
    Letting $\epsilon = k$, $c_i=k/(4 T)$, we have 
    $
        \mathbb E_{\mu_i}[v_i(\delta_{\sigma_{-i}})]-V_i(\mu_i) \leq k/2 + 8 \exp(-2T\delta^2) \|u_i\|_\infty.
    $
    It is then straightforward to see that, for $T$ large enough (and $c_i>0$ small enough), 
    $\mathbb E_{\mu_i}[v_i(\delta_{\sigma_{-i}})]-V_i(\mu_i)<k=\mathbb E_{\mu_i}[v_i(\delta_{\sigma_{-i}})]-v_i(\mu_i)$, proving the result.
\end{proof}

Now consider all possible posteriors following any possible $t$ realizations, $\{\mu_i\mid y_i^t, y_i^t \in \cup_{t \in [1 .. T]}Y^t\}$, observing that this is a finite set.
Given the nature of the information process, since the prior $\mu_i$ has full support, so do the posterior beliefs.
By the above claim, for each $\mu_i\mid y_i^t$, there is a cost $c_i>0$ such that player $i$ would find it optimal to acquire at least one more signal. 
Taking the lowest of all such costs implies that under such cost, player $i$ would deem it optimal to acquire at least $T$ signals, concluding the proof.

\subsubsection{Proof of \hyref{theorem:monotonicity-actions}[Theorem]}
\label{appendix:proofs:theorem:monotonicity-actions}
Label player $i$'s actions so that $u_i(1,\sigma_{-i})- u_i(0,\sigma_{-i})$ is increasing in $\sigma_{-i}$.

\subsubsection*{Proof of \hyref{theorem:monotonicity-actions}[Theorem][(i)]}
We first prove a more general comparative statics result:
\begin{proposition}
    \label{proposition:monotonicity-actions-payoffs}
    $\mathbb P_{\sigma_{-i}}(a_i \in A_i^*(\mu_i\mid y_i^{\tau_i})\text{ and }\tau_i\leq t)$ is increasing with respect to $\geq_{a_i}$, and 
    $\mathbb P_{\sigma_{-i}}(a_i \notin A_i^*(\mu_i\mid y_i^{\tau_i})\text{ and }\tau_i\leq t)$ is decreasing with respect to $\geq_{a_i}$,
    for any $t \in \mathbb N$ and $\sigma_{-i}\in \Delta(A_{-i})$.
\end{proposition}
\vspace*{-1em}
\begin{proof}
    Let $\tilde{u}_i\geq_{a_i}u_i$ and denote the respective 
    (i) value functions, 
    (ii) earliest optimal stopping times, 
    (iii) optimal choices at given beliefs,
    and 
    (iv) 
    selections of optimal choices by
    (i) $\tilde{V}_i$ and $V_i$, 
    (ii) $\tilde{\tau}_i$ and $\tau_i$, 
    (iii) $\tilde{A}_i^*$ and $A_i^*$,
    and 
    (iv) $\tilde{\sigma}_i^*$ and $\sigma_i^*$, 
    respectively.
    With $g:=\tilde{u}_i(a_i,\cdot)-u_i(a_i,\cdot)$, by definition we obtain
    \begin{align*}
        &V_i(\mu_i)
        +
        \mathbb E_{\mu_i}[\tilde{\sigma}_i^*(\mu_i\mid y_i^{\tilde{\tau}_i})(a_i)g(\sigma_{-i}')]
        \,\geq\,
        \tilde{V}_i(\mu_i)
        \,\geq\,
        V_i(\mu_i)
        +
        \mathbb E_{\mu_i}[{\sigma}_i^*(\mu_i\mid y_i^{{\tau}_i})(a_i)g(\sigma_{-i}')]
        \,\geq\,
        V_i(\mu_i)
    \end{align*}
    \begin{lemma}
        \label{lemma:value-monotonicity-payoff-actions}
        $V_i(\mu_i)= \mathbb E_{\mu_i}[u_i(a_i,\sigma_{-i}')] \Longrightarrow \tilde{V}_i(\mu_i)= \mathbb E_{\mu_i}[\tilde{u}_i(a_i,\sigma_{-i}')]$
        and, for $a_i'\ne a_i$
        $\tilde{V}_i(\mu_i)= \mathbb E_{\mu_i}[\tilde{u}_i(a_i',\sigma_{-i}')] \Longrightarrow {V}_i(\mu_i)= \mathbb E_{\mu_i}[{u}_i(a_i',\sigma_{-i}')]$.
    \end{lemma}
    \vspace*{-1em}
    \begin{proof}
        Since 
        $\mathbb E_{\mu_i}[\tilde{\sigma}_i^*(\mu_i\mid y_i^{\tilde{\tau}_i})(a_i)g(\sigma_{-i}')]\leq \mathbb E_{\mu_i}[g(\sigma_{-i}')]$,
        if $V_i(\mu_i)= \mathbb E_{\mu_i}[u_i(a_i,\sigma_{-i}')]$, then
        $\mathbb E_{\mu_i}[\tilde{u}_i(a_i,\sigma_{-i}')]= \mathbb E_{\mu_i}[u_i(a_i,\sigma_{-i}')] + \mathbb E_{\mu_i}[g(\sigma_{-i}')]
        \geq V_i(\mu_i) + \mathbb E_{\mu_i}[\tilde{\sigma}_i^*(\mu_i\mid y_i^{\tilde{\tau}_i})(a_i)g(\sigma_{-i}')] 
        \geq \tilde{V}_i(\mu_i)\geq \mathbb E_{\mu_i}[\tilde{u}_i(a_i,\sigma_{-i}')]$.
        Moreover, 
        if $\tilde{V}_i(\mu_i)= \mathbb E_{\mu_i}[\tilde{u}_i(a_i',\sigma_{-i}')]=\mathbb E_{\mu_i}[{u}_i(a_i',\sigma_{-i}')]$,
        then
        $\mathbb E_{\mu_i}[{u}_i(a_i',\sigma_{-i}')]=\tilde{V}_i(\mu_i)\geq {V}_i(\mu_i)\geq \mathbb E_{\mu_i}[{u}_i(a_i',\sigma_{-i}')]$.
    \end{proof}
    \vspace*{-1em}
    Note that, by the contrapositive of \hyref{lemma:value-monotonicity-payoff-actions}[Lemma], $\tilde{V}_i(\mu_i)> \mathbb E_{\mu_i}[\tilde{u}_i(a_i,\sigma_{-i}')] \Longrightarrow V_i(\mu_i)> \mathbb E_{\mu_i}[u_i(a_i,\sigma_{-i}')]$ 
    and 
    ${V}_i(\mu_i)> \mathbb E_{\mu_i}[{u}_i(a_i',\sigma_{-i}')] \Longrightarrow \tilde{V}_i(\mu_i)> \mathbb E_{\mu_i}[\tilde{u}_i(a_i',\sigma_{-i}')]$ for $a_i'\ne a_i$.
    This implies
    \begin{align*}
        &\left\{\omega \in \Omega \;\bigg\vert\; \tau_i(\omega)\leq t\text{ and }a_i \in A_i^*(\mu_i\mid y_i^{\tau_i}(\omega))\right\}
        \\
        &
        =
        \left\{\omega \in \Omega \;\Bigg\vert\; 
        \exists t'\leq t:
        \begin{array}{l}
        V_i(\mu_i \mid y_i^{t'}(\omega)) = \mathbb E_{\mu_i}[u_i(a_i,\sigma_{-i}')\mid y_i^{t'}(\omega)]\\
        V_i(\mu_i \mid y_i^{\ell}(\omega)) > \mathbb E_{\mu_i}[u_i(a_i',\sigma_{-i}')\mid y_i^{t'}(\omega)],\, \forall \ell<t',\forall a_i'
        \end{array}
        \right\}
        \\
        &\subseteq
        \left\{\omega \in \Omega \;\Bigg\vert\; 
        \exists t'\leq t:
        \begin{array}{l}
        \tilde{V}_i(\mu_i \mid y_i^{t'}(\omega)) = \mathbb E_{\mu_i}[\tilde{u}_i(a_i,\sigma_{-i}')\mid y_i^{t'}(\omega)]\\
        \tilde{V}_i(\mu_i \mid y_i^{\ell}(\omega)) > \mathbb E_{\mu_i}[u_i(a_i',\sigma_{-i}')\mid y_i^{t'}(\omega)],\, \forall\ell<t', \forall a_i'\ne a_i
        \end{array}
        \right\}
        \\
        &=\left\{\omega \in \Omega \;\bigg\vert\; \tilde{\tau}_i(\omega)\leq t\text{ and }a_i \in \tilde{A}_i^*(\mu_i\mid y_i^{\tilde{\tau}_i}(\omega))\right\}
    \end{align*}
    and
    \begin{align*}
        &
        \left\{\omega \in \Omega \;\bigg\vert\; \tilde{\tau}_i(\omega)\leq t\text{ and }a_i \notin \tilde{A}_i^*(\mu_i\mid y_i^{\tilde{\tau}_i}(\omega))\right\}
        \\
        &
        =
        \left\{\omega \in \Omega \;\Bigg\vert\; 
        \exists t'\leq t:
        \begin{array}{l}
        \tilde{V}_i(\mu_i \mid y_i^{t'}(\omega)) = 
        \tilde{v}_i(\mu_i \mid y_i^{t'}(\omega)) >
        \mathbb E_{\mu_i}[\tilde{u}_i(a_i,\sigma_{-i}')\mid y_i^{t'}(\omega)]\\
        \tilde{V}_i(\mu_i \mid y_i^{\ell}(\omega)) > \mathbb E_{\mu_i}[u_i(a_i',\sigma_{-i}')\mid y_i^{t'}(\omega)],\, \forall\ell<t', \forall a_i'
        \end{array}
        \right\}
        \\
        &\subseteq
        \left\{\omega \in \Omega \;\Bigg\vert\; 
        \exists t'\leq t:
        \begin{array}{l}
        V_i(\mu_i \mid y_i^{t'}(\omega)) = v_i(\mu_i \mid y_i^{t'}(\omega))>\mathbb E_{\mu_i}[u_i(a_i,\sigma_{-i}')\mid y_i^{t'}(\omega)]\\
        V_i(\mu_i \mid y_i^{\ell}(\omega)) > \mathbb E_{\mu_i}[u_i(a_i,\sigma_{-i}')\mid y_i^{t'}(\omega)],\, \forall \ell<t'
        \end{array}
        \right\}
        \\
        &=
        \left\{\omega \in \Omega \;\bigg\vert\; \tau_i(\omega)\leq t\text{ and }a_i \notin A_i^*(\mu_i\mid y_i^{\tau_i}(\omega))\right\}.
    \end{align*}
\end{proof}
\vspace*{-1em}
The above, together with \hyref{lemma:never-stop-indifferent}[Lemma]---which is proved independently from \hyref{theorem:monotonicity-actions}[Theorem][(i)]---delivers the result.
Note that \hyref{proposition:monotonicity-actions-payoffs}[Proposition] further implies
\begin{corollary}
    \label{corollary:monotonicity-actions-payoffs}
    $\mathbb P_{\sigma_{-i}}(a_i \in (\text{resp. }\notin) A_i^*(\mu_i\mid y_i^{\tau_i})\,\mid\,\tau_i\leq t)$ is increasing (resp. decreasing) with respect to $\geq_{a_i}$, $\forall \sigma_{-i}\in \Delta(A_{-i})$ and $t \in \mathbb N$.
\end{corollary}

\subsubsection{Proof of \hyref{theorem:monotonicity-actions}[Theorem][(ii)]}
For $a_i \in \{0,1\}$, define (i) $u_i^{a_i}(a_i',\sigma_{-i}):=u_i(a_i',\sigma_{-i})-u_i(1-a_i,\sigma_{-i})$, (ii) $v_i^{a_i}(\mu_i):=\max_{a_i'\in A_i}\mathbb E_{\mu_i}[u_i^{a_i}(a_i',\sigma_{-i})]$, and (iii) $V_i^{a_i}(\mu_i):=\sup_{t_i \in \mathbb T_i}\mathbb E_{\mu_i}[v_i^{a_i}(\mu_i \mid y_i^{t_i})-c_i \cdot t_i]$.
Note that, by definition, $u_i^{a_i}(1-a_i,\sigma_{-i})=0$.
Moreover, $v_i^{a_i}(\mu_i)=v_i(\mu_i)-\mathbb E_{\mu_i}[u_i(1-a_i,\sigma_{-i})]$, which also implies that $V_i^{a_i}(\mu_i)=V_i(\mu_i)-\mathbb E_{\mu_i}[u_i(1-a_i,\sigma_{-i})]$.
A useful property of $V_i^{a_i}$ is as follows:
\begin{lemma}
    \label{lemma:mlr-monotonicity-Vi}
    For any $\mu_i'\geq_{SSD}\mu_i$
    $V_i^1(\mu_i')\geq V_i^1(\mu_i)$ and $V_i^0(\mu_i')\leq V_i^0(\mu_i)$.
\end{lemma}
\vspace*{-1em}
\begin{proof}
    Let $B_i^{a_i}:\mathcal C^0(\Delta(\Delta(A_{-i})))\to \mathcal C^0(\Delta(\Delta(A_{-i})))$ be such that 
    $B_i^{a_i}(w)(\mu_i):=\max\{v_i^{a_i}(\mu_i),\mathbb E_{\mu_i}[w(\mu_i\mid y_i)]-c_i\}$.
    As argued in \hyref{sec:sse}[Section], $V_i^{a_i}$ is a fixed-point of $B_i^{a_i}$.
    Moreover, by \hyref{remark:bounded-stopping}[Remark], there is a finite $n\in \mathbb N$, such that 
    $V_i^{a_i}={B_i^{a_i}}^{(n)}(v_i^{a_i})$, where ${B_i^{a_i}}^{(1)}=B_i^{a_i}$ and, for $n\geq 1$,
    ${B_i^{a_i}}^{(n+1)}=B_i^{a_i}\circ {B_i^{a_i}}^{(n)}$.
    
    Note that $v_i^1$ (resp. $v_i^0$ is increasing (resp. decreasing) in $\geq_{SSD}$.
    If $w \in \mathcal C^0(\Delta(\Delta(A_{-i})))$ is increasing in $\geq_{SSD}$, then so is $B_i^1(w)$---a symmetric argument applies to $B_i^0$.
    To see this, note that
    \begin{align*}
        B_i^1(w)(\mu_i')
        &=
        \max\{v_i^1(\mu_i'),\mathbb E_{\mu_i'}[\sigma_{-i}]w(\mu_i'\mid 1)+\mathbb E_{\mu_i'}[1-\sigma_{-i}]w(\mu_i'\mid 0)-c_i\}
        \\
        &\geq
        \max\{v_i^1(\mu_i),\mathbb E_{\mu_i'}[\sigma_{-i}]w(\mu_i\mid 1)+\mathbb E_{\mu_i'}[1-\sigma_{-i}]w(\mu_i\mid 0)-c_i\}
        \\
        &\geq
        \max\{v_i^1(\mu_i),\mathbb E_{\mu_i}[\sigma_{-i}]w(\mu_i\mid 1)+\mathbb
        E_{\mu_i}[1-\sigma_{-i}]w(\mu_i\mid 0)-c_i\}=B_i^1(w)(\mu_i),
    \end{align*}
    where the first inequality follows from monotonicity of $v_i^1$ with respect to $\geq_{SSD}$, by monotonicity of $w$ and the fact that $\mu_i'\geq_{SSD}\mu_i\Longrightarrow \mu_i'|y_i\geq_{SSD}\mu_i|y_i$ for $y_i \in \{0,1\}$, and the second because $\mathbb E_{\mu_i'}[\sigma_{-i}]\geq\mathbb E_{\mu_i}[\sigma_{-i}]$ (by FOSD) and, as can be shown, $\mu_i|1\geq_{SSD}\mu_i|0$, implying $w(\mu_i|1)\geq w(\mu_i|0)$.
\end{proof}
\vspace*{-1em}

\hyref{lemma:mlr-monotonicity-Vi}[Lemma] implies:
\begin{corollary}
    \label{corollary:mlr-monotonicity}
    Let $\mu_i' \geq_{SSD} \mu_i$.
    $V_i(\mu_i'\mid y_i^t)=\mathbb E_{\mu_i'}[u_i(0,\sigma_{-i})\mid y_i^t]\Longrightarrow 
    0=V_i^1(\mu_i'\mid y_i^t)\geq V_i^1(\mu_i\mid y_i^t)\geq 0
    \Longrightarrow 
    V_i(\mu_i\mid y_i^t)=\mathbb E_{\mu_i}[u_i(0,\sigma_{-i})\mid y_i^t]
    $
    and 
    $V_i(\mu_i\mid y_i^t)=\mathbb E_{\mu_i}[u_i(1,\sigma_{-i})\mid y_i^t]\Longrightarrow 
    0=V_i^0(\mu_i\mid y_i^t)\geq V_i^0(\mu_i'\mid y_i^t)\geq 0
    \Longrightarrow 
    V_i(\mu_i'\mid y_i^t)=\mathbb E_{\mu_i'}[u_i(1,\sigma_{-i})\mid y_i^t]
    $.
\end{corollary}

In order to conclude the proof of \hyref{theorem:monotonicity-actions}[Theorem][(ii)], let $\tau_i$ and $\tau_i'$ denote the earliest optimal stopping times associated with $\mu_i$ and $\mu_i'$. 
Then, by \hyref{corollary:mlr-monotonicity}[Corollary],
$\left\{\omega \in \Omega \;\bigg\vert\; \tau_i(\omega)\leq t\text{ and }1 \in (\text{resp.} \notin) A_i^*(\mu_i\mid y_i^{\tau_i}(\omega))\right\} \subseteq (\text{resp.} \supseteq)
\left\{\omega \in \Omega \;\bigg\vert\; {\tau}_i'(\omega)\leq t\text{ and }1 \in (\text{resp.} \notin) {A}_i^*(\mu_i'\mid y_i^{{\tau}_i'}(\omega))\right\}
$.

\subsubsection{Proof of \hyref{theorem:monotonicity-actions}[Theorem][(iii)]}
Let 
\[\mathcal N:=\left\{n \in \mathbb N_0^2\mid \exists y_i^t: \text{(i) } t=n_0+n_1\text{,\,\, (ii) } \sum_{\ell \in [1\,..\,t]}y_{i,\ell}=n_1\text{,\,\, and \,\,(iii) } \forall \ell\leq t, V_i(\mu_i\mid y_i^\ell)>v_i(\mu_i\mid y_i^\ell)\right\},\]
and, for $j \in \{0,1\}$, let $\mathcal N_j:=\{n \in \mathbb N_0^2\mid n-(j,1-j) \in \mathcal N\}$.
Note that, if $n \in \mathcal N_j$, then there is some sequence $y_i^t$ satisfying $t=n_0+n_1$, $\sum_{\ell\in [1\,..\,t]}y_{i,\ell}=n_1$, and along which player $i$ decides to keep sampling every period (according to $\tau_i$), i.e. $V_i(\mu_i\mid y_i^\ell)>v_i(\mu_i\mid y_i^\ell)$ for all $\ell< t$, and decides to stop at $y_i^t$ and take action $j$---a consequence of \hyref{lemma:never-stop-indifferent}[Lemma].
Let $T_i:=\sup\supp\{\tau_i\}$ (where $\supp$ is defined with respect to $\mu_i$).
By \hyref{remark:bounded-stopping}[Remark], $T_i<\infty$ and thus, $\forall n \in \mathcal N$, $n_0+n_1< T_i$ and $\mathcal N$ is finite.
Below we implicitly rely on the fact that, if $(n_0,n_1),(n_0',n_1') \in \mathcal N_j$ and $n_{1-j}'>n_{1-j}$, then $n_j'\geq n_j$, which is implied by \hyref{corollary:mlr-monotonicity}[Corollary].

We recursively define the probability of stopping and choosing action 1.
Define the asymmetric part of a linear order on $\mathcal N$ given by $n\triangleright n'$ if and only if $n_1'>n_1$ or $n_1'=n_1$ and $n_0'>n_0$.
Let $p:\mathcal N\times [0,1] \to [0,1]$ be given by
$p(n;\sigma_{-i}):=\sigma_{-i}$ if $n+(0,1) \in \mathcal N_1$ and $p(n;\sigma_{-i}):=\sigma_{-i}p(n+(0,1);\sigma_{-i})+(1-\sigma_{-i})p(n+(1,0);\sigma_{-i})$ if otherwise; $p$ can be recursively defined on $n \in \mathcal N$ increasing with respect to $\triangleright$.
Extend $p$ to $n \in \mathcal N_j$ by letting $p(n;\sigma_{-i})=j$ if $n \in \mathcal N_j$.
Note that $p((0,0); \sigma_{-i})=\mathbb P_{\sigma_{-i}}(1=A_i(\mu_i\mid y_i^{\tau_i}))=\mathbb E_{\sigma_{-i}}[\sigma_i^*(\mu_i\mid y_i^{\tau_i})]$.

We now show by induction that, if $(0,0) \in \mathcal N$, $p((0,0);\bullet)$ is $\mathcal C^\infty$ and strictly increasing.
Note that for $n:n'\triangleright n$ for all $n'\ne n$ in $\mathcal N$, $n_0+n_1=T_i-1$ and $p(n;\sigma_{-i})=\sigma_{-i}$ is $\mathcal C^\infty$ and strictly increasing in $\sigma_{-i}$ and $0=p(n+(0,1);\sigma_{-i})< p(n+(0,1);\sigma_{-i})=1$, $\forall \sigma_{-i}\in [0,1)$.

Suppose that, for all $n'\in \mathcal N:n\triangleright n'$, $p(n';\bullet)$ is $\mathcal C^\infty$ and strictly increasing, and that $p(n'+(1,0);\bullet)<p(n'+(0,1);\bullet)$.
As $p(n;\sigma_{-i})=\sigma_{-i}p(n+(0,1);\sigma_{-i})+(1-\sigma_{-i})p(n+(1,0);\sigma_{-i})$ and $n\triangleright n+(1,0),n+(0,1)$, then $p(n+(0,1);\bullet)$ and 
$p(n+(1,0);\bullet)$ are $\mathcal C^\infty$, strictly increasing, and satisfy $p(n+(1,0);\bullet)<p(n+(0,1);\bullet)$.
Then $p(n;\bullet)\in \mathcal C^\infty$ and $\frac{\partial}{\partial \sigma_{-i}}p(n;\sigma_{-i})= p(n+(0,1);\sigma_{-i})-p(n+(1,0);\sigma_{-i})+\sigma_{-i}\frac{\partial}{\partial \sigma_{-i}}p(n+(0,1);\sigma_{-i})+(1-\sigma_{-i})\frac{\partial}{\partial \sigma_{-i}}p(n+(1,0);\sigma_{-i})>0$ for all $\sigma_{-i}\in [0,1)$.

To obtain that $\mathbb P_{\sigma_{-i}}(1=A_i^*(\mu_i\mid y_i^{\tau_i})\text{ and }\tau_i\leq t)=\mathbb E_{\sigma_{-i}}[\sigma_i^*(\mu_i\mid y_i^{\tau_i})1_{\tau_i\leq t}]$ is also $\mathcal C^\infty$ and strictly increasing in $\sigma_{-i}\in [0,1)$ for $t\geq n_1(0)$, 
it is necessary to restrict $\mathcal N$.
Define $\underline n_1(m):=n_1$ if $(m,n_1)\in \mathcal N_1$ and let
$\mathcal N^t:=\{n\in \mathcal N\mid n_0+\underline n_1(n_0)\leq t\}$, 
$\mathcal N_j^t:=\{n\mid n-(1-j,j) \in \mathcal N\}$.
Let 
$p^t:\mathcal N\times [0,1] \to [0,1]$ be given by
$p^t(n;\sigma_{-i}):=\sigma_{-i}$ if $n+(0,1) \in \mathcal N_1^t$ and $p^t(n;\sigma_{-i}):=\sigma_{-i}p(n+(0,1);\sigma_{-i})+(1-\sigma_{-i})p(n+(1,0);\sigma_{-i})$ if otherwise.
Extend $p^t$ to $n \in \mathcal N_j^t$ by letting $p^t(n;\sigma_{-i})=j$ if $n \in \mathcal N_j^t$.
An analogous inductive argument applied to $p^t$ delivers the result.
For $t< n_1(0)$, $\mathbb P_{\sigma_{-i}}(1=A_i^*(\mu_i\mid y_i^{\tau_i})\text{ and }\tau_i\leq t)=0$, for all $\sigma_{-i}\in [0,1]$.

\subsubsection{Proof of \hyref{lemma:never-stop-indifferent}[Lemma]}
\label{appendix:proofs:lemma:never-stop-indifferent}
Label player $i$'s actions so that $u_i(1,\sigma_{-i})- u_i(0,\sigma_{-i})$ is increasing in $\sigma_{-i}$.
If, (i) $u_i(1,\sigma_{-i})- u_i(0,\sigma_{-i})\leq 0$ $\forall \sigma_{-i}$ or (ii) $u_i(1,\sigma_{-i})- u_i(0,\sigma_{-i})\geq 0$ $\forall \sigma_{-i}$, then $\tau_i =0$.
Suppose then that $\tau_i>0$ and note $a_{-i}=\argmax_{a_i \in A_i}u_i(a_i,a_{-i})$.
Let $V_i(\mu_i)>v_i(\mu_i)$ and $V_i(\mu_i\mid a_{-i})=\mathbb E_{\mu_i}[u_i(1-a_{-i},\sigma_{-i})\mid a_{-i}]$ and observe that, from  \hyref{lemma:mlr-monotonicity-Vi}[Lemma],
$0=V_i^{a_{-i}}(\mu_i\mid a_{-i})\geq V_i^{a_{-i}}(\mu_i)\geq 0$, which implies $V_i(\mu_i)=\mathbb E_{\mu_i}[u_i(1-a_{-i},\sigma_{-i})]\leq v_i(\mu_i)$, a contradiction.

\subsubsection{Proof of \hyref{proposition:unique-sse-ne}[Proposition]}
\label{appendix:proofs:proposition:unique-sse-ne}
Note that in a binary action game there are multiple unique Nash equilibria if and only if for any $i \in I$, $u_i(1,1)-u_i(0,1),u_i(0,0)-u_i(1,0)>0$.
This implies that, for both players, both actions are undominated and so, by \hyref{lemma:lower-bound-stopping}[Lemma], players sample at least once whenever the sampling costs are sufficiently low.
By \hyref{lemma:never-stop-indifferent}[Lemma], if $\sigma_{-i}=1$ (=0) and $\tau_i>0$, then $\mathbb E_{\sigma_{-i}}[\sigma_i^*(\mu_i\mid y_i^{\tau_i})]=1$ (=0).
Hence, (0,0) and (1,1) are both Nash equilibria and sequential sampling equilibria.

Recall that, by assumption, $u_i(1,\sigma_{-i})-u_i(0,\sigma_{-i})$ is strictly monotone.
If there is a unique Nash equilibrium, either (i) both players have a weakly dominant action; (ii) one player has both actions undominated, and the other has a weakly dominant action; or (iii) both players have undominated actions.
In (i), uniqueness of a sequential sampling equilibrium follows as both players always choose their weakly dominant action.
In (ii), the player with the weakly dominant action chooses it with probability 1, and the opponent, whenever they sample at least once, by \hyref{lemma:never-stop-indifferent}[Lemma], will choose the best response to the weakly dominant action with probability 1, thus entailing a unique sequential sampling equilibrium.
In (iii), uniqueness of a Nash equilibrium implies a payoff structure akin to a matching pennies game: one player, $i$, wants to match the other, i.e. $u_i(1,\sigma_{j})-u_i(0,\sigma_{j})$ is strictly increasing, and their opponent $j$ seeks to mismatch, i.e. $u_j(1,\sigma_{i})-u_j(0,\sigma_{i})$ is strictly decreasing.
Consequently, by \hyref{theorem:monotonicity-actions}[Theorem], $\sigma_j\mapsto b_i(\sigma_j)=\mathbb E_{\sigma_j}[\sigma_i^*(\mu_i\mid y_i^{\tau_i})]\in [0,1]$ is increasing, and $\sigma_i \mapsto b_j(\sigma_i)=\mathbb E_{\sigma_i}[\sigma_j^*(\mu_j\mid y_j^{\tau_j})]\in [0,1]$ is decreasing, both are continuous, and therefore their graph has a unique intersection.
\hyref{lemma:never-stop-indifferent}[Lemma] clarifies that the sequential sampling equilibrium is unaffected by the selection of tie-breaking $\sigma_i^*$, $\sigma_j^*$.

\subsubsection{Proof of \hyref{proposition:collapsing-bounds}[Proposition]}
\label{appendix:proofs:proposition:collapsing-bounds}
Let $B(\sigma_{-i},t)$ to denote a Beta distribution with parameters $\alpha,\beta\geq 0$, such that $t=\alpha+\beta>0$ and $\sigma_{-i}=\alpha/t$, with the convention that $B(1,t)$ and $B(0,t)$ correspond to Dirac measures on $1$ and $0$, respectively.
Label player $i$'s actions so that $u_i(1,\sigma_{-i})- u_i(0,\sigma_{-i})$ is increasing in $\sigma_{-i}$.

Let $V_i^{a_i}$ be as defined in the proof of \hyref{lemma:mlr-monotonicity-Vi}[Lemma].
We show the following properties of $V_i^{a_i}$:
\begin{lemma}
    $V_i^1(B(\sigma_{-i},t))$ (resp. $V_i^0(B(\sigma_{-i},t))$) is 
    (1) increasing (resp. decreasing) in $\sigma_{-i}$;
    (2) convex in $\sigma_{-i}$;
    (3) decreasing in $t$;
    (4) continuous in $(\sigma_{-i},t)$.
\end{lemma}
\vspace*{-1em}
\begin{proof}
    (1) follows immediately from \hyref{lemma:mlr-monotonicity-Vi}[Lemma], since for $\sigma_{-i}>\sigma_{-i}'$, $B(\sigma_{-i},t)\geq_{SSD} B(\sigma_{-i}',t)$.
    
    For (2), note that $V_i^1(B(\sigma_{-i},t))=V_i^0(B(\sigma_{-i},t))+u_i(1,\sigma_{-i})-u_i(0,\sigma_{-i})$, and thus it suffices to show convexity of $V_i^1(B(\sigma_{-i},t))$ in $\sigma_{-i}$.
    Let $z(\sigma_{-i},t)$ be a random variable that delivers $\frac{1}{t+1}(t\sigma_{-i}+1)$ with probability $\sigma_{-i}$, and $\frac{1}{t+1}(t\sigma_{-i}+0)$ with probability $1-\sigma_{-i}$.
    Note that, for any $w:[0,1]\times \mathbb R_{++}$ that is increasing and convex in the first argument, for any $\sigma_{-i}\geq \sigma_{-i}'$, $\lambda\in (0,1)$, and $\sigma_{-i}'':=\lambda \sigma_{-i}+(1-\lambda)\sigma_{-i}'$, 
    straightforward algebra shows that $\mathbb E[w(z(\sigma_{-i}'',t))]\leq \lambda \mathbb E[w(z(\sigma_{-i},t))]+(1-\lambda)\mathbb E[w(z(\sigma_{-i}',t))]$.
    
    Observe that, by definition, $v_i^1(B(\sigma_{-i},t))=\max_{a_i \in A_i}u_i(a_i,\sigma_{-i})-u_i(0,\sigma_{-i})$, which is increasing and convex in $\sigma_{-i}$ and invariant with respect to $t$.
    
    Now we show that if $w(B(\sigma_{-i},t))$ is increasing and convex in $\sigma_{-i}$, so is $B_i^1(w)(B(\sigma_{-i},t))$:
    \begin{align*}
        &B_i^1(w)(B(\sigma_{-i}'',t))
        =
        \max
        \left\{
            v_i^1(B(\sigma_{-i}'',t))\,,\,
            \mathbb E[w(z(\sigma_{-i}'',t+1))]
            -c_i
        \right\}
        \\
        &\leq 
        \max
        \{
            \lambda v_i^1(B(\sigma_{-i},t)) + (1-\lambda) v_i^1(B(\sigma_{-i}',t))\,,\,
            \lambda \mathbb E[w(z(\sigma_{-i},t+1))]
            +
            (1-\lambda) \mathbb E[w(z(\sigma_{-i}',t+1))]
            -c_i
        \}
        \\
        &
        \leq
        \lambda B_i^1(w)(B(\sigma_{-i},t))+(1-\lambda) B_i^1(w)(B(\sigma_{-i}',t)).
    \end{align*}
    By similar arguments as in \hyref{lemma:mlr-monotonicity-Vi}[Lemma]---
    $V_i^1$ is a fixed point of $B_i^1$ which can be obtained by applying the $n$-th composition of $B_i^1$ with itself to $v_i^1$
    ---
    we have that $V_i^1(B(\sigma_{-i},t))$ is convex in $\sigma_{-i}$, and thus so is $V_i(B(\sigma_{-i},t))$.
    
    For (3), note that $\mathbb E[z(\sigma_{-i},t)]=\sigma_{-i}$ and that, for $t<t'$, $z(\sigma_{-i},t)$ is a mean-preserving spread of $z(\sigma_{-i},t')$; hence for any convex function $f$, $\mathbb E[f(z(\sigma_{-i},t))]\geq \mathbb E[f(z(\sigma_{-i},t'))]$.
    Take any $w$ such that $w(B(\sigma_{-i},t))$ is convex in $\sigma_{-i}$ and decreasing in $t$.
    Then, for any $t'>t$,
    \begin{align*}
        &B_i(w)(B(\sigma_{-i},t'))
        =
        \max
        \left\{
            v_i(B(\sigma_{-i},t'))\,,\,
            \mathbb E[w(B(z(\sigma_{-i},t'),t'))]
            -c_i
        \right\}
        \\
        &=
        \max
        \left\{
            v_i(B(\sigma_{-i},t))\,,\,
            \mathbb E[w(B(z(\sigma_{-i},t'),t'))]
            -c_i
        \right\}
        \leq 
        \max
        \left\{
            v_i(B(\sigma_{-i},t))\,,\,
            \mathbb E[w(B(z(\sigma_{-i},t'),t))]
            -c_i
        \right\}
        \\
        &
        \leq
        \max
        \left\{
            v_i(B(\sigma_{-i},t))\,,\,
            \mathbb E[w(B(z(\sigma_{-i},t),t))]
            -c_i
        \right\}
        =
        B_i(w)(B(\sigma_{-i},t)).
    \end{align*}
    By the same argument as before, $V_i(B(\sigma_{-i},t))$ is then decreasing in $t$.
    
    Finally, (4) follows immediately from the fact that for any $(\sigma_{-i}^n,t^n)\to (\sigma_{-i},t)$, $B(\sigma_{-i}^n,t^n)\to B(\sigma_{-i},t)$ (with respect to $\|\cdot\|_{LP}$), and therefore by \hyref{proposition:properties-value-stopping}[Proposition], $V_i(B(\sigma_{-i},t))$ is continuous in $(\sigma_{-i},t)$.
\end{proof}

By \hyref{corollary:mlr-monotonicity}[Corollary], if $V_i(B(\sigma_{-i},t))=u_i(a_i,\sigma_{-i})$ for $a_i=1$ (resp. $a_i=0$), then $V_i(B(\sigma_{-i}',t))=u_i(a_i,\sigma_{-i}')$ for any $\sigma_{-i}'\geq \sigma_{-i}$ (resp. $\leq$), since $B(\sigma_{-i}',t)\geq_{SSD}B(\sigma_{-i},t)$.
Define $\overline \sigma_{-i}(t):=\min \{\sigma_{-i}\in [0,1]\mid V_i(B(\sigma_{-i},t))=u_i(1,\sigma_{-i})\}$ and 
$\underline \sigma_{-i}(t):=\max \{\sigma_{-i}\in [0,1]\mid V_i(B(\sigma_{-i},t))=u_i(0,\sigma_{-i})\}$.

That $\overline \sigma_{-i}(t)$ is continuous and decreasing in $t$ follows from continuity of $V_i(B(\sigma_{-i},t))$ in $(\sigma_{-i},t)$, continuity of $u_i(1,\sigma_{-i})$ in $\sigma_{-i}$, and the fact that $V_i(B(\sigma_{-i},t))$ is decreasing in $t$ and $u_i(1,\sigma_{-i})-u_i(0,\sigma_{-i})$ is increasing in $\sigma_{-i}$.
An analogous argument applies to show that $\underline \sigma_{-i}(t)$ is continuous and increasing in $t$.

Finally, we prove that both these functions converge to $\tilde \sigma_{-i}$.
To see this, note that for $\mu_i$ given by $B(\sigma_{-i},t)$, simple algebra shows that
$\mathbb E_{\mu_i}[v_i(\mu_i\mid y_i)]-v_i(\mu_i)=\mathbb E[v_i(z(\sigma_{-i},t))]-v_i(\sigma_{-i})$ 
is maximized for any $t$ at $\sigma_{-i}=\tilde \sigma_{-i}:\tilde \sigma_{-i}u_i(1,1)+(1-\tilde \sigma_{-i})u_i(1,0)=\tilde \sigma_{-i}u_i(0,1)+(1-\tilde \sigma_{-i})u_i(0,0)$,
with maximum value $\frac{1}{t+1}(u_i(1,1)-u_i(0,1)+u_i(0,0)-u_i(1,0))(1-\tilde\sigma_{-i})\tilde \sigma_{-i}$. 
Therefore, it is always optimal to keep sampling at belief $B(\tilde \sigma_{-i},t)$ if
$t<T:=(u_i(1,1)-u_i(0,1)+u_i(0,0)-u_i(1,0))(1-\tilde\sigma_{-i})\tilde \sigma_{-i}/c_i-1$.
From here, one can deduce that $\tau_i\leq T$ for any Beta prior and that $\overline \sigma_{-i}(t)>\tilde \sigma_{-i}>\underline \sigma_{-i}(t)$ for all $t<T$ and $\overline \sigma_{-i}(t)=\tilde \sigma_{-i}=\underline \sigma_{-i}(t)$, for all $t\geq T$.

\subsubsection{Proof of \hyref{theorem:convergence-ne}[Theorem]}
\label{appendix:proofs:theorem:convergence-ne}
We prove the result when priors allow for correlation; adjusting the proof to accommodate the case in which they do not is tedious but straightforward.

Let $\Sigma_i^*(\mu_i):= \argmax_{\sigma_i \in \Delta(A_i)}\mathbb E_{\mu_i}[u_i(\sigma_i,\sigma_{-i})]$ denote the set of maximizers at belief $\mu_i$.

We first prove that if a sequence of probability measures $\mu_i^m \in \Delta(\Delta(A_{-i}))$ weak$^*$ converges to $\delta_{\sigma_{-i}}$ and $a_i \in A_i$ is not a best response to $\sigma_{-i}$, then for any sequence of distributions $\sigma_i^m \in \Sigma_i^*(\mu_i^m)$, $\sigma_i^m(a_i)\to 0$.
Note that $\mathbb E_{\mu_i}[u_i(\sigma_i,\sigma_{-i})]$ is jointly continuous in $(\sigma_i,\mu_i)$ with respect to the product metric, where $\Delta(A_i)$ is endowed with the standard Euclidean metric and $\Delta(\Delta(A_{-i}))$ with the Lévy-Prokhorov metric.
Then, by Berge's maximum theorem, $\Sigma_i^*$ is upper-hemicontinuous and compact-valued.
Supposing that $\sigma_i^m(a_i)$ does not converge to zero, implies that for any convergent subsequence of $\sigma_i^m$, its limit assigns strictly positive probability to $a_i$ being chosen, while also belonging to $\Sigma_i^*(\delta_{\sigma_{-i}})$, a contradiction.

Now take any sequence of profiles of action distributions $\{\sigma^n\}_n$ as in the statement of the theorem and let $\{\sigma^{m}\}_{m}$ be a convergent subsequence of $\{\sigma^n\}_n$ with limit $\sigma^k$.
Suppose that $\sigma_i^k$ is not a best response to $\sigma_{-i}^k$.
This implies that $\exists a_i \in A_i$ such that $\sigma_i^k(a_i)\geq \delta>0$ for some $\delta>0$ and $a_i$ is not a best response to $\sigma_{-i}^k$.
By continuity of $u_i$, note that if $a_i$ is not a best response to $\sigma_{-i}^k$, then it is not a best response to any $\sigma_{-i}\in B_\epsilon(\sigma_{-i}^k)$ for small enough $\epsilon>0$.
By convergence of $\sigma^m$, we then obtain that for all large enough $m$, $a_i$ is not a best response to $\sigma_{-i}^m$ and $\sigma_i^m(a_i)\geq \delta/2$.
That is, 
$\mathbb E_{\sigma_{-i}^m}[\sigma_i^*(\mu_i\mid y_i^{\tau_i^m})(a_i)]\geq \delta/2$ $\forall$ large enough $m$, from which we deduce
$\mathbb P_{\sigma_{-i}^m}(\sigma_i^*(\mu_i\mid y_i^{\tau_i^m})(a_i)\geq \delta/4)\geq \delta/4$.
In turn, from the above, this implies that there must be $\epsilon>0$ such that
$\mathbb P_{\sigma_{-i}^m}(\|\mu_i\mid y_i^{\tau_i^m} - \delta_{\sigma_{-i}^m}\|_{LP}\geq \epsilon)\geq \delta/4$.
We now prove that this cannot be the case, that is, we show that $\lim_{m\to \infty}\mathbb P_{\sigma_{-i}^k}(\|\mu_i\mid {y_i}^{\tau_i^m}- \delta_{\sigma_{-i}^k}\|_{LP}>\epsilon)=0$.

It would be natural to expect that, with sampling costs going to zero, optimal stopping time grows unboundedly and, by the law of large numbers, players learn the true distribution of actions of their opponents, best respond to it, and sequential sampling equilibrium converges to a Nash equilibrium.
But, \emph{conditional on stopping}, the set of signals are neither independent or identically distributed, so we cannot apply the law of large number directly.
We then need to take a detour.

Denote player $i$'s the associated earliest optimal stopping time by $\tau_i^m$ and their value function (which depends on $c_i^m$) as $V_i^m$.
From \hyref{lemma:lower-bound-stopping}[Lemma], there is $\{T^{m}\}_{m}$ such that $\tau_i^{m}\geq T^m$ and $T^m \uparrow \infty$.

By \citet{DiaconisFreedman1990AnnStat}, there is $\epsilon(t)$ nonincreasing and such that $\epsilon(t)\to 0$ as $t \to \infty$ such that 
$\|\mu_i\mid y_i^{t} - \delta_{\overline{y_i}^{t}}\|_{LP}\leq \epsilon(t)$ uniformly over sequences of $t$ observations, $y_i^t \in \mathcal Y_i$.
Since, taking $y_{i,\ell}\sim \sigma_{-i}$,
$\|\delta_{\overline{y_i}^{t}}-\delta_{\sigma_{-i}}\|_{LP}=\|{\overline{y_i}^{t}}-{\sigma_{-i}}\|$ is a (bounded) supermartingale with respect to $\sigma_{-i}$,
by the optional stopping theorem, for $\tau_i\geq t$,
$\mathbb E_{\sigma_{-i}}[\|\delta_{\overline{y_i}^{\tau_i}}-\delta_{\sigma_{-i}}\|_{LP}
]\leq \mathbb E_{\sigma_{-i}}[\|\delta_{\overline{y_i}^{t}}-\delta_{\sigma_{-i}}\|_{LP}]$.
Hence,
for any $\sigma_{-i}$,
$\mathbb E_{\sigma_{-i}}[\|\mu_i\mid {y_i}^{\tau_i^m}-\delta_{\sigma_{-i}}\|_{LP}
]
\leq
\mathbb E_{\sigma_{-i}}[\|\delta_{\overline{y_i}^{\tau_i^m}}-\delta_{\sigma_{-i}}\|_{LP}\mid y_i^{T^m}]+\epsilon(T^m)
\leq
\mathbb E_{\sigma_{-i}}[\|\delta_{\overline{y_i}^{T^m}}-\delta_{\sigma_{-i}}\|_{LP}]+\epsilon(T^m)
\leq
\mathbb E_{\sigma_{-i}}[\|{\overline{y_i}^{T^m}}-{\sigma_{-i}}\|]+\epsilon(T^m)
$.

Let $x_m:=\mathbb E_{\sigma_{-i}^m}[\|{\overline{y_i}^{T^m}}-{\sigma_{-i}^m}\|]\in [0,2]$.
Suppose that $\{x_m\}_m$ does not converge to 0.
Take any convergent subsequence $x_\ell\to \gamma>0$.
For all large enough $\ell$, $x_\ell\geq \gamma/2$.
That is,
$\mathbb E_{\sigma_{-i}^\ell}[\|{\overline{y_i}^{T^\ell}}-{\sigma_{-i}^\ell}\|]\geq \gamma/2$,
implying that 
$\mathbb P_{\sigma_{-i}^\ell}(\|{\overline{y_i}^{T^\ell}}-{\sigma_{-i}^\ell}\|\geq \gamma/4)\geq \gamma/4$,
as otherwise 
$\mathbb E_{\sigma_{-i}^\ell}[\|{\overline{y_i}^{T^\ell}}-{\sigma_{-i}^\ell}\|]\leq (1-\gamma/4)\gamma/4+\gamma/4\leq \gamma/4<\gamma/2$.
However, by the Dvoretzky--Kiefer--Wolfowitz--Massart inequality \citep{Massart1990AnnProb}
$\mathbb P_{\sigma_{-i}^\ell}(\|{\overline{y_i}^{T^\ell}}-{\sigma_{-i}^\ell}\|\geq \gamma/4)\leq 2\exp(-T^\ell\gamma^2/8)\to 0$, a contradiction.

We conclude that 
$\lim_{m\to \infty}\mathbb E_{\sigma_{-i}^m}[\|\mu_i\mid {y_i}^{\tau_i^m}-\delta_{\sigma_{-i}^m}\|_{LP}
]\leq \mathbb E_{\sigma_{-i}^m}[\|{\overline{y_i}^{T^m}}-{\sigma_{-i}^m}\|]+\epsilon(T^m)\to 0$.

\subsubsection{Proof of \hyref{proposition:robust-reachability}[Proposition]}
\label{appendix:proofs:proposition:robust-reachability}
By \citet{DiaconisFreedman1990AnnStat}, for any $\mu_i$, there is $T_i<\infty$ such that $\forall t\geq T_i$, 
$\mathbb E_{\mu_i}[\sigma_{-i}'\mid a_{-i}^t]\in B_{\epsilon_i}(\delta_{a_{-i}})$.
Take $T:=\max_{i\in I}T_i$.
By \hyref{lemma:lower-bound-stopping}[Lemma], there is $\overline c$ such that, $\tau_i\geq T$ for every player $i$ for which $a_i$ is not always a best response (where $\overline c$ may depend on $\mu$).
Hence, for any $\mu$ there is an $N$ such that $\forall n\geq N$, $a$ is a sequential sampling equilibrium of $\langle \Gamma, \mu, c^n\rangle$.

\newpage

\section*{Online Appendix}
\renewcommand{\thesection}{Appendix \Alph{section}}
\renewcommand{\thesubsection}{\Alph{section}.\arabic{subsection}}

\section{General Information Structures}
\label{appendix:information-structures}
In this section, we extend sequential sampling equilibrium to accommodate analogy partitions and more general information structures.
For brevity, let us consider the case in which players' beliefs allows for correlation.

\subsection{Existence of a Sequential Sampling Equilibrium}
Let us simplify the notation in our baseline setup: $Y_i:=A_{-i}$, $P_i:=\Delta(Y_i)$, $\mu_i \in \Delta(P_i)$ with full support, and $u_i:A_i\times P_i \to \mathbb R$, a continuous function.

An analogy partition for player $i$ can be represented by a surjective function 
$f_i:Y_i\to Z_i$, where $|Z_i|< |Y_i|<\infty$.
Naturally, it defines a garbling: instead of observing $y_i$, the player has access to coarser information $f_i(y_i)$.
More generally, one can consider $|Z_i|\times |Y_i|$ stochastic matrices $B_i$, where $B_i(z,y) \in [0,1]$ and $\sum_{z \in Z}B_i(z,y)=1$, and such that $B_i$ has rank $|Z_i|<|Y_i|$.\footnote{
	Allowing for $|Z_i|=|Y_i|$ is possible, but makes the proofs more cumbersome.
}

Let $Q_i:=\Delta(Z_i)$ and $\nu_i$ be the pushforward measure on $Q_i$ given $\mu_i$ and $B_i$, where for every measurable set $S \subseteq Q_i$,
$\nu_i(S):=\mu_i(\{p_i \in P_i\mid B_ip_i = \in S\})$.
We assume that the player now has access to iid draws from a fixed $B_i p_{i,0} \in Q_i$.
It is straightforward to adjust the definition of the optimal stopping problem, expand the definition of an extended game with the additional primitives $\{\pi_i\}_{i\in I}$, where $\pi_i$ denotes the information structure defined by ($B_i$, $Z_i$), and have sequential sampling equilibrium accommodate such more general information structures.

We provide the following sufficient condition for existence of a sequential sampling equilibrium:
\begin{theorem} 
    Let $G:=\langle \Gamma, \mu, c, \pi\rangle$ be an extended game such that for every player $i$, $\mu_i$ admits a continuous density, and $\text{rank}(B_i)=|Z_i|<|A_{-i}|$.
    Then $G$ admits a sequential sampling equilibrium.
\end{theorem}
\vspace*{-1em}
\begin{proof}
    Let $J_{B_i}:=\text{det}(B_iB_i^T)$, which is strictly positive, as $\text{rank}(B_i)=|Z_i|$.
    For convenience, define $\pi_i:P_i\to Q_i$ as $\pi_i(p_i):=B_ip_i$.
    Denote by $g_{\mu_i}$ the density of $\mu_i$.
    Denoting $\lambda^n$ the $n$-dimensional Lebesgue measure, by the coarea formula \citep[see][Theorem 3.10]{EvansGariepy2015Book} we have
    \begin{align*}
        \mathbb E_{\mu_i}[u_i(a_i,p_i)\mid z_i^t]
        &=
        \int_{P_i}
        \prod_{\ell \in [1 .. t]}(B_i p_i)(z_{i,\ell}^t)
        u_i(a_i,p_i)g_{\mu_i}(p_i)
        \diff\lambda^{|Y_i|-1}(p_i)
        \\
        &=
        \int_{Q_i} 
        \prod_{\ell \in [1 .. t]}q_i(z_{i,\ell}^t)
        \int_{\pi_i^{-1}(q_i)}u_i(a_i,p_i)g_{\mu_i}(p_i) {J_{B_i}}^{-1/2}
        \diff\lambda^{|Y_i|-|Z_i|}(p_i)
        \diff\lambda^{|Z_i|-1}(q_i).
    \end{align*}
    Define 
    \[
        u_i(a_i,q_i):=\int_{\pi_i^{-1}(q_i)}u_i(a_i,p_i)g_{\mu_i}(p_i)
        \diff\lambda^{|Y_i|-|Z_i|}(p_i){J_{B_i}}^{-1/2}.
    \]
    We will now prove the following:
    \begin{lemma}
        $u_i(a_i,q_i)$ is continuous in $q_i$.
    \end{lemma}
    \vspace*{-1em}
    \begin{proof}
        We first show that $\pi_i^{-1}$ is a continuous correspondence.
        
        Let $K(P_i)$ denote the set of nonempty, compact, and convex subsets of $P_i$.
        Take any $q_{i,n} \to q_i$ and $p_{i,n} \in \pi_i^{-1}(q_{i,n})$ converging to $p_i$ and note that $B_ip_i = \lim_n B_ip_{i,n}=\lim_n q_{i,n}=q_i$, and thus $\pi_i^{-1}$ is upper-hemicontinuous (uhc).
        
        To show that it is lower-hemicontinuous (lhc) take any open set $U\subseteq P_i$ such that $U\cap \pi_i^{-1}(q_i)\ne \emptyset$.
        This implies that there is $p_i \in \interior(U\cap \pi_i^{-1}(q_i))$ and $\epsilon>0$ such that $B_\epsilon(p_i)\subset U\cap \pi_i^{-1}(q_i)$.
        As $\pi_i$ is a linear mapping, then by the open mapping theorem \citep[Theorem 2.11]{Rudin1973Book}, $\pi_i(B_\epsilon(p_i))$ is open and, therefore, $\exists \delta>0$ such that $B_\delta(q_i)\subseteq \pi_i(B_\epsilon(p_i))$.
        Consequently, $\forall q_i' \in B_\delta(q_i)$, $\emptyset \ne \pi_i^{-1}(q_i')\cap B_\epsilon(p_i) \subseteq \pi_i^{-1}(q_i')\cap U$.
        
        We highlight that $\pi_i^{-1}$ is not only continuous, but also compact- and convex-valued correspondence, from $Q_i$ to $P_i$.
        When restricted to $\pi_i(P_i)$, it is also nonempty, and thus 
        then $\pi_i^{-1}:\pi_i(P_i) \to K(P_i)$ is continuous with respect to the Hausdorff metric \citep[see][Theorem 7.15]{AliprantisBorder2006Book}.
        
        Let $h:P_i \to \pi_i^{-1}(q_i)$ be such that
        $h(p_i):= \argmin_{p_i' \in \pi_i^{-1}(q_i)}\|p_i-p_i'\|_\infty$.
        By continuity of $\pi_i^{-1}$, for any $\epsilon$, there is an $N$ such that $\forall n\geq N$, $\pi_i^{-1}(q_{i,n})\subset B_\epsilon(\pi_i^{-1}(q_i))$.
        Hence, for large $n$, for any point in $\pi_i^{-1}(q_{i,n})$, there is a point in $\pi_i^{-1}(q_{i})$ that is at most $\epsilon$ away.
        As $u(a_i,p_i)g_{\mu_i}(p_i)$ is continuous in $p_i$ and, by Heine-Cantor theorem, uniformly so ($P_i$ is compact).
        Hence, for any $q_i'$ close enough to $q_i$, the difference in the payoff function will be well-approximated by the the difference in measure (up to a constant scaling factor),
        $|u_i(a_i,q_{i}')-u_i(a_i,q_{i})|\approx |\lambda^{|Y_i|-|Z_i|}(\pi_i^{-1}(q_{i}'))-\lambda^{|Y_i|-|Z_i|}(\pi_i^{-1}(q_{i}))|$.
        
        In the sequel, we show that the measure is continuous in $q_i$.
        Take any sequence ${(q_{i,n})}_n \subseteq \pi_i(P_i)$ (a compact set) satisfying $q_{i,n} \to q_i$.
        As $\pi_i^{-1}$ is continuous, and, in particular, upper hemicontinuous, 
        \[\limsup_{n}\lambda^{|Y_i|-|Z_i|}(\pi_i^{-1}(q_{i,n}))\leq \lambda^{|Y_i|-|Z_i|}(\pi_i^{-1}(q_i)),\]
        since for any open set containing $\pi_i^{-1}(q_i)$, it will contain $\pi_i^{-1}(q_{i,n})$ for $q_{i,n}$ sufficiently close to $q_i$.
        
        We now argue that the above inequality holds with equality.
        To see this, fix an arbitrary $\epsilon>0$ and take a collection of points on the boundary of $\pi_i^{-1}(q)$ such that any two points are not closer than $\epsilon/2$ and not farther away than $\epsilon>0$.
        This implies that we have a finite collection of such points.
        As $\pi_i^{-1}$ is lhc, there is a $\delta>0$ such that for every $q_i' \in B_\delta(q_i)$, $\pi_i^{-1}(q_i')$ contains a point in an $\epsilon/4$ neighborhood of every point in our collection, and by convexity, their convex hull.
        This implies that we can approximate arbitrarily well the interior of $\pi^{-1}(q_i)$ taking any $n$ sufficiently large; i.e.
        there is a $\gamma(\epsilon)>0$ such that $\lambda^{|Y_i|-|Z_i|}(\pi_i^{-1}(q_i))-\gamma(\epsilon)\leq \lambda^{|Y_i|-|Z_i|}(\pi_i^{-1}(q_i'))$, with $\gamma(\epsilon)\to 0$ as $\epsilon \to 0$.
        
        Hence, 
        as $q_{i,n}\to q_i$, 
        for any $q_{i,n}\ne q_i$, 
        $\lambda^{|Y_i|-|Z_i|}(\pi_i^{-1}(q_{i,n}))\to 
        \lambda^{|Y_i|-|Z_i|}(\pi_i^{-1}(q_{i}))$.
    \end{proof}
    \vspace*{-1em}
    We can then redefine the problem by considering $\nu_i$ to be uniform on $Q_i$ and take $u_i(a_i,q_i)$ as the utility function.
    
    Redefining 
    \begin{enumerate}[label=\textbullet]
        \item $v_i:\Delta(Q_i)\to \mathbb R$ as $v_i(\nu_i'):=\max_{a_i\in A_i}\mathbb E_{\nu_i}[u_i(a_i,q_i)]$;
        \item $V_i:\Delta(Q_i)\to \mathbb R$ as
        $V_i(\nu_i):=\sup_{\tau' \in \mathbb T_i}\mathbb E_{\nu_i}[v_i(\nu_i\mid z_i^{\tau'}-c_i\tau_i]$;
        \item $\tau_i(\omega):=\inf\{t\mid V_i(\nu_i\mid z_i^t(\omega))=v_i(\nu_i\mid z_i^t(\omega))\}$;
        \item $b_i(p_i):=\mathbb E_{\pi_i(p_i)}[\sigma_i^*(z_i^{\tau_i})]$, for some fixed 
        $z_i^t \mapsto \sigma_i^*(z_i^t)\in \argmax_{\sigma_i \in \Delta(A_i)}\mathbb E_{\nu_i}[u_i(a_i,q_i)\mid z_i^t]$.
    \end{enumerate}
    we obtain---by analogous arguments to \hyref{proposition:properties-value-stopping}[Proposition]---that $V_i$ is continuous and, by \citet{Berk1966AnnMathStat}, that $\nu_{i,t}$ weak$^*$ converges to $B_i p_{i}$, $p_{i}$-a.s., when $z_{i,t}\sim \pi_i(p_i)$, for all $t$.
    Therefore, $\tau_i$ is finite $p_{i}$-a.s., for any $p_i\in P_i$.
    Finally, an analogous version of \hyref{lemma:stopping-implies-continuity}[Lemma] applies and
    $b_i$ is continuous in $p_i$ and maps to $\Delta(A_i)$.
    Hence, by essentially the same arguments as in \hyref{theorem:existence}[Theorem], a sequential sampling equilibrium exists.
\end{proof}

In the above, we restricted to the case in which $|Z_i|<|Y_i|$.
If instead $\text{rank}(B_i)=|Z_i|=|Y_i|$, we have the following:
\begin{proposition}
     Let $G:=\langle \Gamma, \mu, c, B, Z\rangle$ be an extended game such that for every player $i$, $\text{rank}(B_i)=|Z_i|=|A_{-i}|$.
    Then $G$ admits a sequential sampling equilibrium.
\end{proposition}
\vspace*{-1em}
\begin{proof}
    Note that now $B_i$ is invertible and $\pi_i(p_i):=B_i p_i$ is bijective when restricting its domain to $\pi_i(P_i)$.
    Hence, $\pi_i$ admits a continuous inverse (note it is a linear mapping).
    Since for any $p_i \in \Delta(A_{-i})$, and $\mu_i$ has full support, $\pi_i(p_i)$ is in the support of the pushforward measure $\nu_i:=\pi_i\#\mu_i \in \Delta(Q_i)$.
    Thus, for any $p_i$, by \citet{Berk1966AnnMathStat}, $\nu_{i,t}$ weak$^*$ converges to a Dirac on $\pi_i(p_i)$.
    Moreover, if $\nu_i\in \Delta(Q_i)$ is the pushforward measure given $\mu_i\in \Delta(P_i)$ and $\pi_i:P_i\to Q_i$, we now also have that $\mu_i$ is $\nu_i$'s pushforward measure given $\pi_i^{-1}:Q_i\to P_i$.
    Then, weak$^*$convergence of $\nu_{i,t}$ to $\delta_{q_i}$ implies weak$^*$ convergence of the $\mu_{i,t}$ to $\delta_{\pi_i^{-1}(q_i)}$.
    Uniform continuity of $V_i:\Delta(P_i)\to \mathbb R$ (as originally defined on the main text) delivers existence of an equilibrium as in \hyref{theorem:existence}[Theorem].
\end{proof}

Naturally, all the above can also be extended to Bayesian games.

\subsection{Relation to Analogy-Based Expectation Equilibrium}
Finally, we discuss convergence to analogy-based expectation equilibria---see \citet{Jehiel2021WP} for a survey.
In line with the literature, we consider payoffs that are linear in the distribution of actions.

This solution concept when applied to normal-form games (including Bayesian games) can be readily translate to our setup:
$\sigma$ is an analogy-based expectation equilibrium if, for each player $i\in I$,
\begin{enumerate}[label=(\arabic*)]
    \item $q_i(z_i)=\sum_{a_{-i} \in f_i^{-1}(z_{i})}\sigma_{-i}(a_{-i})$ for every $z_i \in Z_i$;
    \item $u_i(a_i,z_i):=\sum_{a_{-i} \in f_i^{-1}(z_{i})}\frac{1}{|f_i^{-1}(z_{i})|}u_i(a_i,a_{-i})$; and
    \item $\sigma_i \in \argmax_{\sigma_i' \in \Delta(A_i)} \mathbb E_{q_i}u_i(a_i,z_i)$.
\end{enumerate}
Recalling that $f_i$ is a surjective mapping from $A_{-i}$ to $Z_i$, condition (1) states that each player $i$ bundles difference action profiles (or players, or types, contingencies) $a_{-i}$ into the same `analogy class' $z_i$.
Condition (2) can be read as a simplification device by player $i$: the player cannot distinguish across the different $a_{-i}$ within the same analogy class $z_i$, they consider the average behavior, as if the probability of each $a_{-i}$ within the same analogy class were the same.
Then (condition (3)), they best respond to the expected payoff given the actual distribution over analogy classes, but assuming that, within the analogy class, distribution over contingencies is uniform.

In the above setup, this is achieved whenever $\mu_i$ is uniform.
The result follows from arguments analogous to those in the proof of \hyref{theorem:convergence-ne}[Theorem], but with a crucial simplification: 
the linearity of payoffs in distributions, the posterior means pin-down the set of best responses.
And, upon stopping, $\mathbb E_{\mu_i}[q_i'\mid z_i^{\tau_i}]=\frac{\tau_i}{\tau_i+|A_{-i}|}\overline{z_i}^{\tau_i}+\frac{1}{\tau_i+|A_{-i}|}$.
Since, for any player $i$ for which no action is always a best response, there is a lower bound to the stopping time $\tau_i\geq T^n$, that grows unboundedly as sampling costs vanish, $c_i^n\to 0$ (\hyref{lemma:lower-bound-stopping}[Lemma]), 
by a similar application of the optional stopping theorem as in the proof of \hyref{theorem:convergence-ne}[Theorem],
$\mathbb E_{q_i}\left[\left\|\mathbb E_{\mu_i}[q_i'\mid z_i^{\tau_i^m}]-q_i\right\|\right]
\leq 
\mathbb E_{q_i}\left[\left\|\mathbb E_{\mu_i}[q_i'\mid z_i^{T^m}]-q_i\right\|\right]
\leq
\frac{T^m}{T^m+|A_{-i}|}\mathbb E_{q_i}[\|\overline{z_i}^{T^m}-q_i\|]+
2\frac{1}{T^m+|A_{-i}|}
$.
Then, by the law of large numbers, $\mathbb E_{q_i}[\|\overline{z_i}^{T^m}-q_i\|]\to 0$.
This provides a shorter route to show that players' posterior means converge to the underlying true distribution.

\section{Experimental Data and Analysis Details}
\label{appendix:experimental-analysis}

In this appendix, we provide details on the experimental data used along with additional analysis.
A total of 164 subjects were recruited for sessions run in the Columbia Experimental Laboratory in the Social Sciences (CELSS) to play matching pennies games as the one depicted in \hyref{figure:matching-pennies-matrix}[Figure]. 
Subjects are randomly and anonymously matched, but their roles are fixed throughout.
Player $M$'s payoff to action $a$, $\delta_M$, took one of six values (here rescaled by a factor of 20 for convenience): 4, 2, 1/2, 1/4, 1/10, and 1/20.

The experiment consisted of two stages.
In the first stage, actions elicited and each game is played twice.
In the second stage, each game is played 5 times and either both actions and beliefs about the probability that the opponent chooses action $a$ are elicited or only actions are elicited.
Beliefs here refer to point estimates reported by the subjects, neglecting any strategic uncertainty.
In other words, belief reports would correspond to posterior means in our framework.
Elicitation of actions and beliefs is incentive-compatible and robust to risk attitudes and game payoffs correspond to probability points towards prizes of \$10.
Throughout, no feedback was provided, game order was randomized and, importantly for our purposes, decision times are recorded.
Other details on the experimental design can be found in \citet{FriedmanWard2022WP}.

There are some important caveats to note.
First, beliefs elicited in the second stage refer to opponent's actions from the first stage.
This, together with the fact that elicitation of actions and beliefs is sequential instead of simultaneous, with beliefs being elicited first, may raise concerns of whether reported beliefs are a good proxy for the beliefs that subjects hold when taking an action.
Second, while decision time was recorded, subjects were forced to wait a minimum of 10 seconds before reporting their beliefs.
As the subjects' decision times will be used as a proxy to test sequential sampling equilibrium's predictions for stopping times, the forced minimum decision time may undermine the exercise.
Finally, the authors highlight there being evidence of ``no-feedback learning'' as the same subject plays the same game multiple times.
This is especially worrying when comparing instances where only actions are elicited with those where both actions and beliefs are.
In order to avoid issues due to experience or learning, and focus on initial response as much as possible, we restrict attention to choice data when beliefs are not elicited.

\begin{table}[!th]\setstretch{1.1}
	\centering
	\small
	\hspace*{-1em}\begin{tabular}{l@{\extracolsep{4pt}}cccc@{}}
\hline\hline
&\multicolumn{2}{c}{$\{$Player $M$ chooses $a$$\}$}  &  \multicolumn{2}{c}{$\{$Player $C$ chooses $a$$\}$} \\
\cline{2-3} \cline{4-5}
               &           OLS &          Logit &            OLS &          Logit \\
& (1) & (2) & (3) & (4) \\
\hline
    $\delta_M$ & 0.230$^{***}$ &  0.949$^{***}$ & -0.772$^{***}$ & -3.430$^{***}$ \\
               &       (0.041) &        (0.169) &        (0.036) &        (0.197) \\ [.25em]
     Intercept & 0.329$^{***}$ & -0.702$^{***}$ &  0.842$^{***}$ &  1.522$^{***}$ \\
               &       (0.018) &        (0.079) &        (0.017) &        (0.090) \\

\hline
(Pseudo) R$^2$ &          0.02 &           0.01 &           0.20 &           0.15 \\
  Observations &        1782 &           1782 &         1806 &           1806 \\
\hline\hline
\multicolumn{4}{l}{\footnotesize Heteroskedasticity-robust standard errors in parentheses. }\\
\multicolumn{4}{l}{\footnotesize $^{*}$ \(p<0.1\), $^{**}$ \(p<0.05\), $^{***}$ \(p<0.01\).}\\
\end{tabular}
    \begin{minipage}{1\textwidth}
        \small
        \vspace*{.5em}
        \caption{Payoffs and Choices: Own- and Opponent-Payoff Choice Effect} 
        \label{table:table-payoffs-actions}
        \vspace*{-1em}
        \singlespacing \emph{Notes}: 
        The table exhibits the association between player $M$'s payoff to action $a$ and the frequency with which subjects in each role choose action $a$.
        $\delta_M$ parametrizes player $M$'s payoff to action $a$.
        The games in question are generalized matching pennies games as given in \hyref{figure:matching-pennies-matrix}[Figure], for $\gamma_C=1$ (and scaled by 20).
        The sample includes data from rounds in which there is no belief elicitation.
        The data is from \citet{FriedmanWard2022WP}.
    \end{minipage}
\end{table}

\begin{table}[!h]\setstretch{1.1}
	\centering
	\small
	\hspace*{-1em}\begin{tabular}{@{\extracolsep{4pt}}ccc@{}}
\hline\hline
\multicolumn{3}{c}{Player $C$ Beliefs} \\
\multicolumn{3}{c}{$\sigma_M^{\tau_C} \mid \delta_M^{High} \geq_{FOSD} \sigma_M^{\tau_C} \mid \delta_M^{Low}$}\\
High & Low & KS-Statistic\\
\cline{1-1} \cline{2-2} \cline{3-3}
 (1)  &  (2)  & (3)\\
\hline
4 & 2 & 0.33$^{***}$\\
2 & 1/2 & 0.76$^{***}$\\
1/2 & 1/4 & 0.40$^{***}$\\
1/4 & 1/10 & 0.23$^{***}$\\
1/10 & 1/20 & 0.23$^{***}$\\
\hline\hline
\multicolumn{3}{l}{\footnotesize $^{*}$ \(p<0.1\), $^{**}$ \(p<0.05\), $^{***}$ \(p<0.01\).}\\
\end{tabular}
    \begin{minipage}{1\textwidth}
        \small
        \vspace*{.5em}
        \caption{Opponent Payoff and Beliefs: FOSD Tests}
        \label{table:table-clasher-beliefs-fosd}
        \vspace*{-1em}
        \singlespacing \emph{Notes}: 
        The table exhibits the results of two-sample Kolmogorov-Smirnov tests for first-order stochastic dominance of the distribution of reported beliefs by subjects in the role of player $C$ in games with different values of $\delta_M$, High (col (1)) and Low (col (2)).
        Column (3) presents the test statistic, with number of observations $(n,m)=(280,280)$.
        $\delta_M$ parametrizes player $M$'s payoff to action $a$.
        The games in question are generalized matching pennies games as given in \hyref{figure:matching-pennies-matrix}[Figure], for $\gamma_C=1$ (and scaled by 20).
        The data is from \citet{FriedmanWard2022WP}.
    \end{minipage}
\end{table}

\hyref{table:table-payoffs-actions}[Table] documents the own- and opponent-payoff choice effects mentioned on \hyref{sec:implications:comparative-statics-actions}[Section]: as player $M$'s payoffs to action $a$ increase, subjects in that tend to choose the action more often and action $b$ less often, while the opposite is true for subjects in the opponent's role, player $C$.
In the main text, \hyref{table:table-time-distance-indifference}[Table] provides support for the collapsing boundaries result presented in \hyref{sec:implications:time-revealed-intensity}[Section]: a negative association between the distance of subjects' reported beliefs to their indifference point and the decision time.
Related to this result, the main text discussed a first-order stochastic dominance shift of beliefs for player $C$ as $\delta_M$ (player $M$'s payoffs to action $a$) increases.
\hyref{figure:figure-fosd-beliefs-clasher}[Figure] exhibits such lawful relation; appropriate statistical tests are now given in \hyref{table:table-clasher-beliefs-fosd}[Table].

\section{Misspecified Priors}
\label{appendix:misspecification}

As in finite dimensional spaces, the Bayesian learning is consistent for any distribution if and only if the prior has full support \citep{Freedman1963AnnMathStat}, \hyref{remark:bounded-stopping}[Proposition] uncovers an important consequence of Bayesian learning for optimal stopping: Not only is the decision-makers' optimal stopping time finite with probability one, for any true distribution of their samples, it is also bounded uniformly across all distributions of samples.
This effectively transforms the optimal stopping problem from infinite to finite horizon, allowing for a solution to be obtained by backward recursion, simplifying the problem significantly.

The intuition underlying the result is that if the prior has full support, the posterior accumulates around the empirical mean.
Then, one can guarantee a bound on the rate at which the posterior accumulates around the empirical mean, depending on the number of observations but not on the sample path itself \citep{DiaconisFreedman1990AnnStat}.
With this, it is possible to bound the gains in expected payoff of sampling further regardless of the realized sample path and show that there is a number of observations after which the cost of an additional observation dwarfs the expected gain, regardless of realizations.
Hence, one concludes that the decision-maker necessarily stops after such number of samples and we can find an explicit upper bound for the stopping time that depends only on the prior $\mu_i$, payoffs $u_i$, and sampling cost $c_i$.

This stands in contrast to the canonical problem in \citet{ArrowBlackwellGirshick1949Ecta} where the prior has finite support, and optimal stopping time is not bounded.\footnote{
    Similarly, optimal stopping time is also not bounded in the continuous-time version of the canonical problem, with Gaussian noise, be it with \citep{MoscariniSmith2001Ecta} or without experimentation concerns \citep{Chernoff1961}.
    In some cases with finite support prior, however, stopping time can be bounded, as in the case with Poisson arrival of conclusive information, but not when the decision-maker can choose from different information sources \citep{CheMierendorff2019AER}.
}
Further, it stands in sharp contrast to the case in which beliefs are misspecified.

We now provide an example in which misspecification leads to a player never stopping with probability 1 (with respect to the true distribution of opponents' actions), and sequential sampling equilibrium fails to exist.

Let $\Gamma$ be a two-player game in which player $i$'s opponent has three possible actions, $a$, $b$, and $c$, and always chooses $c$ (e.g. because $c$ is dominant, or because their sampling cost is too high and $c$ is uniquely optimal under their prior).
Denote $\sigma_{-i}=(\sigma_{-i}(a),\sigma_{-i}(b),\sigma_{-i}(c))\in \Delta(\{a,b,c\})$.
Suppose player $i$'s prior beliefs about $\sigma_{-i}$, $\mu_i$, are such that player $i$ assigns probability 1/2 to $(1/2,1/6,1/3)$ and probability 1/2 to $(1/6,1/2,1/3)$.
Player $i$ can choose either $a$ or $b$ and player $i$'s payoffs are given by $u_i(a_i,a_{-i})=1$ if $a_i=a_{-i}$, and $0$ if otherwise.
Then, if $y_i^t$ is such that $y_{i,\ell}=c$ for all $\ell \in [1\,..\,t]$, $\mu_i\mid y_i^t=\mu_i$.
Under their prior, $v_i(\mu_i)=1/4$, 
$v_i(\mu_i\mid a)=v_i(\mu_i\mid b)=3/4$, and $v_i(\mu_i\mid c)=v_i(\mu_i)=1/4$, hence
$\mathbb E_{\mu_i}[v_i(\mu_i\mid y_i)]=\frac{2}{3}\frac{3}{4}+\frac{1}{3}\frac{1}{4}=\frac{7}{12}$.
Note that, a necessary condition for player $i$ to stop is that $\mathbb E_{\mu_i}[v_i(\mu_i\mid y_i)]-v_i(\mu_i)\leq c_i$. 
But, since at any sequential sampling equilibrium player $i$'s opponent chooses $c$ with probability 1, we have that $\mu_i\mid y_i^t=\mu_i$ and, for any $c_i<1/3$, 
we always obtain 
$\mathbb E_{\mu_i}[v_i(\mu_i\mid y_i)]-v_i(\mu_i)=\frac{1}{3}> c_i$.
Therefore, since $\sigma_{-i}(c)=1$, $\mathbb P_{\sigma_{-i}}(\tau_i=\infty)=1$.

\end{document}